\RequirePackage{fix-cm}
\documentclass[smallextended]{svjour3}       
\smartqed  
\usepackage{graphicx}
\usepackage{natbib}
\usepackage[colorlinks=true,citecolor=blue,urlcolor=blue,breaklinks]{hyperref}
\def\araa{ARA\&A}             
\def\apj{{ApJ\ }}                 
\def\apjl{{ApJ\ }}                
\def\apjs{{ApJS\ }}               
\def\aap{{A\&A\ }}                
\def\aapr{{A\&A~Rev\ }}          
\def\aaps{{A\&AS\ }}              
\def\pasj{{PASJ\ }}               
\def\solphys{{Sol~Phys\ }}      
\def\ssr{{Space~Sci~Rev\ }}     
\def\nat{{Nature}}              
\def\grl{{Geophys~Res~Lett\ }} 
\def\jgr{{J~Geophys~Res\ }}    

\journalname{Living Reviews in Solar Physics}
\begin{document}
 
\title{Space weather: the solar perspective}

\subtitle{An update to Schwenn (2006)}


\author{Manuela Temmer}


\institute{M. Temmer \at
              Institute of Physics, University of Graz \\
              Tel.: +43 316-380-8610\\
              Fax: ++43 316-380-7255\\
              \email{manuela.temmer@uni-graz.at}           
           }

\date{Received: date / Accepted: date}

\maketitle

\begin{abstract}
The Sun, as an active star, is the driver of energetic phenomena that structure interplanetary space and affect planetary atmospheres. The effects of Space Weather on Earth and the solar system is of increasing importance as human spaceflight is preparing for lunar and Mars missions. This review is focusing on the solar perspective of the Space Weather relevant phenomena, coronal mass ejections (CMEs), flares, solar energetic particles (SEPs), and solar wind stream interaction regions (SIR). With the advent of the STEREO mission (launched in 2006), literally, new perspectives were provided that enabled for the first time to study coronal structures and the evolution of activity phenomena in three dimensions. New imaging capabilities, covering the entire Sun-Earth distance range, allowed to seamlessly connect CMEs and their interplanetary counterparts measured in-situ (so called ICMEs). This vastly increased our knowledge and understanding of the dynamics of interplanetary space due to solar activity and fostered the development of Space Weather forecasting models. Moreover, we are facing challenging times gathering new data from two extraordinary missions, NASA's Parker Solar Probe (launched in 2018) and ESA's Solar Orbiter (launched in 2020), that will in the near future provide more detailed insight into the solar wind evolution and image CMEs from view points never approached before. The current review builds upon the Living Reviews paper by Schwenn from 2006, updating on the Space Weather relevant CME-flare-SEP phenomena from the solar perspective, as observed from multiple viewpoints and their concomitant solar surface signatures.

\end{abstract}

\setcounter{tocdepth}{3}
\tableofcontents

\section{Introduction}\label{intro}
\begin{figure}
  \includegraphics[width=\textwidth]{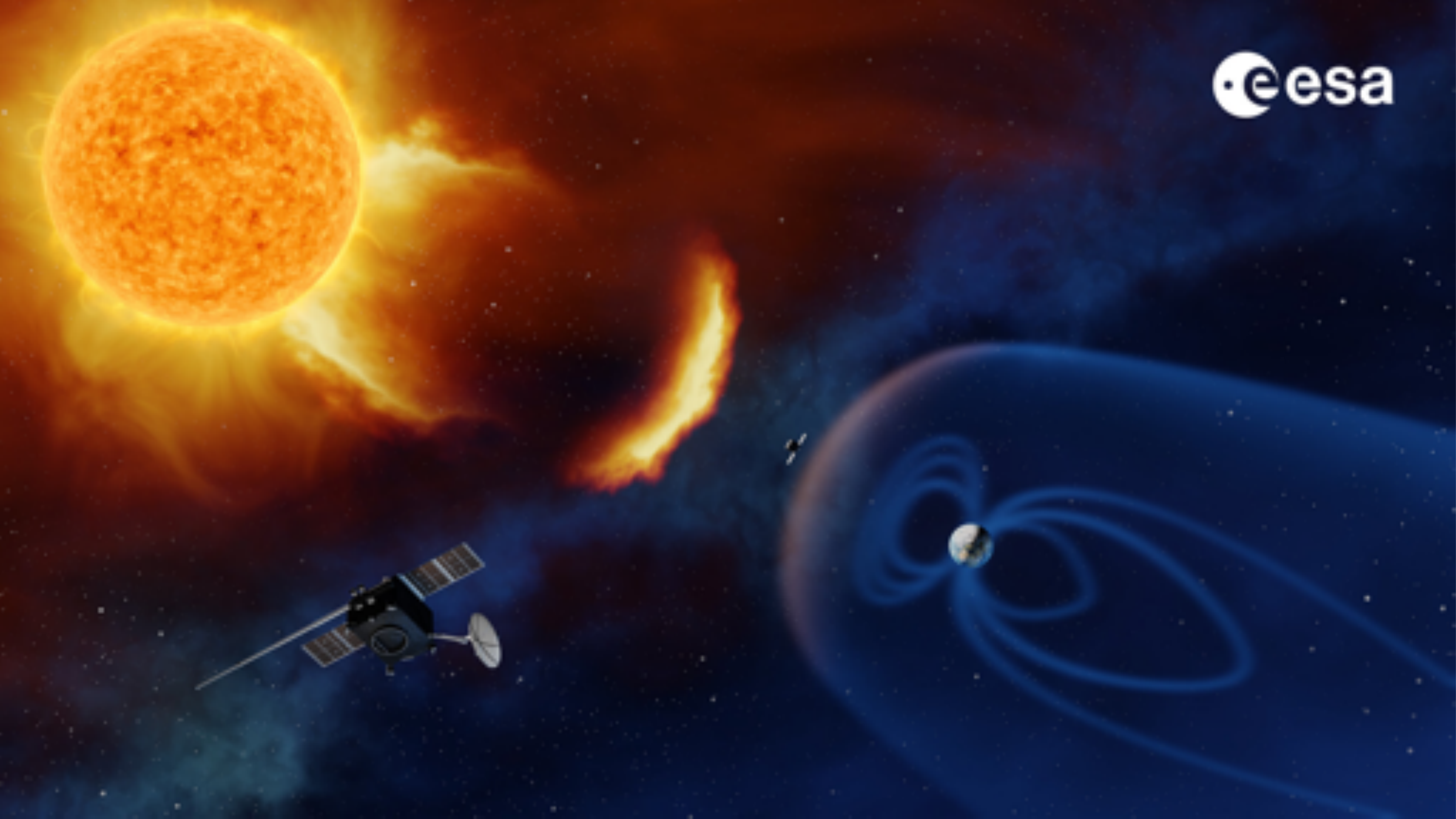}
\caption{Solar activity phenomena (depicted here as CME) affect Earth and near-Earth space and therefore, need to be permanently monitored. Space Weather forecasting is of global interest and funded by international agencies. In the near future, satellites will observe the Sun and its dynamic phenomena from different viewpoints, such as a combined L1 and L5 position. Image courtesy: ESA}
\label{fig:esa}       
\end{figure}

Our Sun is an active star and as such undergoes cyclic variations, which are related to more or less frequently occurring activity phenomena observed at the solar surface. High energetic activity phenomena, produced due to changes in the Sun's magnetic field, propagate through our solar system where they interact with the planet's atmospheres. At Earth, these interactions are well documented and known to cause geomagnetic disturbances having consequences for modern society. The influence by the Sun on our solar system is termed \textit{Space Weather}. Therefore, solar activity needs to be permanently monitored from space and ground in order to assess times of increased influence. International space agencies created programs, such as ESA \textit{Space Situational Awareness} (SSA) or NASA \textit{Living with a star} (LWS) (cf. Figure~\ref{fig:esa}), to enhance Space Weather awareness and with that support and fund on a long-term basis fundamental research and development of Space Weather forecasting tools. 

This review article focuses on the following Space Weather phenomena: 
\begin{enumerate}
    \item Coronal mass ejections
    \item Flares
    \item Solar Energetic Particles  
    \item Solar wind stream interaction regions  
\end{enumerate}

To properly describe these phenomena from the solar perspective, a number of processes need to be understood, such as active region and magnetic field evolution, energy build-up and release, as well as the global structuring of inner heliospheric space. Space Weather is a topic of broad interest and sustains an exciting and wealthy interdisciplinary research community\footnote{For example, the SCOSTEP effort that resulted in excellent publications via CAWSES \url{http://www.terrapub.co.jp/onlineproceedings/ste/CAWSES2007/index.html}, the VarSITI programs \cite[e.g., ISEST][see \url{http://www.varsiti.org}]{zhang18} or the international Space Weather Action Teams, iSWAT, where interdisciplinary groups gather together under \url{https://www.iswat-cospar.org}.}. With that it fosters information and knowledge exchange between international research groups on solar-, heliospheric- and geo-space (Sun-to-impact disciplines) in order to enhance scientific knowledge for improving existing and developing new models for Space Weather forecasting.

Coronal mass ejections (CMEs) are a rather recent phenomenon, discovered just about 50 years ago, but in the meantime are known as the main drivers of the most severe Space Weather disturbances \citep[see e.g.,][]{howard06,Gopalswamy16_CME_history}. They are huge structures that manifest themselves within some tens of minutes as clouds of magnetized plasma impulsively expelled from the Sun and subsequently propagating into interplanetary space \citep[see e.g.,][]{forbes00}. CMEs arise from usually complex and closed magnetic field structures in equilibrium that is disrupted due to some instability causing its eruption \cite[e.g., emerging magnetic flux, remote reconfiguration of large scale magnetic field, or field rotation; see e.g.,][]{torok13,schmieder15,green18}. Instabilities in the solar magnetic field and their occurrence frequency are modulated by the 11-year activity cycle of the Sun. The most strong CME events may propagate the 1AU distance within a day \citep[e.g.,][]{cliver90,gopalswamy05_extreme,liu14_nature}. Less strong events, on average, propagate the same distance in up to 4 days \citep[see e.g.,][]{shanmugaraju14}. CMEs may be linked to large geomagnetic disturbances, due to shock compression and reconnection with the Earth’s magnetic field. They may lead to ionospheric and geomagnetically-induced currents \citep[see e.g.,][]{pirjola05}.  Usually the most severe geomagnetic storms are caused by fast and massive CMEs, erupting from the central region of the visible solar disk and carrying a strong southward magnetic field component that reconnects with the Earth’s magnetic field \citep[see e.g.,][]{pulkkinen07}. Consequently, CMEs are a major topic of solar and Space Weather research.

The power for making a CME energetic (i.e., being fast and wide) undoubtedly stems from the free magnetic energy which is released as consequence of magnetic reconnection processes. Magnetic reconnection enables to impulsively drive plasma and to accelerate particles to high energies causing on the one hand flare emission, which is observed in the solar atmosphere, and on the other hand solar energetic particles (SEPs), which are measured in interplanetary space. Energetic particles from strong SEP events may reach almost speed of light and travel the 1AU distance within about 10 minutes. High energy SEP events (about 1 GeV) may lead to enhanced proton fluxes even at ground level. Hence, most intense events can endanger life and technology on Earth and in space. Further consequences of CMEs and SEPs are disruptions of satellite operations, radio communications and ground power systems \citep[e.g.,][]{bothmer07}. Unlike CMEs, having lead times of some tens of hours between first observational signatures and impact at Earth, flares and SEP events occur and impact almost simultaneously \citep[see e.g.,][]{lugaz17,cairns18,Malandraki18_book}. Accordingly, to predict the occurrence of flares and SEPs one needs to predict the instabilities leading to the onset of magnetic reconnection processes, one of the big challenges in solar physics.

The continuous solar wind flow in a quiet state (usually termed background solar wind) is represented by an alternation of slow and fast solar wind streams that interact and form stream interaction regions (SIRs). If steady in their existence and persisting over more than one solar rotation, they are called co-rotating interaction regions (CIRs). During times of low solar activity, Space Weather is dominated by CIR induced storms \citep{tsurutani06}. Different flow speeds of the background solar wind also change the propagation behavior of CMEs in interplanetary space. This has consequences on the CME transit time and impact speed at planetary atmospheres \citep[drag force; see][]{gopalswamy00_accelCME,vrsnak01_solph,Cargill2004OnEjections,vrsnak06}. Moreover, CMEs disrupt the continuous outflow of the solar wind and reconfigure the magnetic field on large spatial and short temporal scales altering the background solar wind. For Space Weather and CME modeling/forecasting purposes, these ever changing conditions in interplanetary space are very challenging to tackle.  

 \begin{figure}
  \includegraphics[width=\textwidth]{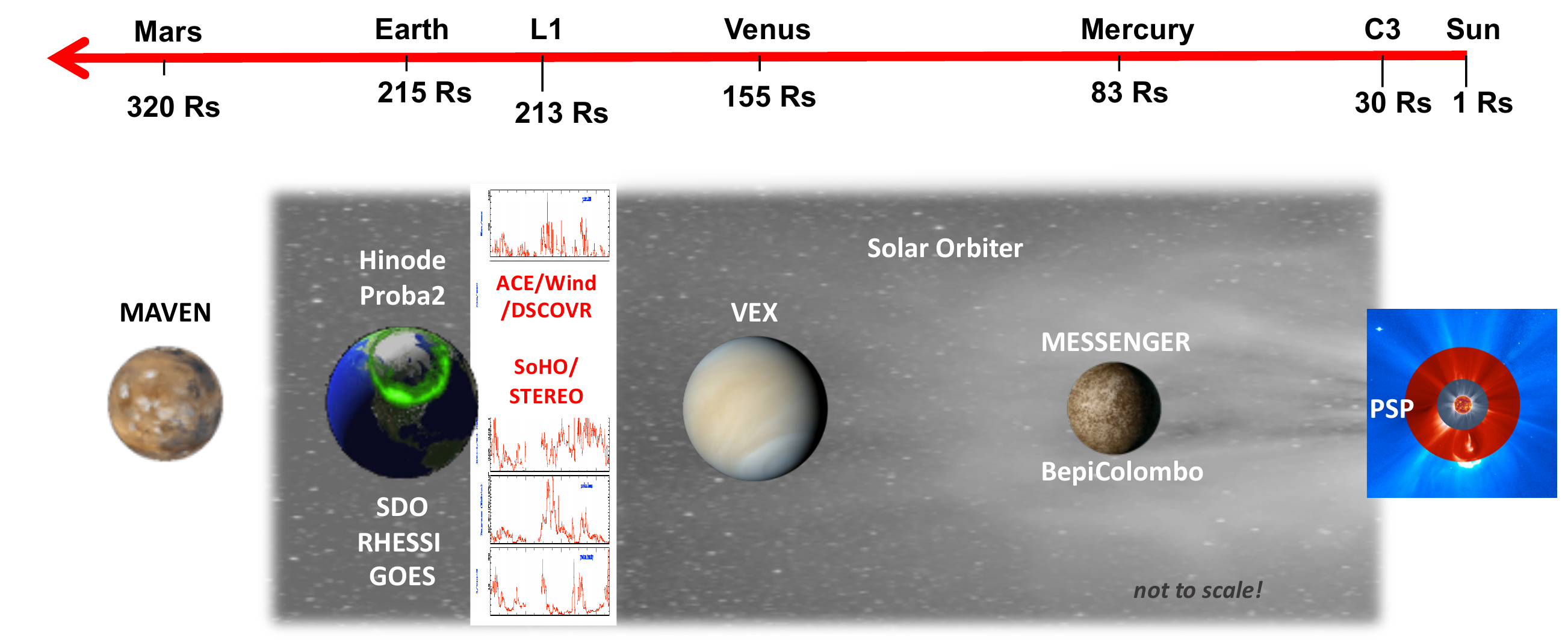}
\caption{Current and past space missions carrying instruments for gathering remote sensing image data and in-situ plasma and magnetic field measurements. The majority of spacecraft is located in the ecliptic plane orbiting planets or at the Lagrangian point L1. The coronagraph field of view of the SoHO/LASCO instrument C3 covers 30 solar radii. The background white-light image is taken from STEREO/HI1+2 data covering about 90 degrees in the ecliptic. Not to scale.
}
\label{fig:insitu}       
\end{figure}

For comprehensive investigations a rich source of observational data is currently available from many different instruments located at multiple viewpoints and different radial distances (see Figure~\ref{fig:insitu}). In Earth orbit current operational missions are e.g., GOES (Geostationary Operational Environmental Satellite), SDO \citep[SDO:][]{Pesnell2012TheSDO}, Proba-2 \citep{santandrea13}, located at L1 - 1.5 million km upstream of Earth - there is the Solar and Heliospheric Observatory \citep[SoHO:][]{Domingo95_SOHO}, the Advanced Composition Explorer \cite[ACE:][]{Stone98_ACE}, the WIND spacecraft \cite{ogilvie95}, and DSCOVR \cite[][]{burt12}. At $\sim$1AU with variable longitudinal angles from Earth, there is the Solar TErrestrial RElations Observatory \cite[STEREO:][]{kaiser08_STEREO} consisting of two identical spacecraft named STEREO-Ahead and STEREO-Behind (lost signal end of 2014). The combination of remote sensing image data and in-situ measurements is found to be optimal for enhancing our knowledge about the physics of Space Weather phenomena. For better understanding large eruptive activity phenomena, multi-viewpoint and multi-wavelength data are exploited (e.g., combined L1, STEREO as well as ground-based instruments). The various available data from spacecraft orbiting around planets (e.g., VEX (2006--2014), MESSENGER (2011--2015), MAVEN (2014--), BepiColombo (2018--)) also enable to analyze the evolution of Space Weather phenomena as function of distance and longitude.

\begin{figure}
  \includegraphics[width=\textwidth]{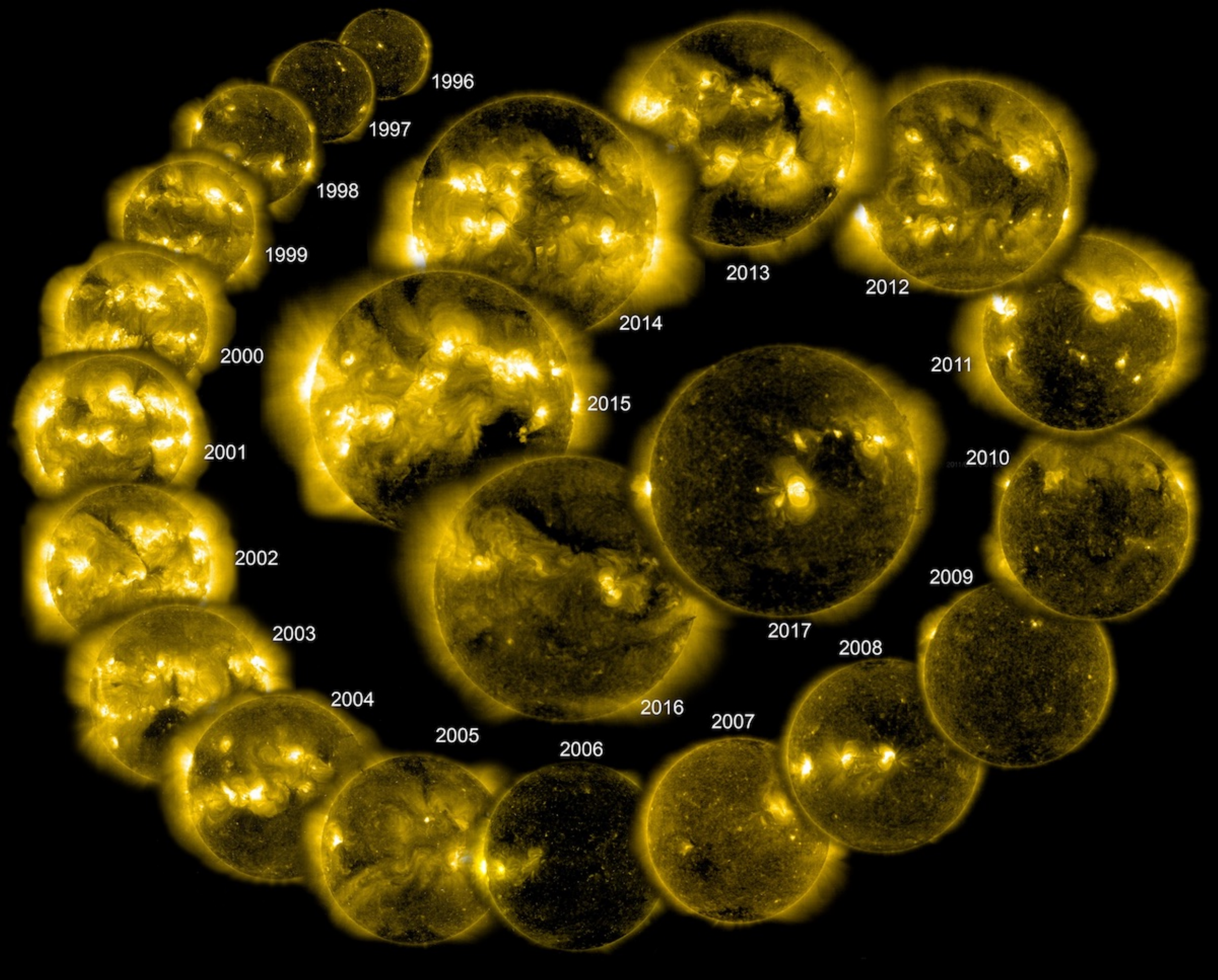}
\caption{Each image shown here is a snapshot of the Sun taken every spring with the SOHO Extreme ultraviolet Imaging Telescope (EIT) in the 284\AA\ wavelength range. It shows the variations of the solar activity in terms of increasing and decreasing number of bright active regions visible in the corona. Courtesy: NASA/ESA}
\label{fig:soho22}       
\end{figure}

A flagship of international collaboration and boost for Space Weather research, is SoHO which now achieved 25 years in space. Figure~\ref{fig:soho22} shows SOHO/EIT \citep{delaboudiniere95} EUV image data covering the variations of the solar corona over a full magnetic solar cycle (Hale cycle). Long-term observations are of utmost importance for monitoring and learning about the interaction processes of solar activity phenomena with Earth and other planets as well as for improving our capabilities in Space Weather forecasting. Most recent and unprecedented missions are Parker Solar Probe, launched in August 2018 \citep{fox16}, and Solar Orbiter, launched in February 2020 \citep{mueller20} having on-board imaging and in-situ facilities with the goal to approach the Sun as close as never before ($\sim$0.05AU and $\sim$0.3AU) and investigating the Sun out of the ecliptic ($\sim$30 degrees). To support space missions and for providing valuable complementary data, we must not forget the importance of ground based observatories that observe the Sun over broad wavelength and energy ranges allied in international networks such as the Global high-resolution H$\alpha$ network\footnote{\url{http://bbso.njit.edu/Research/Halpha/}}, the Global Oscillation Network Group\footnote{\url{https://gong.nso.edu}},  the database for high-resolution Neutron Monitor measurements\footnote{E.g., \url{http://www01.nmdb.eu/nest/search.php}}, muon telescope networks, or the Worldwide Interplanetary Scintillation Stations Network \footnote{\url{http://helios.mexart.unam.mx/pruebas/wipss/index.html}}.

From the derived research results based on the observational data, over the past years a plethora of models could be developed for predicting Space Weather and their geomagnetic effects. The permanent monitoring of the Sun and provision of data in almost real-time enabled to apply those results and even to install operational services that produce forecasts mostly in an automatic manner (e.g., facilitated by ESA/SSA\footnote{\url{http://swe.ssa.esa.int}}; NASA/CCMC\footnote{\url{https://ccmc.gsfc.nasa.gov}}; NOAA/SWPC\footnote{\url{https://www.swpc.noaa.gov}}). However, the operational services also clearly demonstrated the limitations in the forecasting accuracy as on average the errors are large and get worse with increasing solar activity. This is mainly due to the large uncertainties coming from the model input, namely observational parameters at or close to the Sun. It also reveals the complexity of the interplay between the different driving agents of Space Weather, that makes it difficult to fully capture the physics behind and to improve models. Reliable Space Weather forecasting is still in its infancy.

\section{Space Weather}

From the historical perspective, the so-called ``Carrington-event'' from September 1, 1859 is the reference event for referring to extreme Space Weather and with that the beginning of Space Weather research \citep[see also][]{Schwenn06}. At that time only optical observations of the solar surface were performed and the observed emitted radiation in white-light for that event showed impressively the vast amount of energy that was distributed to the dense lower atmospheric layers of the Sun where it heated the photosphere. At Earth, the associated geomagnetic effects were observed in terms of aurora occurring from high to low latitudes (e.g., Honolulu at 20 degrees northern latitude) and ground-induced currents in telegraph wires \citep[see][]{Eastwood17}. The associated SEP event is thought to be about twice as large as the huge SEP events from July 1959, November 1960, or August 1972 \citep{cliver13}. Only several years after the Carrington event, the usage of spectroscopes enabled to regularly observe prominence eruptions revealing the dynamic changes of the solar corona and material ejections with speeds exceeding hundreds of km/s \citep{tandberg95}. The continuous monitoring of the Sun was intensified in the 1940’s, when solar observations in radio, white-light and in the H$\alpha$ wavelength range were performed. At that time also galactic cosmic rays were studied and found that they are anti-correlated with solar activity \citep[so-called Forbush decrease, measured as sudden drop in the cosmic ray flux due to interplanetary disturbances; see also][]{cane00}. In the early 1960’s magnetic structures driving shocks were inferred from observations in the metric radio observations and geomagnetic storm sudden commencements \citep{gold62,fokker63}. The \textit{transient events with mass moving through the solar corona and actually leaving the Sun}, i.e., CMEs, that were associated with the prominence/filament eruptions were discovered only in the early 1970's with the advent of the space era \citep[see][]{Tousey71,macqueen74}. Recent reviews on the history of prominences and their role in Space Weather can be found in \cite[][and references therein]{vial15,Gopalswamy16_CME_history}. While most of the extreme space weather events happen during the solar cycle maximum phase, occasionally strong geoeffective events may occur close to the solar cycle minima and also during weak solar activity, provided there are appropriate source regions on the Sun \citep[see also e.g.,][]{Vennerstrom16,hayakawa20}. For more details about the solar cycle see the Living Reviews by \cite{hathaway10}.

Nowadays, a wealth of space and ground-based instruments are available, delivering valuable observational data, as well as modeling facilities. This enables to study in rich detail the manifold processes related to Space Weather events and to better understand the physics behind. To forecast the geomagnetic effects of an impacting disturbance at Earth (e.g., by the Dst\footnote{The disturbance storm time (Dst) index monitors variations in the Earth's equatorial ring current.} or Kp index\footnote{The planetary K index (Kp) monitors variations in the horizontal component of the Earth's magnetic field.}), the most common parameters we need to know in advance - and various combinations of these - are the amplitude/orientation and variation of the north-south component of the interplanetary magnetic field ($B_{\rm z}$), speed ($v$), and density ($n$). Especially, the electric field $vB_{\rm s}$ ($B_{\rm s}$ = $B_z<0$) is found to show a high correlation with the Dst storm index \citep[see e.g.,][]{baker81,wu02,gopalswamy08}. For details on the geomagnetic effects of Space Weather phenomena as described here, see the Living Reviews by \cite{pulkkinen07}.

The Space Weather ``chain of action'' from the solar perspective is described best by the recent example of the multiple Space Weather events that occurred in September 2017 (see Section~\ref{sec:active}). But before that, we elaborate the physical basis.

\section{Magnetic reconnection: common ground}
The commonality that unites everything and yet produces such different dynamic phenomena is \textit{magnetic reconnection} and the release of free magnetic energy. This leads to particle acceleration, heating, waves, etc. and to a restructuring of the (local) magnetic field in the corona by newly connecting different magnetic regimes and with that changing magnetic pressure gradients. Especially the latter shows to affect the solar corona globally.

\begin{figure*}
  \includegraphics[width=1.8\textwidth,angle=90]{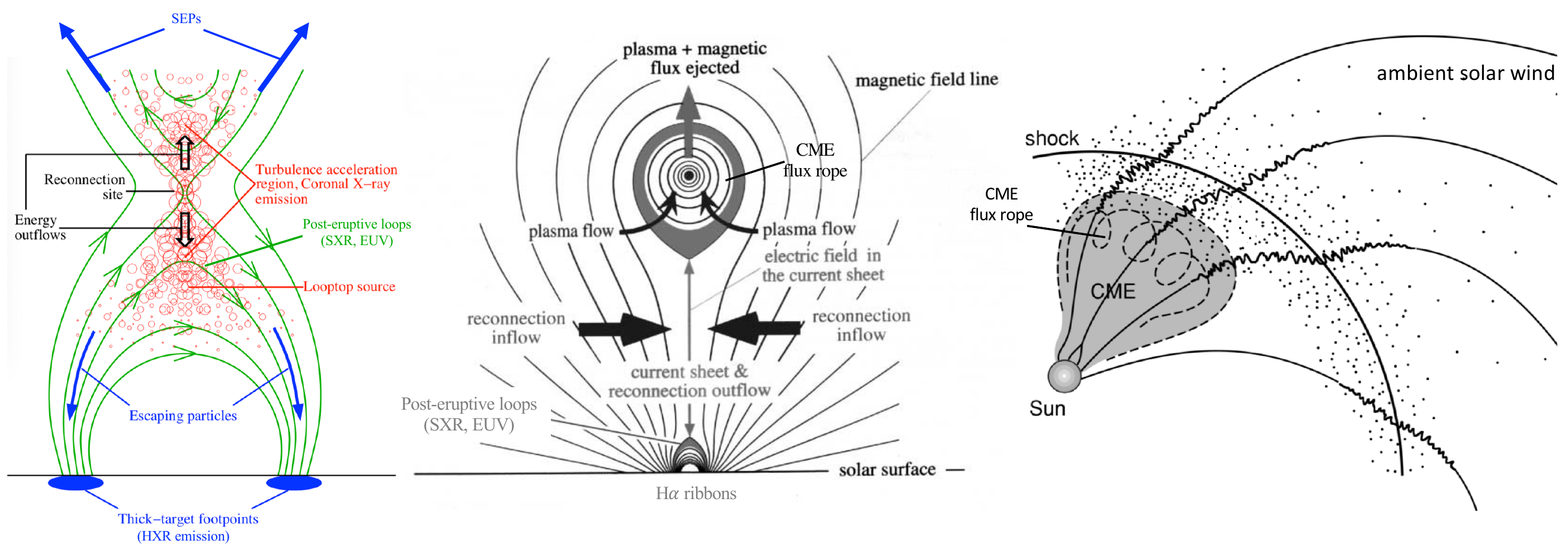}
\caption{Left: Stochastic acceleration model for solar flares. Magnetic field lines (green) and turbulent plasma or plasma waves (red circles) generated during magnetic reconnection. Blue arrows and areas mark accelerated particles impinging on the lower denser chromosphere where they produce Bremsstrahlung and on the upside may escape to interplanetary space where they are detected as SEPs \citep[adapted from][]{Petrosian04,Vlahos19_SEP}. Middle: CME-flux rope configuration in the classical scenario (CSHKP) covering also the post eruptive arcade usually observed in SXR and EUV wavelength range \citep[adapted from][]{Lin00}; Right: CME flux rope acting as driver of a bow shock (black arc) may accelerate SEPs (black dots) in the corona or heliosphere via diffusive shock acceleration \citep[adapted from][]{Mikic06}. }
\label{fig:flare_CME_SEP}       
\end{figure*}

In order to derive a complete picture about Space Weather, we first need to understand the interrelation between these many individual processes starting at the Sun. This covers a cascade of small and large scale phenomena varying over different time scales. The primary source of Space Weather producing phenomena, i.e., CMEs-flares-SEPs (note that in the following eruptive phenomena are considered and not stealth CMEs), are active regions representing the centers of strong magnetic field and energy \citep[more details on the evolution of active regions, see the Living Reviews by][]{van_driel15,toriumi19}. However, in detail the energy build-up and release processes are not well understood. The key-driver certainly is the magnetic field configuration below the visible surface (photosphere), that cannot be directly observed and characterized for giving reliable predictions of its status and further development. The lack of magnetic field information is also given in the upper atmospheric layers. There are currently no instruments enabling measurements of the magnetic field in the corona, hence, we need to rely on models simulating the coronal and, further out, interplanetary magnetic field \citep[see, e.g., the Living Reviews by][on coronal and solar wind MHD modeling ]{gombosi18}. While active regions are characterized by closed magnetic field, coronal holes cover mainly open magnetic field from which high speed solar wind streams emerge. They structure interplanetary space and set the coupling processes between continuous solar wind flow and transient events. To better understand the propagation behavior of transient events, we also need to study the evolution and characteristics of the solar wind flow, and hence, the interplay between open and closed magnetic field.

Figure~\ref{fig:flare_CME_SEP} sketches three different time steps in the evolution of an eruptive flare event, causing a CME and SEPs, as a consequence of magnetic reconnection \citep[see][]{Petrosian04,Lin00,Mikic06}. The left panel of Figure~\ref{fig:flare_CME_SEP} focuses on the early evolution stage of the eruptive event, introducing stochastic acceleration processes causing high energetic particles to precipitate along magnetic field lines towards and away from the Sun. Flare emission is observed on the solar surface due to the acceleration of particles towards the Sun. Particles that escape into interplanetary space along the newly opened magnetic field, produce SEPs. The middle panel of Figure~\ref{fig:flare_CME_SEP} shows the creation of the CME body, i.e., the production of a closed magnetic field structure (flux rope), as well as the related post-eruptive arcade which is formed below. The exact acceleration mechanism(s) of SEPs is still an open issue, hence, cartoons as shown here usually present both possible driving agents, the flare and the CME shock. To complete the picture for a flare-CME-SEP event, the right panel of Figure~\ref{fig:flare_CME_SEP} depicts the interplanetary magnetic field and its behavior which differs from the typical Parker spiral orientation due to the propagating CME shock component causing SEP acceleration in interplanetary space. The deviation of the interplanetary magnetic field from the nominal Parker spiral is an important issue when dealing with magnetic connectivity for studying SEPs and propagation behavior of CMEs. 

In the following, we will discuss in more detail the characteristics of the different manifestations occurring in an eruptive flare event.\\


\section{Solar Flares} \label{sec:flares}

\subsection{Eruptive capability of an active region}

Active regions may be classified either by the morphology of an active region using the McIntosh classification \citep{mcintosh90} or the magnetic structure using Hale's/K\"unzel's classification \citep{kunzel60}. Due to the emergence of magnetic flux the degree of complexity in the magnetic field of an active region grows, which increases the likeliness to create strong flares and CMEs \citep[e.g.,][]{sammis00,toriumi17}. The probability that an X-class flare is related to a CME is found to be larger than 80\% \citep{yashiro06}, however, there are well observed exceptions reported. So-called confined flares are neither accompanied by a CME nor a filament eruption \citep[e.g.,][]{moore01}. Their special magnetic field configuration allows particle acceleration (observed as flare), but they do not escape into interplanetary space and, hence, do not produce SEPs \citep{gopalswamy09}. Therefore, confined flares may produce strong X-ray emission but, presumably due to a strong bipolar overlying coronal magnetic field configuration, are not related to the opening of the large-scale magnetic field \citep[e.g.,][]{wang07,sun15,Thalmann15}. The electromagnetic radiation of confined flares can still instantaneously cause sudden changes in the ionospheric electron density profile (disturbing radio wave communication or navigation), also known as \textit{solar flare effect} or geomagnetic crochets \citep{campbell03} but occurring rather rarely. However, confined flares are also potential candidates for false Space Weather alerts in terms of an erroneous forecast of geomagnetic effects due to the magnetic ejecta that would have arrived tens of hours later at Earth.

Therefore, the manifestation of the eruptive capability of an active region is one of the prime targets for prospective forecasting of SEPs and CMEs. For example, the length of the main polarity inversion line of an active region or the magnetic shear and its sigmoidal morphology, is obtained to be highly indicative of the potential to open large scale magnetic field and to produce CMEs and SEPs \citep[e.g.,][]{Canfield99}. Studies also showed that active regions, for which the polarity inversion line quickly changes with height into a potential field configuration, are more favorable for producing non-eruptive events \citep{Baumgartner18}. Likewise, the decay index of the horizontal magnetic field (ratio of the magnetic flux in the lower corona to that in the higher corona) is found to be lower for failed eruptions compared to that for full eruptions \citep[cf.,][]{torok05,fan07,guo10,olmedo10}.

For more details on the issue of flare-productive active regions I refer to the Living reviews by \cite{toriumi19}. See also \cite{forbes00}, \cite{webb12}, \cite{parenti14}, or \cite{chen17} for a more theoretical approach on that issue.

\subsection{Eruptive solar flares: general characteristics}\label{sec:flare}

\begin{figure*}
  \includegraphics[width=0.6\textwidth]{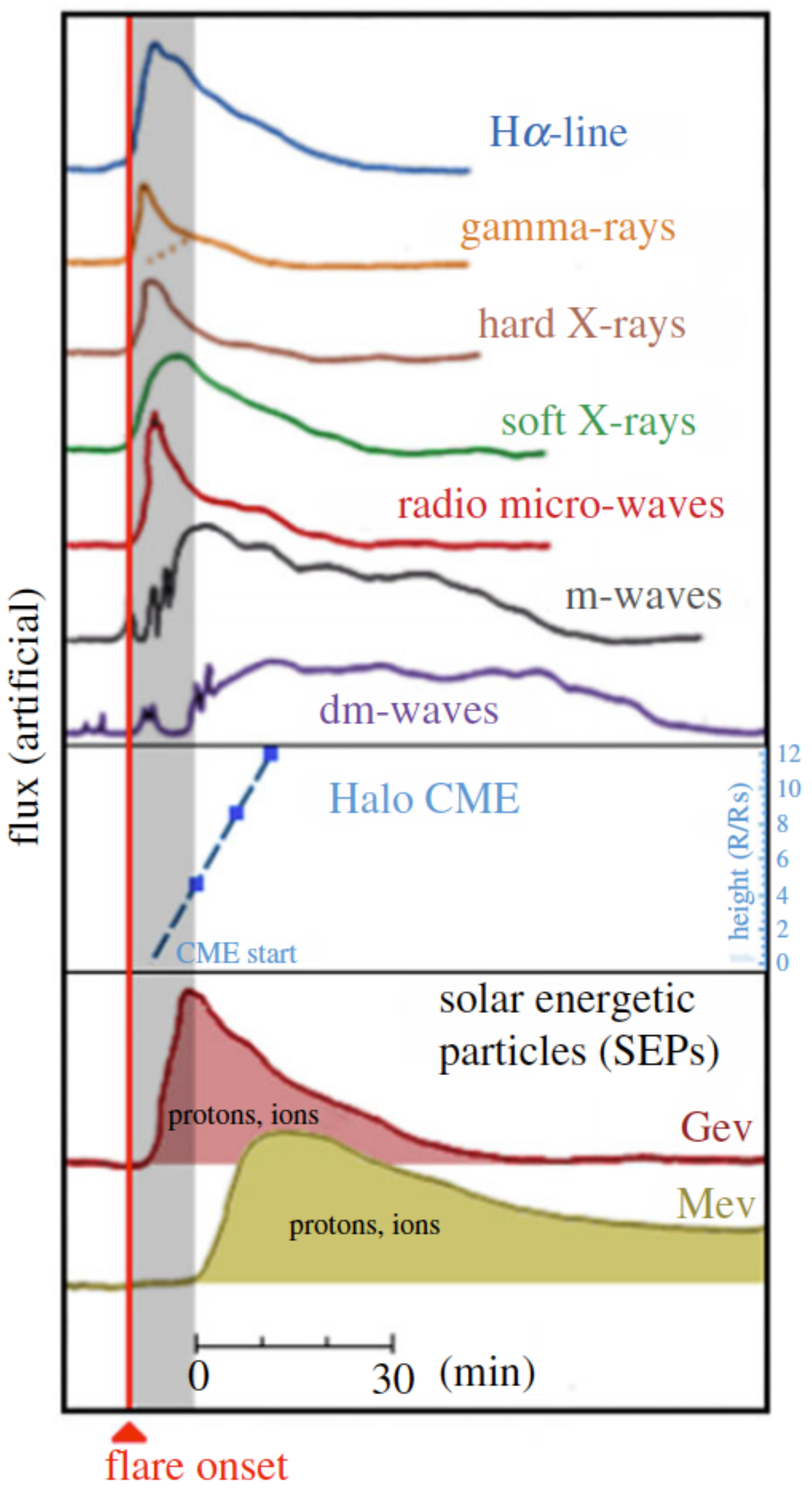}
\caption{Flare-CME-SEP relation in time. The onset of the solar flare is indicated by the vertical red line. The grey
shaded area marks the time difference between flare start and SEP flux increase for MeV energies. Taken from \cite{Anastasiadis19} who adapted it from \cite{miro03}.}
\label{fig:timing}       
\end{figure*}

Flares are observed to release a huge amount of energy from 10$^{19}$ up to 10$^{32}$ erg over a timescale of hours\footnote{An automatically updated list of flares is available under \url{http://www.lmsal.com/solarsoft/latest_events/} or \url{https://www.solarmonitor.org}}. With the advent of modern ground-based and space-borne instruments, our small optical window was massively enlarged and it is now well-known that this energy is radiated over the entire electromagnetic spectrum from decameter radio waves to gamma-rays beyond 1 GeV. Figure~\ref{fig:timing} depicts the temporal relation between flare emission, observed in different wavelength ranges, CME kinematics and SEP flux profiles in the GeV and MeV energy range. The flare activity profile consists of a so-called pre-flare phase, showing thermal emission in SXR and EUV, as well as H$\alpha$ kernel brightenings. If related to a filament eruption, this phase partly coincides with the slow rise phase of the filament\footnote{Filament detection and eruption catalogues can be found e.g., under \url{http://cesar.kso.ac.at/sn_iv/filaments.php} or \url{http://aia.cfa.harvard.edu/filament/}}. This is followed by the impulsive flare phase during which most of the energy is released and non-thermal emission in terms of hard X-ray (HXR) footpoints appears due to particles accelerating out of the localized reconnection area and bombarding the denser chromosphere where they emit Bremsstrahlung \citep[for a review on solar flare observations see e.g.,][]{fletcher11}. At this point also the CME body forms as consequence of the closing of the magnetic field lines in the upper part of the reconnection area revealing a flux rope structure (note that the most compelling argument for an already existing flux rope is actually a filament). As the flare emission increases also the SEP flux in the GeV energy range starts to rise. After the flare reaches a maximum in intensity, the decay phase is observed during which the intensity level goes back to the background level from before the flare start. The exact timing of the rise and decay phase is dependent on the energy release and the energy range in which the flare is observed which is known as the so-called Neupert effect \citep[the HXR flux rise phase time profile corresponds to the derivative of the SXR flux time profile; see][]{neupert68}. The last phase may have a duration of several hours or longer. During that phase also post-eruptive arcades (or loops) start to form, that may still grow over 2--20 hours. The growth of the post-eruptive arcade is hinting towards an ongoing reconnecting process, which is not energetic enough to produce a significant emission in EUV or SXR \citep[see e.g.,][]{tripathi04}. For more details on the global properties of solar flares I refer to the review by \cite{hudson11}.

\begin{figure*}
  \includegraphics[width=1.\textwidth]{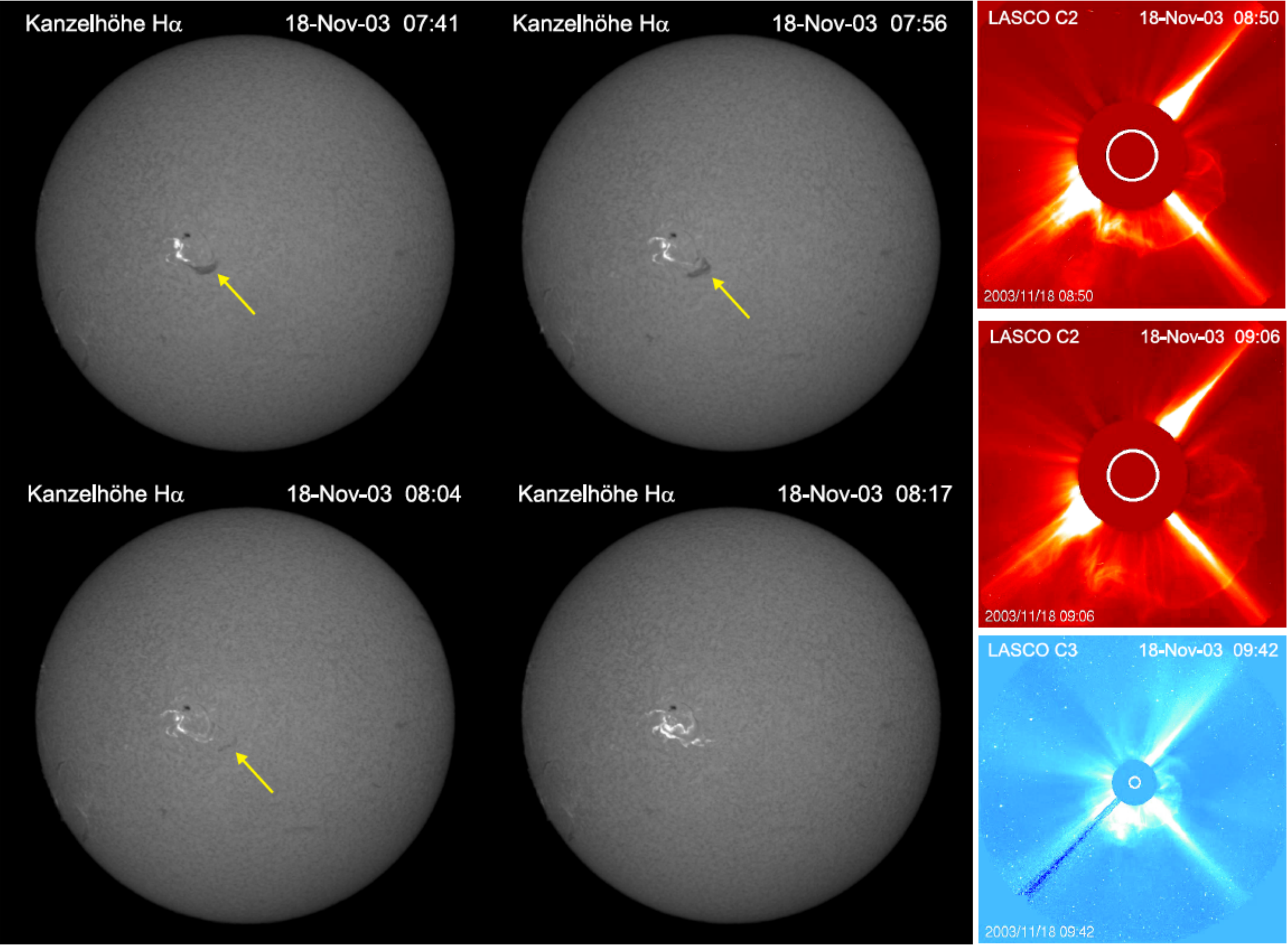}
\caption{Global flare evolution and relation to CME from the November 18, 2003 event. Left panels: H$\alpha$ filtergrams from the Kanzelh{\"o}he Solar Observatory (Austria). The associated erupting filament is indicated by arrows. Right panels: Temporal evolution of the CME in coronagraph images from SOHO/LASCO. Taken from \cite{mostle08}.}
\label{fig:moestl08}       
\end{figure*}

Figure~\ref{fig:moestl08} shows the temporal evolution of a flare and erupting filament observed in H$\alpha$ and the associated CME observed in white-light coronagraph image data. The event is classified in the emitted SXR flux as GOES M3.9 flare (corresponding to the measured power of $3.6\times10^{-5}$ W/m$^{2}$) which occurred on November 18, 2003 in a magnetically complex $\beta\gamma$ active region. The associated CME caused two days later one of the strongest geomagnetic storms of solar cycle 23 having a minimum Dst value of $-$472~nT \citep{gopalswamy05_nov}. Inspecting the time stamps on the image data of that event, about one hour after the appearance of the flare signatures, the CME became visible in the coronagraph. The filament started to rise some tens of minutes before the flare emission occurred.

\begin{figure*}
  \includegraphics[width=1.\textwidth]{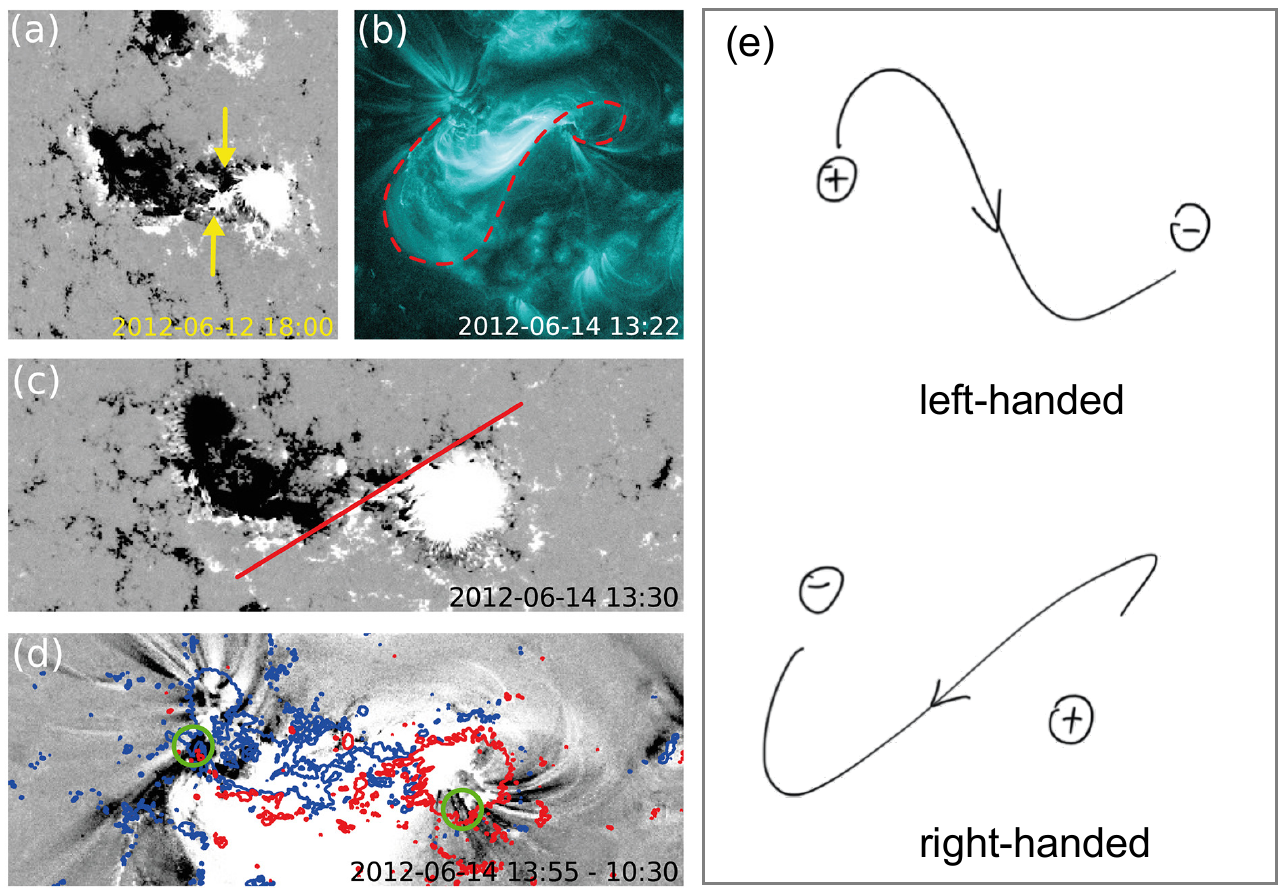}
\caption{(a) SDO/HMI data from June 14, 2012 showing the magnetic tongues of the erupting active region revealing a positive chirality. (b) Forward‐S sigmoidal structure from the coronal loops observed by SDO/AIA 131\AA, indicating a right‐handed flux rope. (c) SDO/HMI magnetogram showing the approximated polarity inversion line (red line). (d) Base‐difference SDO/AIA 131\AA~image overlaid with the HMI magnetogram contours saturated at $\pm$200~G (blue = negative polarity; red = positive polarity). The dimming regions indicating the flux rope footpoints are marked by green circles. Panels (a--d) are adapted from \cite{palmerio18}. (e) The cartoon shows the handedness inferred from the magnetic field and sigmoidal structures or orientation of post eruptive loops.}
\label{fig:palmerio17}       
\end{figure*}

The orientation of the magnetic structure, especially of the $B{\rm z}$ component, of an ICME is key to forecast its geoeffectiveness and poses the Holy Grail of Space Weather research. Knowing the flux rope orientation already at the Sun could provide information on the impact of CMEs early in advance, hence, as soon as they erupt or even before. While the handedness of flux ropes can be well observed from in-situ measurements \citep{Bothmer98,mulligan98}, on the Sun observational proxies need to be used. Figure~\ref{fig:palmerio17} shows several surface signatures from which the magnetic helicity (sense of twist of the flux rope: right-handed or left-handed) can be inferred. Typically, in EUV observations these are sigmoidal structures (S- or reverse S-shaped) or post-eruptive arcades (skewness of EUV loops and polarity of the underlying magnetic field), in H$\alpha$ the fine structures of filaments are used (orientations of barbs) or statistical relations like the \textit{hemispheric helicity rule} \cite[see][]{wang13}. However, strong coronal channeling, latitudinal and also longitudinal deflection and/or rotation, that the magnetic component of the CME undergoes as it evolves through the low solar corona, may change those parameters as shown in various studies by e.g., \citet{shen11,Gui11,bosman12,Panasenco13,WangY14,kay15,moestl15}, or \citet{Heinemann19}. Recent approaches in ICME $B{\rm z}$ forecasting can be found in, e.g., \cite{savani15}, \cite{palmerio17}, or \cite{kay17}.

For more details about the energetics and dynamics of solar flares I refer to the Living Reviews by \cite{benz17} and for the magnetohydrodynamic processes in active regions responsible for producing a flare to the Living Reviews by \cite{Shibata11}.

\section{Coronal mass ejections (CMEs)}\label{subsec:cme}

\subsection{General characteristics}
 
CMEs are optically thin large-scale objects, that quickly expand, and are traditionally observed in white-light as enhanced intensity structures. The intensity increase is due to photospheric light that is Thomson scattered off the electrons forming the CME body and integrated over the line-of-sight \citep{Hundhausen93}. Due to strong projection effects their apparent morphology greatly depends on the viewpoint and, hence, makes CMEs a rather tricky object to measure  \citep[see e.g.,][]{burkepile04,cremades04}.

\begin{figure*}
  \includegraphics[width=1.\textwidth]{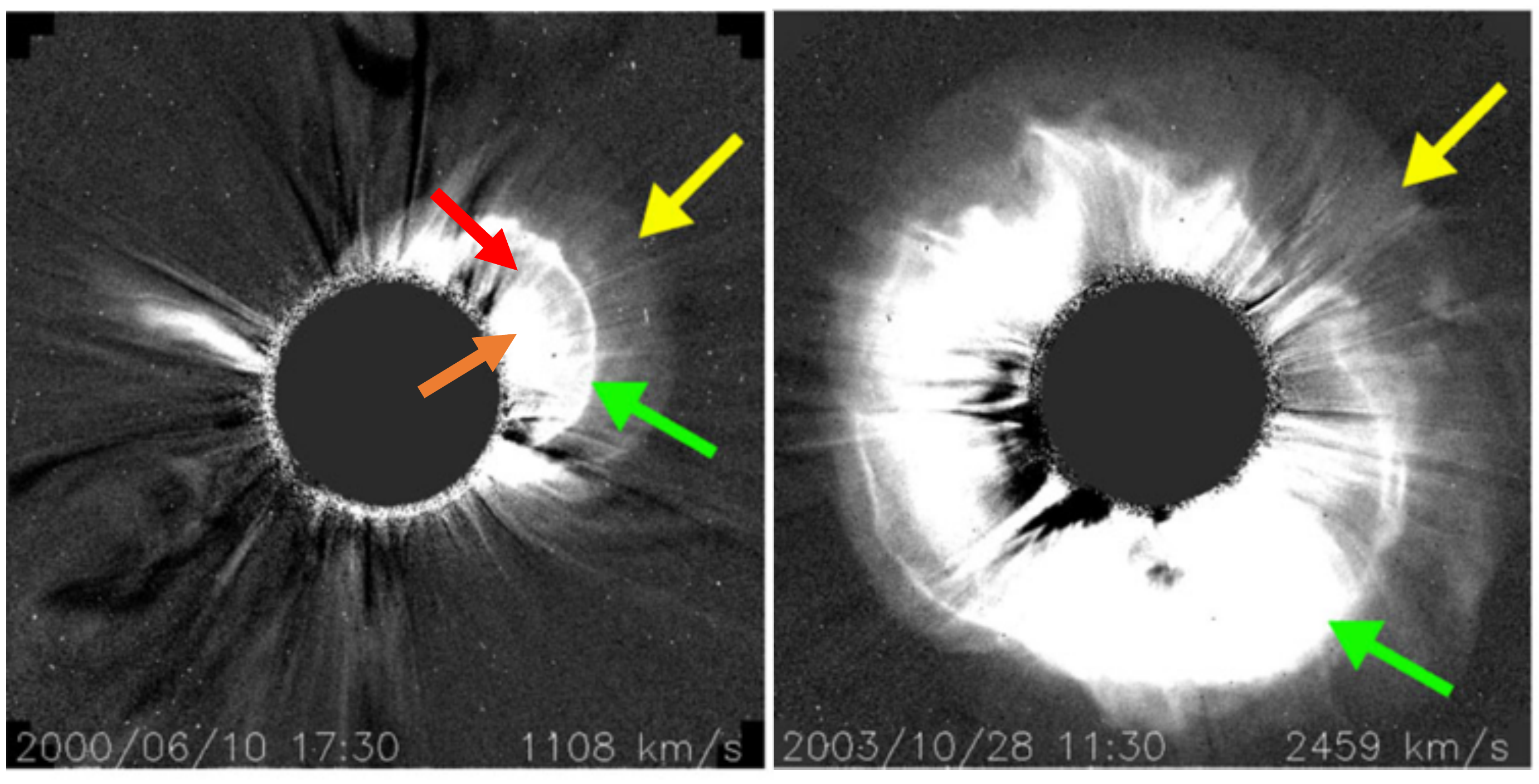}
\caption{LASCO CME excess mass images showing the expanding shock wave front (yellow arrow) and the CME leading edge density enhancement (green arrow) for two different events. For the CME propagating rather in the plane of sky (left panel), typical structures such as the cavity due to the expanding magnetic ejecta} (red arrow) followed by some intensity enhancement (orange arrow) can be observed, that is less well visible for the halo CME (right panel). The projected LASCO CME speeds are given in the legend \citep[adapted from][]{Vourlidas13}.
\label{fig:CME_struct}       
\end{figure*}

By using coronagraphs, CMEs are visible with teardrop-like shapes that are characterized by multiple structures. Figure~\ref{fig:CME_struct} shows SOHO/LASCO \citep{Brueckner1995TheLASCO} coronagraph white-light images of two CMEs having different propagation directions. For the CME that leaves the Sun in a rather perpendicular angle to the observer (left panel of Figure~\ref{fig:CME_struct}), the various CME structures are well visible. In general, we distinguish between the shock (yellow arrow) and CME body (green arrow) that are followed by some cavity created by the expanding magnetic flux rope (red arrow) and an increased brightness structure (orange arrow). Partly these structures are detected also from in-situ measurements for the interplanetary counterparts of CMEs (ICMEs; see Section~\ref{sec:sub3-5}). The increased brightness structure consists of prominence material \citep{Vourlidas13} or is suggested to appear due to a brightness increase of the two overlapping CME flanks \citep{howard17}. The sheath region behind the shock has less clear signatures in coronagraph images taken close to the Sun as it is generated later when the solar wind plasma gets piled-up in interplanetary space \citep[see e.g.,][]{kilpua17,salman20}. For CMEs propagating in the line-of-sight towards or away from the observer (right panel of Figure~\ref{fig:CME_struct}), the different structures are less well visible. As these CMEs are launched close to the central meridian of the observed disk, they most severely suffer from projection effects. Energetic ones are frequently observed as so-called \textit{halo} CMEs, revealing extensive white-light signatures made of compressed plasma material surrounding the occulting disk of a coronagraph. For halo CMEs, evidence that the CME is actually moving towards the observer is given from the associated activities observed on the solar disk (such as filament eruptions, flare emission, dimming regions, or coronal wave signatures). Highly relevant for Space Weather, halo CMEs are of special interest and are diversely studied mostly by using single spacecraft data from the coronagraphs aboard SoHO.

Up to the distance of about 30 solar radii (LASCO/C3 field of view), statistical studies showed that CMEs undergo several phases in their dynamics. Before the actual launch a slow rising phase occurs (initiation phase), continued by the acceleration phase over which a rapid increase in speed is observed in the inner corona, that is followed by a rather smooth propagation phase as the CME leaves the Sun \citep{Zhang2006AEjections}. On average, over the coronagraphic field of view, CME fronts reveal radial speeds in the range of 300--500~km/s with maximum values observed up to 3000~km/s, accelerations of the order of 0.1--10~km/s$^2$, angular widths of about 30--65 degrees and masses of $\sim$10$^{14}$--10$^{16}$\,g \citep[e.g.,][]{vourlidas10,Lamy19}. The ratio in density between the CME body and surrounding solar wind decreases from $\sim$11 at a distance of 15 solar radii to $\sim$6 at 30 solar radii \citep{temmer21}.  However, CMEs vary in their occurrence rate as well as in their characteristics over different solar cycles. While flare rates and their properties have not changed much over the past solar cycles, the CME properties for solar cycle 24 are significantly different as given in recent statistics \citep{Lamy19,dagnew20Cycle}. CMEs were found to be more numerous and wide compared to solar cycle 23. Close to the Sun, the CME expansion is driven by the increased magnetic pressure inside the flux rope, while further out they most probably expand due to the decrease of the solar-wind dynamic pressure over distance \citep{lugaz20}. Therefore, the increased width for CMEs of cycle 24 may be explained by the severe drop ($\sim$50\% ) in the total (magnetic and plasma) heliospheric pressure \citep[see e.g.,][]{mccomas13,gopalswamy14,gopalswamy15,dagnew20}. Interestingly, also the maximum sunspot relative number in cycle 24 reached only 65\% of that from cycle 23\footnote{\url{http://sidc.be/silso/cyclesminmax}}. The different expansion behaviors have consequences also for Space Weather effects in terms of their abilities in driving shocks \citep[see e.g.,][]{lugaz17_radial}.

\subsection{CME early evolution}\label{sec:sub2-5}

\begin{figure*}
  \includegraphics[width=\textwidth]{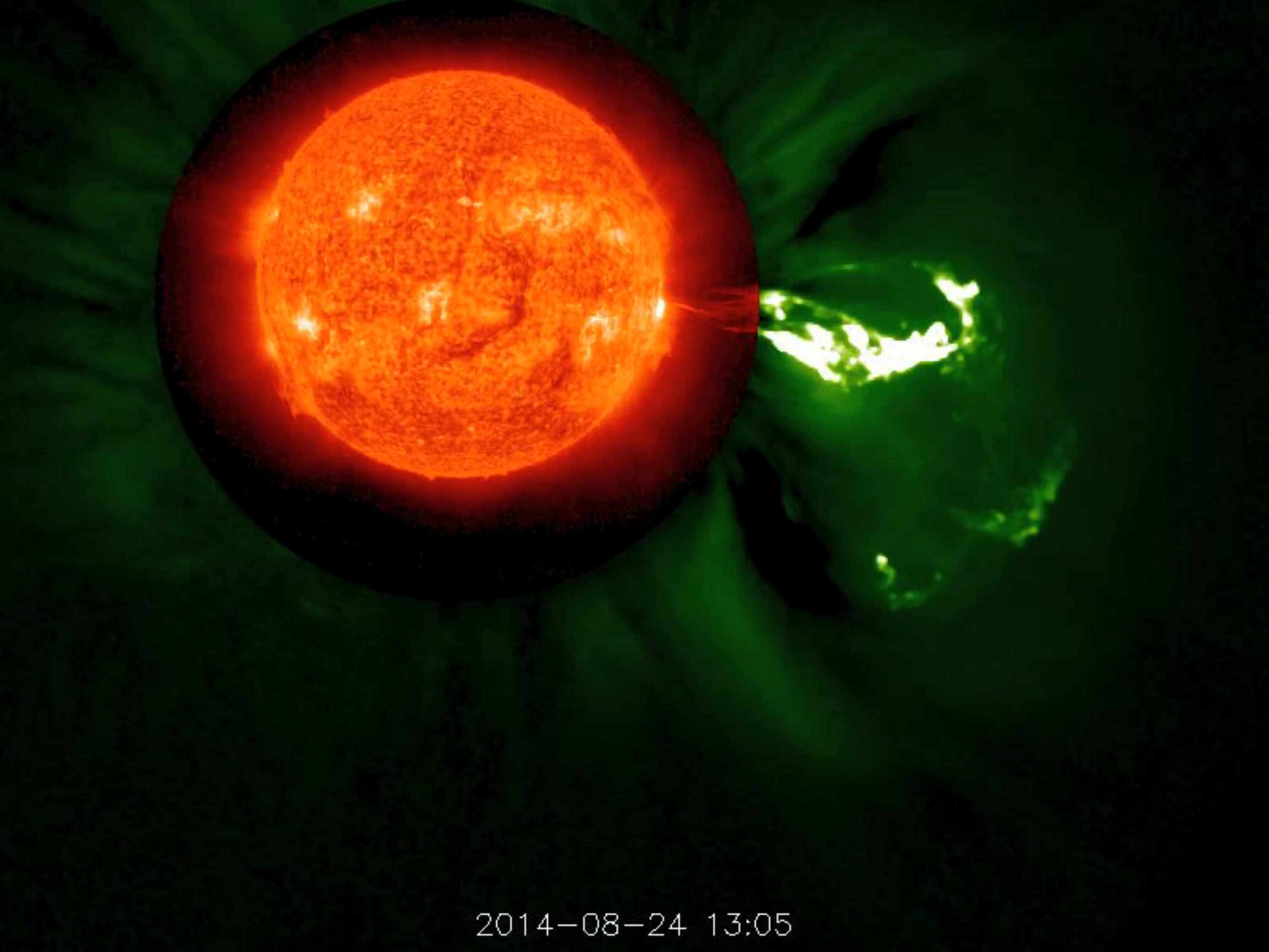}
\caption{STEREO-B observations of the CME from August 24, 2014. The images show combined EUVI (304\AA) and COR1 image data. Filament plasma material is ejected into space forming the bright CME core following the cavity. Plasma that is lacking sufficient kinetic energy to escape from the Sun's gravity, falls back onto the solar surface. Credit: STEREO/NASA. The movie is online available. 
}
\label{fig:euv_cor1}       
\end{figure*}

Besides the traditional observations in white-light images, also EUV or SXR imagery reveal CME signatures, presumably due to compression and heating that makes it visible in filtergrams sensible for high temperatures \citep[see e.g.,][]{glesener13}. Satellite missions that carry EUV instruments having large field of views can be effectively used with combined white-light coronagraph data to track CME structures for deriving the kinematical profile over their early evolution covering the CME main acceleration phase. The SECCHI instrument suite \citep{Howard08_secchi} aboard STEREO provides EUV and white-light data that seamlessly overlap\footnote{SoHO EIT and C1 also provided that possibility but C1 was lost in June 1998 due to spacecraft failure. For a couple of events the usage of combined EIT-C1 data could be shown \citep[see][]{Gopalswamy00_earlyCME,Zhang01,cliver04b}.}. as shown in Figure~\ref{fig:euv_cor1}. For such studies one needs to keep in mind that the observational data image different physical quantities (density and temperature in EUV, and density in white-light), hence, dark and bright features in both image data do not necessarily match. 

\begin{figure*}
  \includegraphics[width=1.6\textwidth, angle=90]{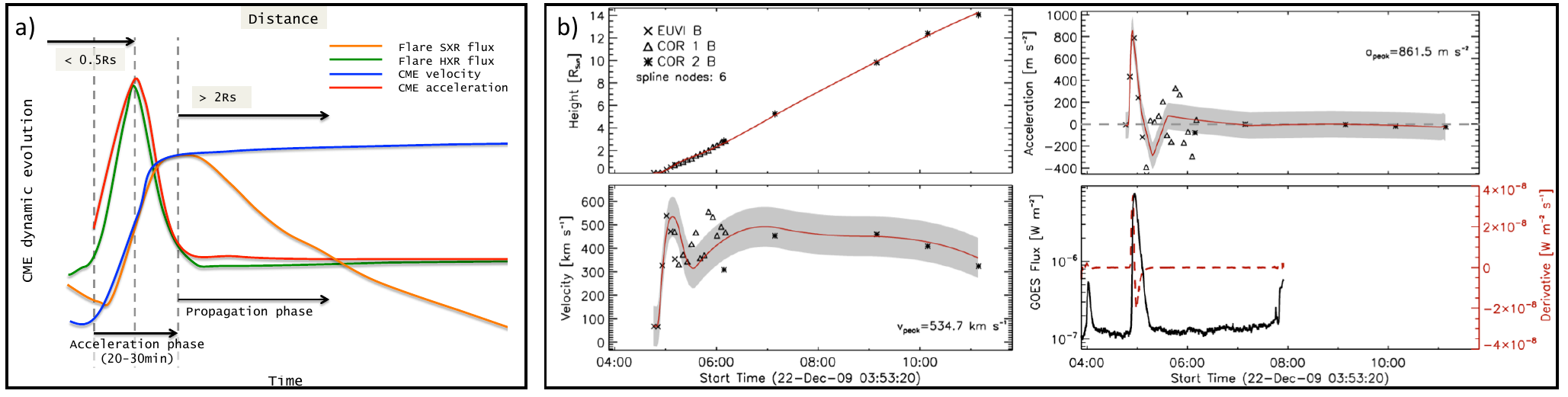}
\caption{CME-flare relation. a) schematics of the thermal (SXR) and non-thermal (HXR) flare energy release in comparison to the CME kinematical evolution close to the Sun. It is found that CME acceleration and HXR emission as well as the CME speed and SXR emission, respectively, are closely related. Taken from \cite{Temmer16}. b) observational results for the December 22, 2009 CME event revealing the early evolution from combined EUV and coronagraph data (STEREO-B spacecraft) and GOES SXR flux profile for the related flare and derivative (proxy for HXR emission). Taken from \cite{Bein2012ImpulsiveEruptions}.}
\label{fig:temmer16}       
\end{figure*}

From combined high temporal resolution EUV and white-light data a more detailed understanding about the energy budget (see also Section~\ref{sec:budget}) and relation between flares, filaments and CMEs is revealed providing relevant information for SEP acceleration and generation of radio type II bursts. It is found that the thermal flare emission observed in SXR and the CME speed profile show similar behavior in timing \citep{Zhang01,Zhang04,chen03,Maricic07}. For strong eruptive events an almost synchronized behavior between flare HXR emission and CME acceleration is obtained through a feedback relation \citep{Temmer08,temmer10}. The CME acceleration is found to peak already as low as about 0.5 solar radii above the solar surface \citep[for statistics see][]{Bein2011ImpulsiveCharacteristics}. Figure~\ref{fig:temmer16} gives the schematic profiles and distances over time between non-thermal (HXR) and thermal (SXR) flare energy release and CME kinematics (acceleration, speed). The flare-CME feedback loop can be well explained by the CSHKP standard model \citep{Carmichael64,Sturrock66,Hirayama74,Kopp76} through the magnetic reconnection process underlying both activity phenomena. In a simplistic scenario, we may summarize that magnetic reconnection drives particle acceleration (neglecting details on the actual acceleration process) leading to flare emission and closes magnetic field increasing the magnetic pressure inside the presumable CME flux rope (neglecting details on the actual magnetic configuration of the active region and surrounding). For strong flares that are related to CMEs of high acceleration values, the available free magnetic energy might be larger. This occurs preferably for CMEs initiated at lower heights where the magnetic field is stronger. With that, particles get accelerated to larger energies, hence, producing stronger flares, and more poloidal flux can be added per unit time, hence, generating a stronger expansion of the flux rope and a faster CME eruption. This is supported by theoretical investigations on the feedback process, covering magnetic reconnection with the ambient coronal magnetic field \citep[reconnective instability][]{welsch18}. More details are found in, e.g.,  \cite{chen03,Vrsnak2007AccelerationScales,vrsnak08,jang17}.

Associated to the erupting CME, we frequently observe coronal dimming regions that evolve over a few tens of minutes \citep{1996hudson,webb00}. Core dimming regions are assumed to be located at the anchoring footpoints of the associated magnetic flux rope and reveal the loss of plasma from the corona into the CME structure adding mass to the CME body \citep[see][]{Temmer17_flar-cme}. Secondary dimming regions most probably refer to mass depletion in the wake of the large-scale magnetic field opening as the CME fully erupts \citep[for more details on core and secondary dimming regions, see][]{mandrini07}. Recent studies discovered a strong relation between dimming intensity and flare reconnected flux as well CME speed \citep[e.g.,][]{dissauer18,dissauer19}. Also the final width of the CME can be estimated from the amount of magnetic flux covered by the CME associated post-eruptive flare arcade as the surrounding magnetic field prevents the CME flux rope from further expansion \citep{moore07}. On the contrary, the CME surrounding shock as well as associated coronal waves on the solar surface, that are ignited by the lateral CME expansion, are freely propagating and are not limited in their spatial extend \citep[for more details on globally propagating coronal waves I refer to the Living Reviews by][]{Warmuth15}.

\begin{figure*}
  \includegraphics[width=\textwidth]{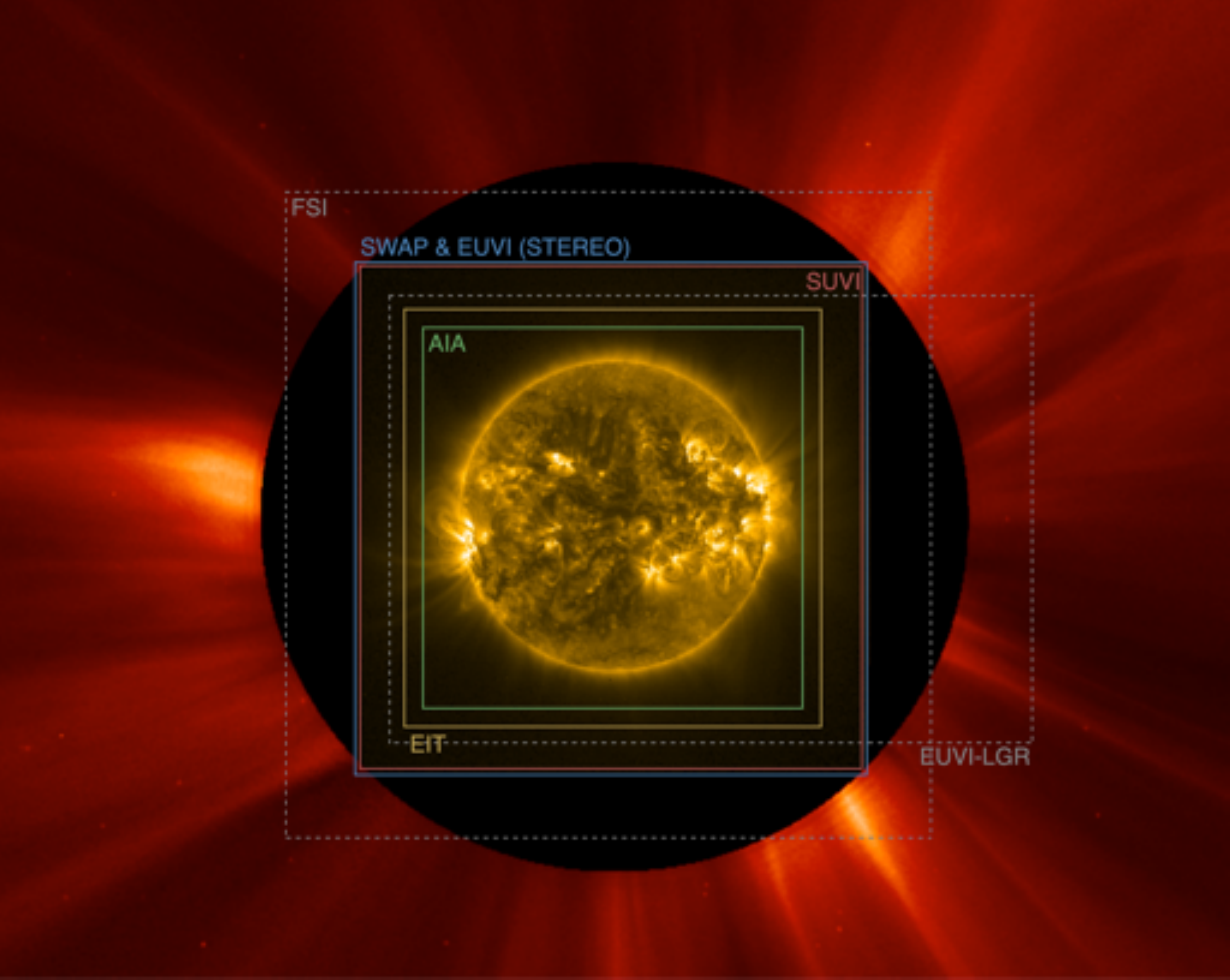}
\caption{EUV image from Proba-2/SWAP combined with a LASCO/C2 coronagraph image covering in total a field of view up to $\sim$4 solar radii. The colored boxes mark the relative nominal field of views of different EUV observing instruments. FSI (Full Sun Imager is part of the EUI suite aboard Solar Orbiter), EUVI-LGR (aboard the  planned L5 \textit{Lagrange} mission), and SoHO/EIT. STEREO/EUVI, Proba-2/SWAP, GOES/SUVI are instruments with the largest field of view of about 1.7 solar radii. Taken from \url{http://middlecorona.com}.
}
\label{fig:mid-corona}       
\end{figure*}

To derive in more detail the temporal linking of flare-CME-SEP events, image data covering large field of views for observing the lower and middle corona is of utmost importance. Figure~\ref{fig:mid-corona} shows the different field of views of currently available and future EUV instruments to observe and study the middle corona (distance up to about 4 solar radii). The Extreme EUV Imager suite aboard Solar Orbiter works at the 174\AA~and 304\AA~EUV passbands \citep[EUI:][]{Rochus20_SoloEUI}. The EUVI-LGR instrument aboard ESA's \textit{Lagrange} L5 mission (launch planned for 2027) has an extended field of view to the West limb of the Sun, that is perfectly suited to track the early evolution of Earth directed CMEs from L5 view (60 degrees separation with Earth). We must not forget the capabilities of ground-based coronagraph instruments such as the COSMO K-Cor at the Mauna Loa Solar Observatory in Hawaii (replaced in 2013 the aging MLSO Mk4 K-coronameter\footnote{Details can be found under: \url{https://www2.hao.ucar.edu/cosmo/documentation}}) observing a field of view starting as low as 1.15 solar radii, however, quite restricted in observational time compared to satellite data.

For more details on CME trigger mechanisms I refer to recent review articles by \cite{schmieder15} or \cite{green18}. For a more specific background on CME initiation models, see, e.g., the Living Reviews by \cite{webb12}.


\subsubsection{Shock formation, radio bursts, and relation to SEPs}\label{shock-form}
Closely related to studies of the CME early evolution and acceleration profiles, are shock formation processes. To generate a shock wave, a short-duration pulse of pressure is needed. Besides the CME, acting as piston, there is also the possibility that a strong flare energy release initiates a blast wave or simple-wave shock \citep[e.g.,][]{vrsnak08b}. At which height shocks are formed by an erupting disturbance is important for understanding particle acceleration processes. The acceleration profile derived from tracking the CME frontal part suggests its formation at rather low coronal heights $<$1.5 solar radii. The  shock formation height itself is also strongly depending on the plasma environment. From model calculations a local minimum of the Alfv{\'e}n speed is derived for a distance of about 1.2--1.8 solar radii and a local maximum around 3.8 solar radii from the Sun \citep{mann99,Gopalswamy01_CME+longwave_typeII,vrsnak02}. Hence, the statistical maximum of CME acceleration profiles is also in accordance with the local minimum of the Alfv{\'e}n speed. The occurrence of such local extrema has major consequences for the formation and development of shock waves in the corona and the near-Sun interplanetary space as well as their ability to accelerate particles.

The most compelling argument for shock formation is the observation of a radio type II burst. In the case of being driven by a CME, they are reported not only to occur at the apex of a CME shock front, but also to originate from the lateral expansion of the CME as observed with LOFAR\footnote{Low Frequency Array \citep{haarlem13}. Recent attempts to use LOFAR for Space Weather purposes are reported under \url{http://lofar4sw.eu}.} \citep[e.g.,][]{zucca18}. Due to the large density in the lower coronal heights a large compression appears with a quasi-perpendicular geometry, favoring the shock formation process. In that respect, moving type IV radio bursts might actually represent shock signatures due to CME flank expansion, that can be used as additional diagnostics for studying the lateral evolution of a CME \citep{morosan20_flank}. The SEP intensity is found to be correlated with the width of a CME, and as such identifies the CME flank region to be an efficient accelerator of particles \citep[see][]{holman83,mann05,richardson15}. Comparing the CME apex and flanks, the field lines are disturbed at different heights that may lead to different onset times for the acceleration of SEPs (cf., Figure~\ref{fig:reames09}). The time needed for shock formation also leads to a temporal delay of the onset of SEP events with respect to both, the initial energy release (flare) and the onset of the solar type II radio burst (evidence of shock formation). Hence, the timing is an important factor and has to be taken into account when relating these phenomena to each other.

\begin{figure*}
  \includegraphics[width=1.\textwidth]{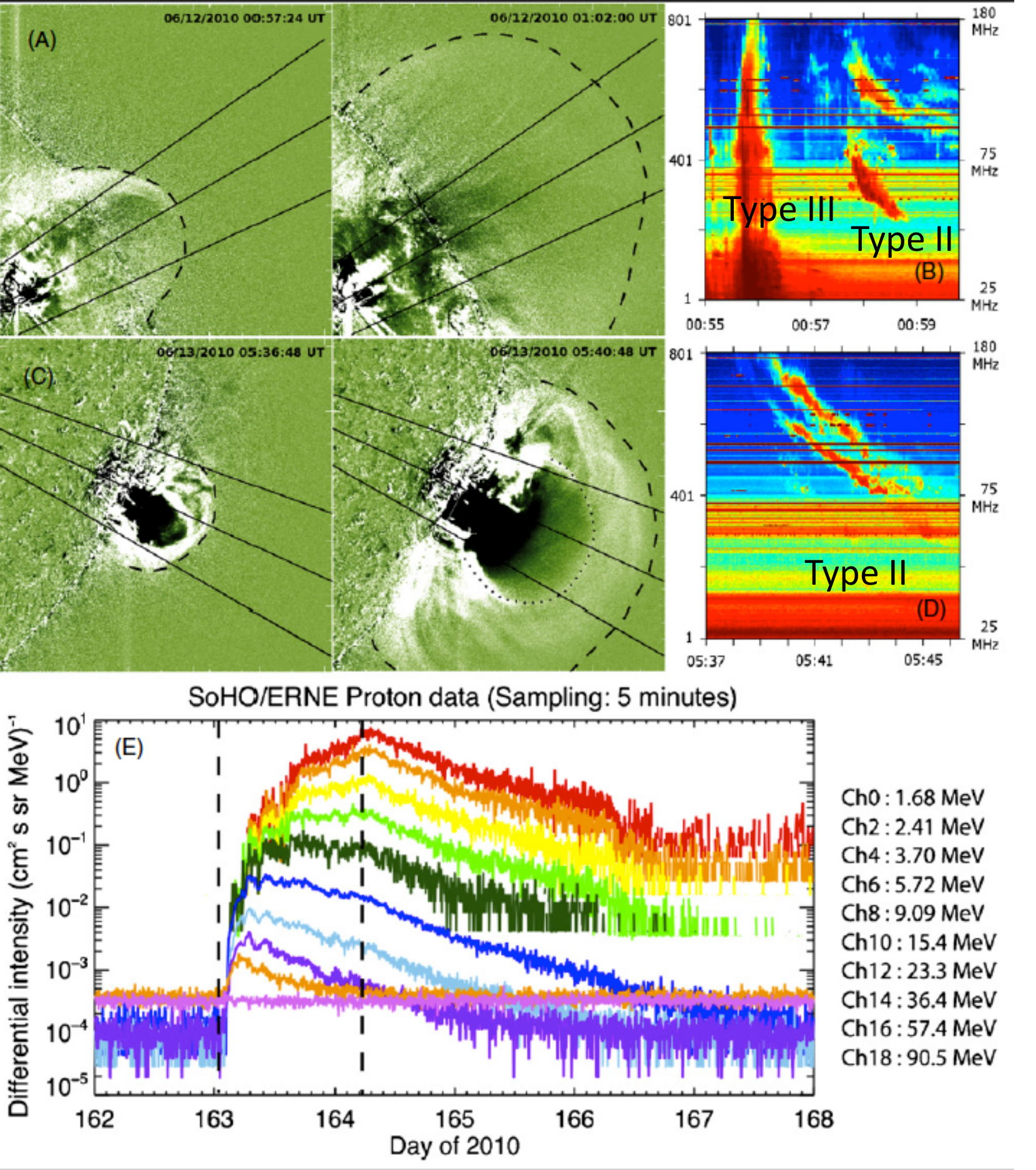}
\caption{Two CME events and associated coronal waves (June 12 and 13, 2010) are investigated with high cadence SDO EUV 211\AA~data. Manually tracked positions of the wavefronts are marked by dashed black lines. The connection between coronal surface wave and CME front is nicely observed. The right panels give radio data (Learmonth observatory, Australia) revealing type III and type II bursts (upper and lower frequency band) and measurements from particle detectors at L1 (bottom panel; vertical dashed lines show AIA waves onsets). Adapted from \cite{Kozarev11}.}
\label{fig:kozarev11}       
\end{figure*}

Shocks may also be formed at larger distances from the Sun (several tens of solar radii), depending on the acceleration phase duration, the maximum expansion velocity and the width of the CME \citep{Zic2008CylindricalShocks}. Due to the declining magnetic field with distance \citep[well defined band-splits in type II bursts can be used to estimate the magnetic field in the corona; see e.g.,][]{vrsnak02}, shocks forming at larger heights are related to softer SEP spectra \citep[e.g.,][]{gopalswamy17_SEP}. As can be seen, CME acceleration, shock formation height and hardness of SEP spectra is closely connected. Compared to SEPs, which strongly depend on the magnetic connectivity with the observer, type II bursts can be observed without connectivity issues and thus, give additional information about particle acceleration processes driven by CME shocks. In that respect, type II radio bursts may be used for predicting SEPs as well as shock arrival times \citep[e.g.,][]{gopalswamy08_typeII,cremades15}. In combination, these parameters have strong implications for Space Weather impact, revealing the importance of monitoring and studying the early evolution phase of solar eruptive events. More details on SEPs are given in Section \ref{sec:4}.


Figure~\ref{fig:kozarev11} presents a case study about the evolution of a CME front close to the Sun by using high cadence EUV images from SDO for June 12 and 13, 2010. The derived kinematics of the CME reveals a fast acceleration of its frontal part with about 1~km/s$^2$ over the distance range 1.1--2.0 solar radii. The almost vertical traces in the radio spectra are type III bursts, identified with streams of electrons (radio emission due to particles moving along open magnetic field lines), followed by diagonal structures of a moving type II bursts, identified with shock waves. The onset of the type II burst appears together with the CME shock front, as observed in EUV, with a bit of a delay with respect to the shock formation that occurs close to the maximum CME acceleration. By the time the CME occurs in the LASCO field of view, the CME speed decreased to the sub-Alfv{\'e}nic regime. The event produced an enhanced proton flux at 1AU. However, the complex magnetic topology related to the active region prevents from making strong conclusions about the possible sites of particle acceleration \citep[see also][]{Kozarev11,suli11,gopalswamy12}. Definitely, more such detailed case studies combined with improved modeling of the magnetic environment is needed for advancing our understanding in the processes that accelerate particles. 

\subsubsection{Stealth CMEs}
In contrast to fast and massive CMEs and their related cascade of solar surface signatures, there exist so-called \textit{stealth CME} events that are most probably caused by some simple (low-energetic) magnetic field reconfiguration in the upper corona releasing magnetic flux ropes of low density that usually do not exceed the solar wind flow speed. Actually, they were recognized already in the mid 1980's and were identified as \textit{spontaneous CMEs} or \textit{unassociated CMEs} (meaning no surface signatures) by \cite{wagner84}. Later studies showed, that they start at very large heights in the corona without noticeable signatures, such as flare emission, filament eruptions, coronal waves, or coronal dimmings \citep{Robbrecht09}. Stealth CMEs are potential candidates to cause \textit{problem} storms and missed Space Weather events, as they are hardly recognized in white-light data and due to the lack of observational imprints on the solar disk. In the recent years several studies have been published on this issue discussing those events \citep[see e.g.,][]{DHuys2014ObservationalSignatures,nitta17,vourlidas18}. 

%

\subsection{Advantages due to multi-viewpoint observations}\label{sec:sub1-5}

\begin{figure}
  \includegraphics[width=\textwidth]{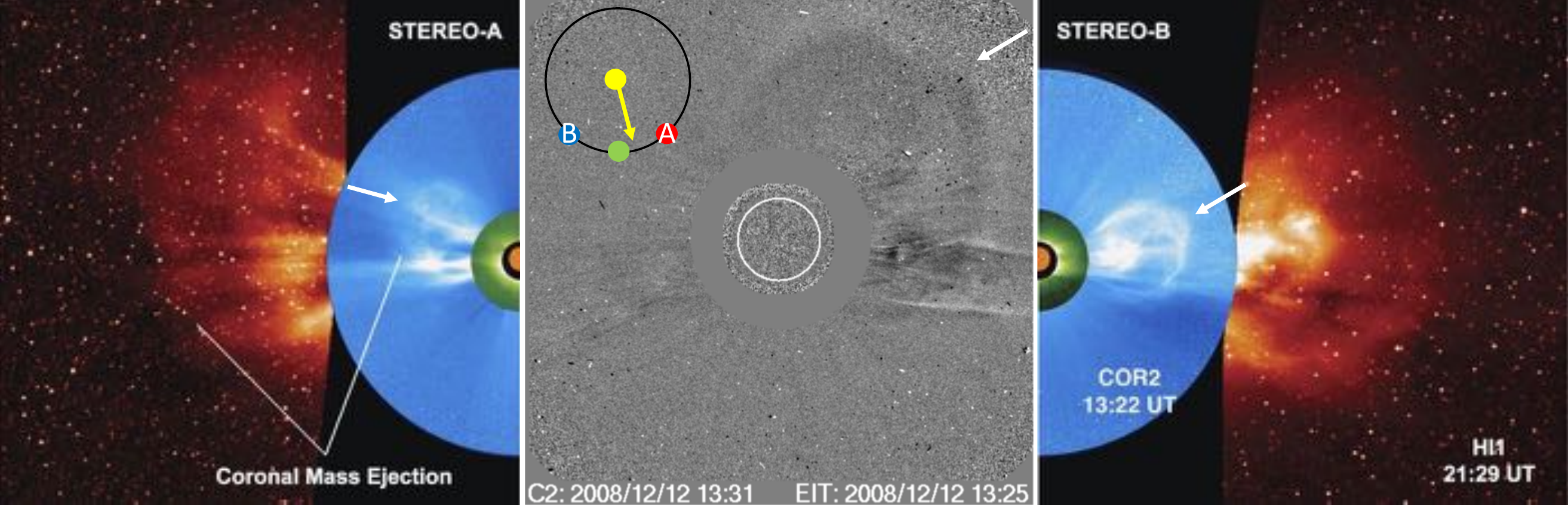}
\caption{Earth-directed CME from December 12, 2008 as observed from multiple perspectives. STEREO-A (left) and STEREO-B (right) are separated from Earth by an angle of about 45 degrees. The running difference image from LASCO/C2 (middle panel) observes the CME as weak partial halo event. The inlay in the middle panel gives the spacecraft location (STEREO-A red filled circle; STEREO-B blue filled circle) with respect to Earth (green filled circle) and the CME propagation direction (yellow arrow). White arrows in each panel point to roughly similar parts of the CME observed with the different instruments.} The closer the CME propagates in the plane-of-sky of the instrument, the higher the intensity in white-light. Adapted from \cite{Byrne10}.
\label{fig:multi_byrne}       
\end{figure}

In contrast to a flare, which is a rather localized phenomenon, the analyzes of CMEs and related coronal waves, propagating over large areas of the solar surface, as well as SEPs profit enormously from at least two viewpoints. The twin-spacecraft STEREO unprecedentedly provides, since its launch end of 2006, image data in EUV and white-light from multiple perspectives. STEREO consists of two identical spacecraft, STEREO Ahead (A) and Behind (B; lost signal October 2014), orbiting the Sun in a distance close to Earth, with STEREO-A being closer and STEREO-B further away from the Sun. The separation angle between the two spacecraft increases by about 45 degrees per year\footnote{Current position of STEREO and other spacecraft can be found under \url{https://stereo-ssc.nascom.nasa.gov/where.shtml}}. There are four instrument packages mounted on each of the two STEREO spacecraft, SECCHI comprising EUV and white-light coronagraphs and imagers \citep{Howard08_secchi}, IMPACT sampling the 3D distribution of solar wind plasma and magnetic field \citep{Luhmann08_IMPACT,Acuna08_IMPACT_MAG}, SWAVES tracking interplanetary radio bursts \citep{Bougeret08_SWAVES}, and PLASTIC measuring properties of the solar wind plasma characteristics \citep{galvin08_PLASTIC}. Conjoined with instruments from Earth perspective, such as SoHO (1995--), Hinode (2006--) and SDO (2010--), as well as ground based observatories (covering the radio and visual wavelength range, e.g., chromospheric H$\alpha$ and Ca II lines), the evolution of active regions together with flares, CMEs, and SEPs could be for the first time stereoscopically observed. Unfortunately, a big drawback for multi-viewpoint magnetic field investigations was the lack of magnetographs onboard STEREO \citep[this might be overcome by the ESA \textit{Lagrange} L5 mission planned to be launched in 2027; see also  e.g.,][]{gopal11_L5,Lavraud2016AScience}.

Besides having more than one vantage point, STEREO carries the heliospheric (HI) instruments, enabling to seamlessly observe the entire Sun-Earth line in white-light. They provide a unique long-term, synoptic data-set to be exploited for Space Weather application. Wide-angle image data allow to undoubtedly link CMEs to their interplanetary counterparts (ICMEs) as measured in-situ and to investigate in detail the in-situ signatures caused by the different CME structures and orientations. More details on ICMEs are given in Section~\ref{sec:sub3-5}. 

Figure~\ref{fig:multi_byrne} shows an Earth-directed CME observed from multiple perspectives and over a large distance range using STEREO data. From Earth perspective (shown in the middle panel), the CME is observed as weak halo event which makes it almost impossible to reliably determine a propagation direction and its radial speed. From STEREO perspective, the CME is observed close to the plane of sky of the instruments, lowering the projection effects for deriving its radial kinematics. Hence, the multiple viewpoints and homogeneous dataset of STEREO, enable to do 3D reconstructions of solar structures and to investigate projection effects with the attempt of correcting them, or at least limit and assess the uncertainties of the projected measurements. For SEPs, the identical instruments aboard the two spacecraft bring the advantage of having the same energy threshold, allowing systematic studies of SEPs coming from the same active region but related to a different magnetic connectivity and to probe the longitudinal dependencies.

%


\begin{figure}
  \includegraphics[width=1.\textwidth]{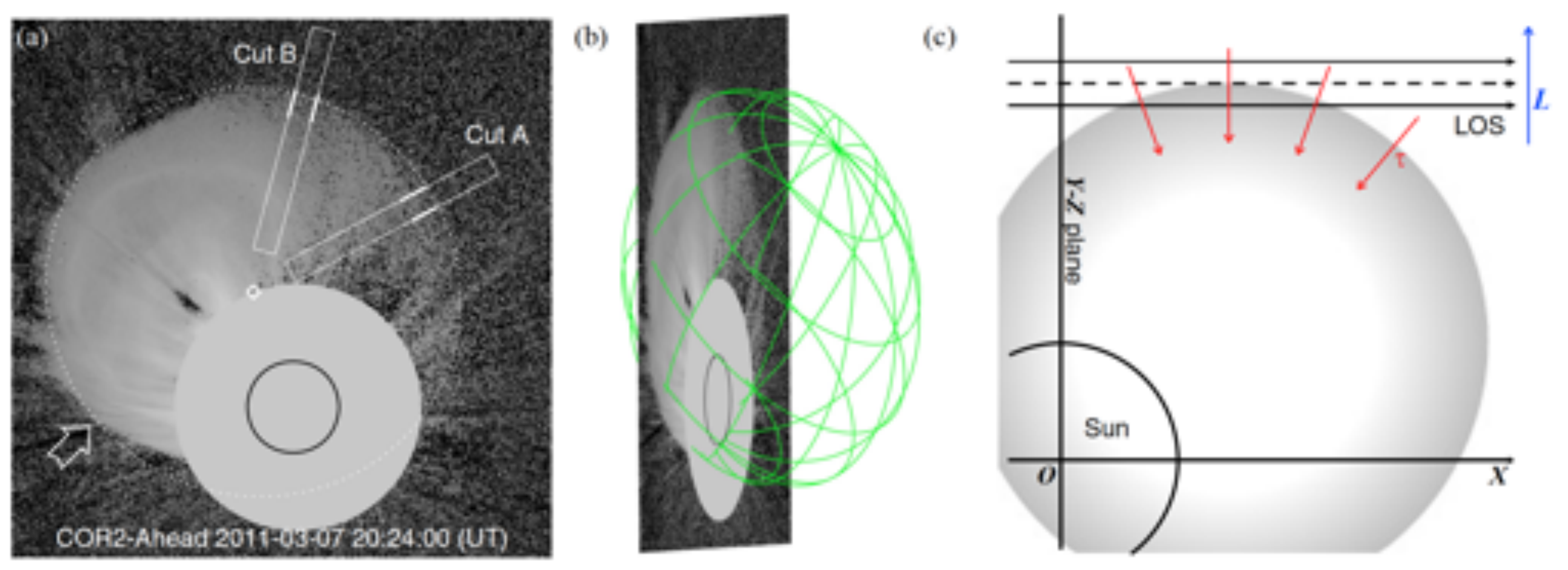}
\caption{CME from March 7, 2011: (a) Excess brightness image from STEREO-A COR2. 3D shock front (green mesh in panel (b)) projected on the image plane is shown with the dotted line. The diamond marks the geometric center of the ellipsoid model projected onto the same plane. (b) Excess brightness in panel (a) with the 3D shock front (green mesh) modeled with the ellipsoid model described in \cite{kwon17}. (c) Geometric relation among the Sun, shell-like sheath, and line-of-sight. A partial circle around the origin O is the solar disk. A shell-like sheath is represented in gray color. Arrows in black, blue and red are the line-of-sight, the projected shock normal on the image plane, and the actual shock normal in 3D, respectively. Taken from \cite{kwon18}.}
\label{fig:kwon}       
\end{figure}

Multi-spacecraft views enable to study especially the CME geometry and its substructures in more detail. With that, the different manifestations of shock and driver could be well confirmed and it is now well acknowledged that the outer envelope of the observed CME presents the expanding shock or compressed shell that encompasses the driver \citep[e.g.,][]{Ciaravella06,Ontiveros09,Vourlidas13}. As the different parts have a different impact on Earth, for forecasting purposes, measurements of the CME's outer front should be clearly specified (e.g., shock front versus magnetic structure). In addition, the long-standing question whether halo CMEs would be different compared to limb events \citep[see e.g., the Living Reviews by][]{chen11} could be solved. The shock shell of the CME can be presented as sphere-like structure expanding over 360 degrees (see Figure~\ref{fig:kwon}). It was found that especially strong events (having a large compression) can be observed as halo CME independent of the viewpoint \citep{kwon15,kwon18}. In that respect STEREO data also showed that the outermost shock component of the CME matches well the solar surface structure of coronal EUV waves \citep{Kienreich09,Patsourakos09,Veronig2010FirstWave,kwon17}. Therefore, observations of the surface signatures of CME related coronal waves give supportive information about the CME expansion and propagation direction and should be closely monitored for early Space Weather forecasting.


The CME speed is actually a mixture of lateral and radial expansion dynamics making it tricky to derive the ``true'' propagation behavior. Multiple viewpoints enable to separately study projected versus deprojected speeds and radial versus lateral expansion behaviors of CMEs. Comparing single and multiple spacecraft data revealed that single viewpoint measurements are definitely valid. However, especially measurements of the CME width (or lateral expansion) and speed for slow CMEs (deprojected speeds below 900km/s) reveal high uncertainties depending on the perspective \citep{Shen13,Balmaceda18}. Models taking into account projection effects showed to significantly decrease the uncertainties in forecasting the arrival times of CMEs \citep[e.g.,][]{colaninno13,Mishra13,shi15,makela16,rollett16}. A well known empirical relation exists between the radial and the lateral expansion speed, $V_{\rm rad} = 0.88 V_{\rm exp}$, as described by \cite{dalLago03} and \cite{schwenn05}. Follow-on studies showed that this relation can also be described by the CME half-width, $w$ (assuming a cone model), given by $f(w)=1/2(1+\cot{w})$ and that the kinematics of extremely fast CMEs is better estimated by $V_{\rm rad} \approx V_{\rm exp}$ \citep{michalek09}. Moreover, statistical studies revealed that the relationship between the radial and lateral expansion speed is a linear function, hinting towards the self-similar expansion behavior of CMEs already close to the Sun \citep{Vourlidas17,balmaceda20}. However, in the low corona, for some events a strong overexpansion is observed \citep[e.g.,][]{Patsourakos10}. The assumption of a rather self similar expansion is found to be valid for most of CME events when propagating in interplanetary space \citep[e.g.,][]{Bothmer98,Leitner07,demoulin08,Gulisano12,vrsnak19}.

\begin{figure}
  \includegraphics[width=1.\textwidth]{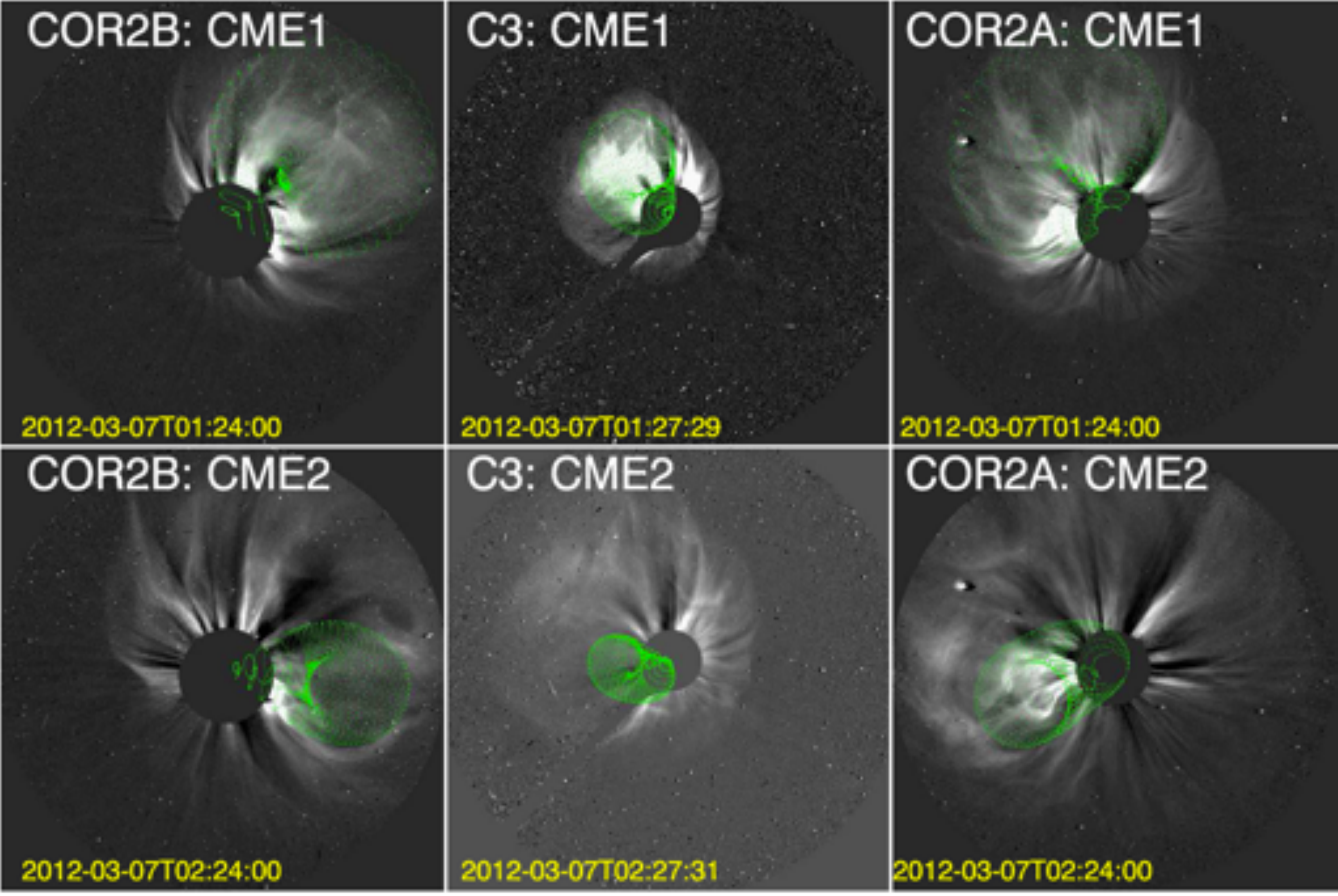}
\caption{March 7--11, 2012: GCS fitting (green mesh) of two CMEs (CME1: top panels. CME2: bottom panels - note that CME1 is visible as extended bright structure in these images) using white-light data from the three spacecraft, STEREO-A, SOHO and STEREO-B. The first, second, and third columns contain coronagraph images from COR2 aboard STEREO-B, C3 aboard SOHO, and COR2 aboard STEREO-A, respectively. Taken from \cite{patsourakos16}.}
\label{fig:patsourakos16}       
\end{figure}

\begin{figure}
  \includegraphics[width=1.\textwidth]{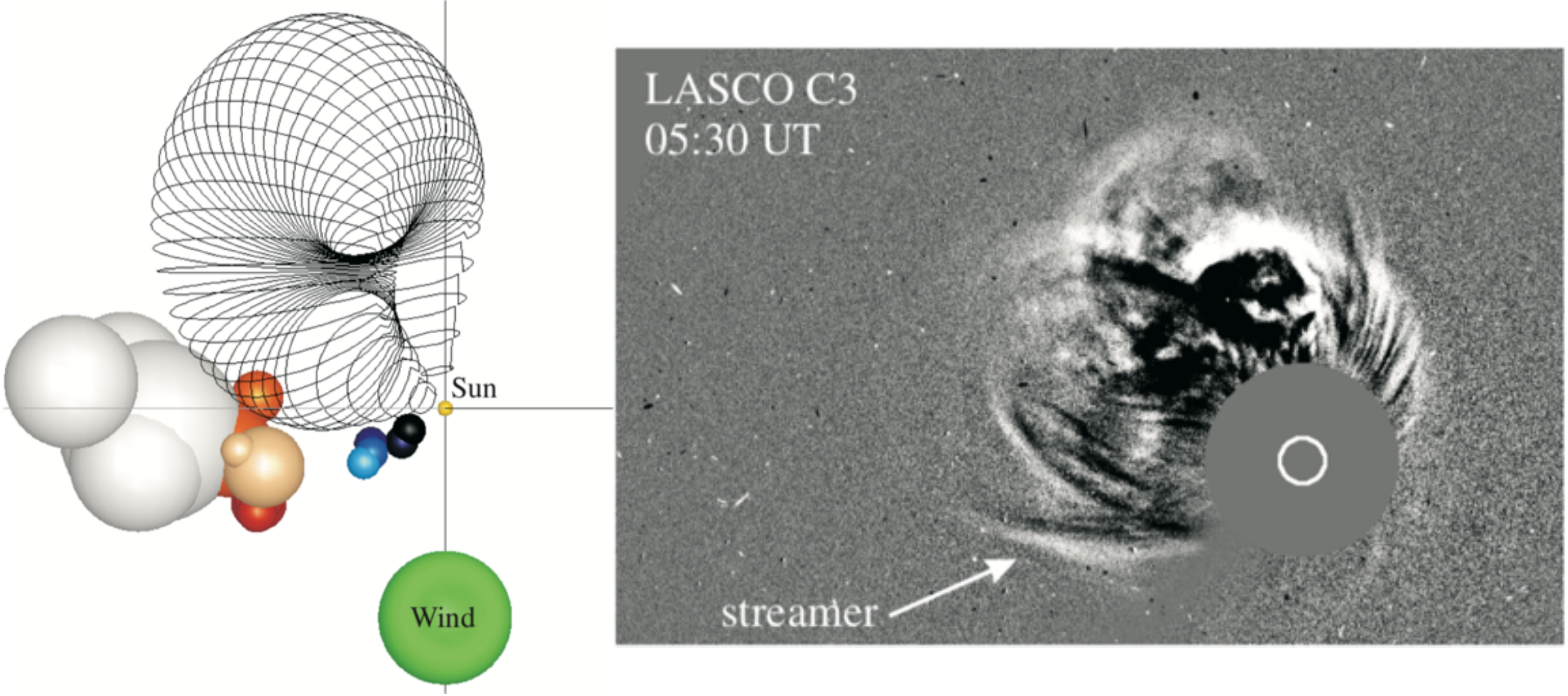}
\caption{The CME flux rope obtained from GCS 3D reconstruction (black grid croissant). The 3D reconstruction of the radio type II burst (dark and light blue spheres) using gonopolarimetric technique. The yellow sphere represents the Sun (left panel). View of the flux rope and radio sources as seen from Earth (right panel) SOHO/LASCO C3 image showing the CME as seen from Earth, for comparison with panel. Adapted from \cite{magdalenic14}.}
\label{fig:gono_radio}       
\end{figure}

Since we cannot gather the full complexity of the CME structure, idealized geometries assuming self-similar expansion, act as basis of many CME models and 3D reconstruction techniques that were developed over the past years. Basic models make use of a simple cone-type geometry \citep[e.g.,][]{St.Cyr00,schwenn05,Michalek06,Xie04}. With the availability of image data from multiple views, those tools were refined and full 3D reconstructions were enabled from which estimates of the deprojected kinematics, geometry, and propagation direction are derived. Methods comprise, e.g., inter-image tie points and triangulation in various wavelength ranges \citep[see e.g.,][]{Harrison08,Howard08,maloney09,Reiner09,Temmer09,Liewer10}, forward models related to white-light data \citep[e.g.,][]{wood09}, or center of mass calculations \citep{Colaninno09}. Also online tools were made available, such as e.g., the CCMC tools StereoCat\footnote{StereoCat \url{https://ccmc.gsfc.nasa.gov/analysis/stereo/}}. A well known and widely used technique is the graduated cylindrical shell (GCS) forward model developed by \cite{Thernisien06,Thernisien09}. Coronagraph image data showing the CME from at least two different vantage points are required, on that an idealized flux rope structure in the form of a croissant is fitted. The GCS model depends on a number of free parameters, such as the flux-rope height and angular width as well as the aspect ratio which determines the rate of self-similar expansion. Figure~\ref{fig:patsourakos16} gives the 3D reconstruction of two CME events using GCS applied on STEREO and LASCO data in a study by \cite{patsourakos16}. Especially for multiple events, the investigation and determination of the cause of geoeffectiveness is rather challenging as the processes happening on the Sun are complex. In that respect, geometrical fitting methods help to derive the propagation direction of a particular solar event in order to reliably link it to a geoeffective event at Earth. In a similar way, stereoscopy can be applied also on radio data. Figure~\ref{fig:gono_radio} shows results from so-called goniopolarimetric observations using WIND and STEREO spacecraft data studying the location of radio type II bursts. That method is used to derive the direction of arrival of an incoming electromagnetic radio wave, its flux, and its polarization.

Besides geometry related information, STEREO coronagraph data can also be used to derive the CME deprojected mass close to the Sun \citep{Colaninno09,Bein2013TheObservations} which, together with the early acceleration phase, are taken for better estimating the energy budget between flares and CMEs (see also Section~\ref{sec:budget}). STEREO and its wide-angle HI instruments also enable to derive the 3D geometry of compressed density structures like CIRs \citep[see][]{Rouillard08,wood10}. Sometimes the disentanglement between CMEs and CIRs in HI is tricky, hence, it needs careful inspection of the data when tracking specific features \citep[see e.g.,][]{Davis10}. 

Using multi-viewpoint data and applying different reconstruction techniques we vastly gained important insight about CME characteristics. Moreover, the results stemming from STEREO observations clearly challenged existing CME and SEP models. However, using idealized geometries, the real 3D structure of a CME or SEP paths can only be approximated and we need to keep in mind that there are strong deviations from these. Especially, in interplanetary space, the geometry of the CME front clearly changes, as flanks and nose interact differently with the non-uniform solar wind \citep[pancaking effect; see e.g.,][]{riley04,Nieves-Chinchilla12}. In addition, the CME shape might vary due to intensity changes as the relative position to the Thomson sphere changes, that makes the tracking of specific white light structures complicated \cite[e.g.,][]{vourlidas06}. Therefore, the derived deprojected values and forecasts based on them still need to be treated with caution \citep[see also][]{Mierla10,riley18}. For more details on the complex interactions of CMEs with their surroundings I refer to the review by \cite{manchester17}.

For more information on solar stereoscopy and tomography techniques, applied to various large-scale structures in the solar corona, I refer to the Living Reviews by \cite{aschwanden11}. For comprehensive investigations, the EU funded project HELCATS\footnote{Heliospheric Cataloguing, Analysis and Techniques Service; \url{https://www.helcats-fp7.eu}} established databases ready to use for analyzing STEREO 3D CME characteristics and HI CME tracks on a statistical basis \citep[see e.g.,][]{Murray18,Barnes19}. Out of that an extensive ICME catalogue\footnote{\url{https://helioforecast.space/icmecat}} was compiled by \cite{moestl17} and \cite{palmerio18}. A conjunction catalogue covering CME in-situ measurements by two or more radially aligned spacecraft (MESSENGER, Venus Express, STEREO, Wind/ACE) is given by \cite{salman20}.


\section{Interplanetary counterparts of CMEs: ICMEs}\label{sec:sub3-5}

\begin{figure*}
  \includegraphics[width=\textwidth]{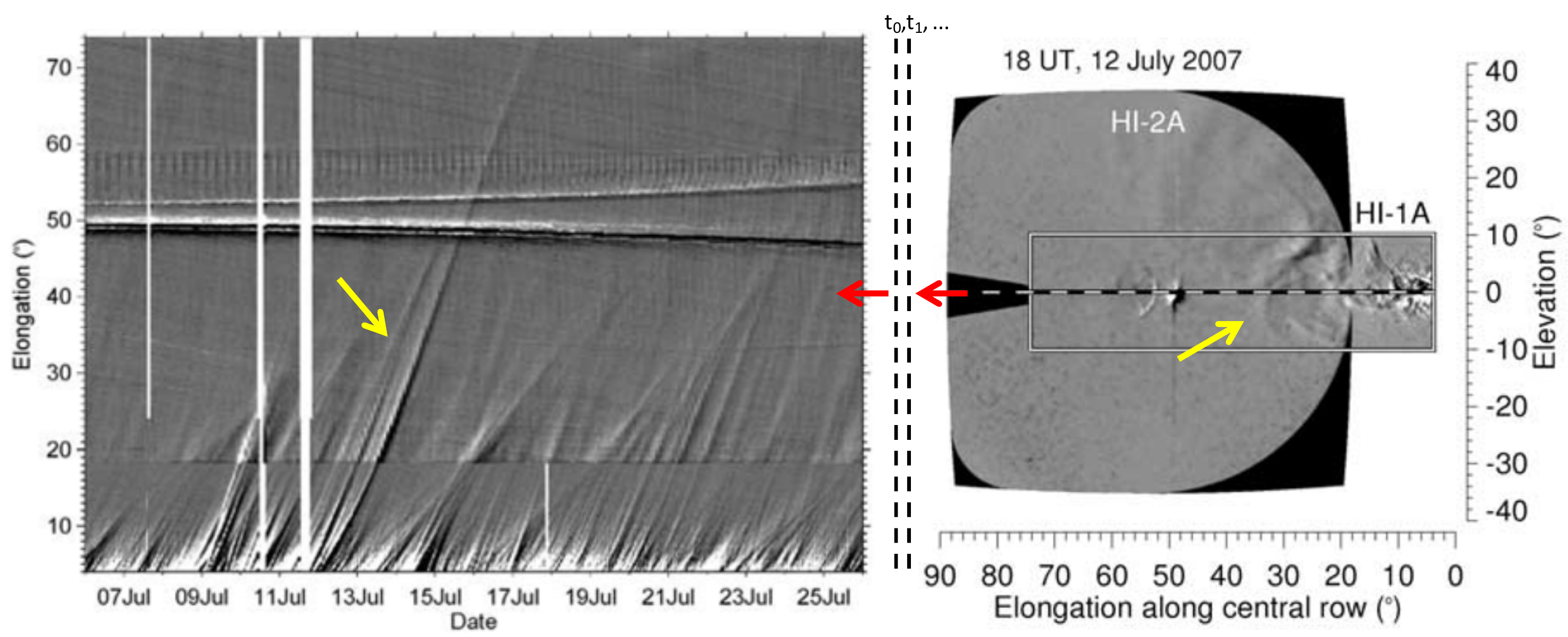}
\caption{From running difference STEREO-A HI1+2 image data the central rows are extracted (right) at each time step and rotated by 90 degrees (middle). From this a time-elongation plot (so-called Jmap) is constructed (left). The CME front is marked by a yellow arrow in the direct image as well as in the Jmap. Adapted from \cite{davies09}.}
\label{fig:jmap}       
\end{figure*}

Newly developed imaging capabilities clearly enhanced our understanding about the relation between solar eruptions, CMEs, and their counterparts in interplanetary space (ICMEs). SMEI, the Solar Mass Ejection Imager \citep[SMEI:][]{Eyles03} on the Earth-orbiting Coriolis spacecraft, was the first heliospheric white-light imaging instrument covering the Sun-Earth space \citep[for more details see the review by][]{howard13}. The successor of SMEI are the heliospheric imagers \citep[HI:][]{Eyles09} aboard STEREO \citep{kaiser08_STEREO}. The WISPR instrument \citep{vourlidas19} aboard the Parker Solar Probe mission and the SoloHI instrument \citep{howard20} aboard Solar Orbiter build upon the STEREO/HI heritage and make similar observations of the inner heliosphere. The observational principle is like a coronagraph, but as these are wide-angle instruments, they observe much larger distances from the Sun enabling the tracking of CMEs throughout interplanetary space. The unprecedented image data facilitated the tracking of CMEs through interplanetary space and with that could unambiguously relate the CME white-light structure to in-situ measurements \citep[see e.g.,][]{moestl09_APJL,moestl14} and moreover, to get better insight on how CMEs interact with the ambient solar wind structures. Figure~\ref{fig:jmap} shows a so-called Jmap which is constructed from running difference white-light HI data covering the Sun-Earth distance range. By extracting the central part of the HI images in the horizontal direction, the ICME front can be rather easily followed as function of the elongation angle. Before further analysis, the elongation-time measurements need to be converted into radial distance. For that, methods assume either a certain CME geometry and apply the propagation direction of the CME \cite[see e.g.,][]{lugaz10} or use fitting functions \cite{Rouillard08}. These procedures cover rather high uncertainties in the derived kinematical profiles, that needs to be taken into account when interpreting CME propagation profiles for interplanetary space \citep[e.g.,][]{rollett12,liu13}.

It is well known that CMEs during their propagation phase tend to get adjusted to the ambient solar wind flow owing to the drag force exerted by the ambient solar wind \citep{Gopalswamy01,wang04}. As consequence, CMEs which are faster than the ambient flow speed get decelerated while those which are slower get accelerated. This alters their speed, hence, travel time and with that has impact on Space Weather forecasting. The adjustment to the ambient flow speed happens most probably in interplanetary space \citep[e.g.,][]{Sachdeva15}. At which distance exactly depends on the competing forces acting on a CME, Lorentz versus drag force \citep[e.g.,][]{vrsnak08}. The longer the CME is driven, hence, the longer the magnetic reconnection process is ongoing (which might be inferred from flare emission and growing post eruptive arcades), the farther away from the Sun the adjustment may occur. Empirical relations found between CME kinematics and flare properties (flare ribbons, coronal dimmings, or post-eruptive arcade regions) actually may be used to estimate the reconnected flux that empowers the CME \citep[e.g.,][]{gopalswamy17,Temmer17_flar-cme,dissauer18,Tschernitz18}\footnote{A database of more than 3000 solar flare ribbon events observed by SDO and reconnection flux is given in \cite{kazachenko17}.}. The amount of drag from the solar wind depends on the relative speed and density between the solar wind and the CME as well as the CME width/size. It is found that wide CMEs of low mass tend to adjust rather quickly to the solar wind speed and, hence, their transit time (i.e., how long a CME needs to traverse a certain distance) is determined primarily by the flow speed in interplanetary space. Narrow and massive CMEs propagating in a fast solar wind have the shortest transit times \citep[see e.g.,][]{vrsnak10}. 

\begin{figure*}
  \includegraphics[width=\textwidth]{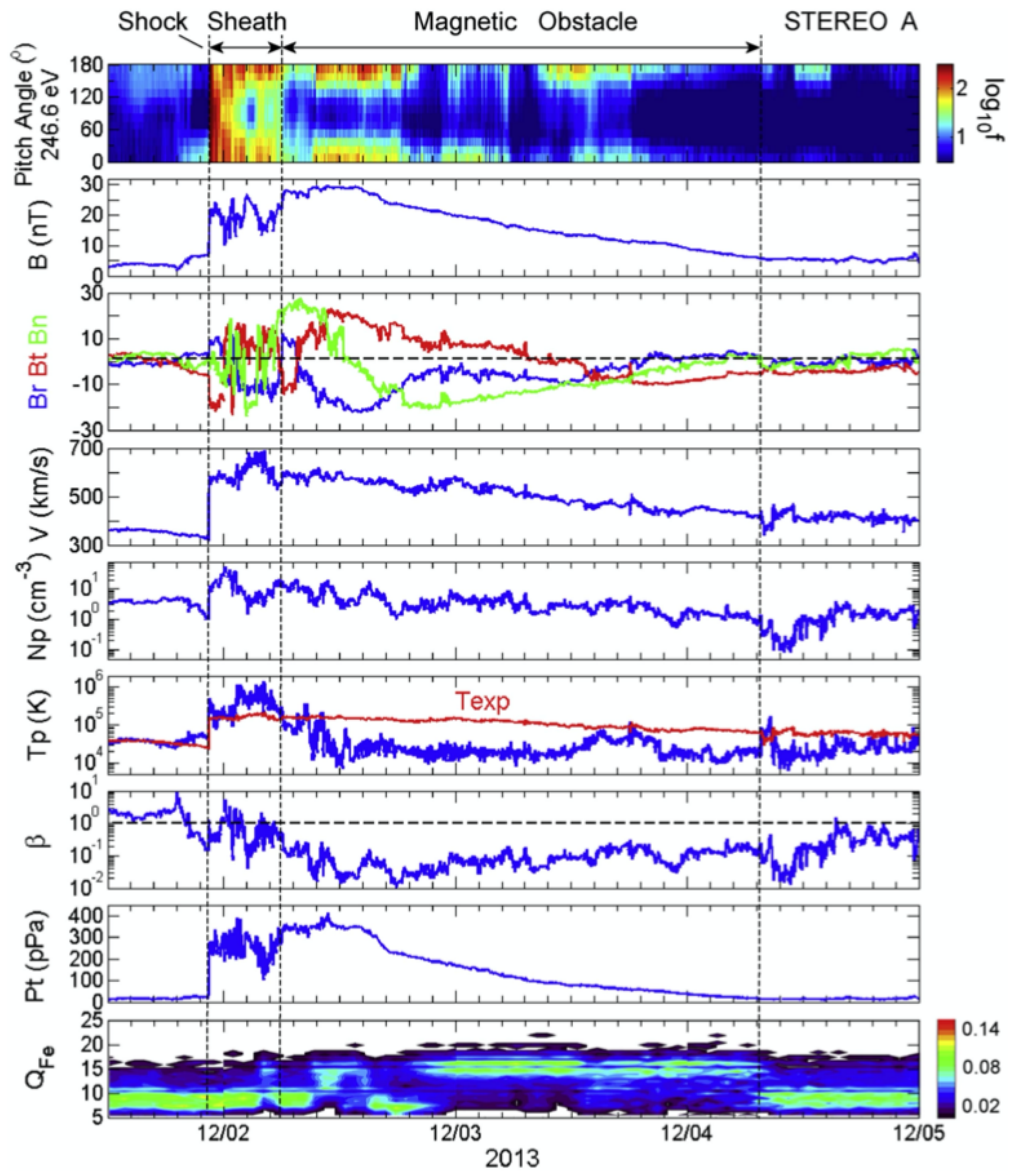}
\caption{STEREO-A in-situ measurements and identification of a CME together with its closed magnetic structure. From top to bottom: pitch-angle distribution data of suprathermal electrons, total magnetic field intensity, magnetic field vectors (in RTN coordinates), solar wind proton bulk speed, proton number density, proton temperature \citep[in red the expected proton temperature is given calculated from an empirical relation to the solar wind speed as given by][]{Richardson95}, plasma-beta, total pressure, distribution of the iron charge state. Vertical dashed lines mark the shock-sheath, and the boundaries of the magnetic structure. Taken from \cite{Jian18}.}
\label{fig:ICME}       
\end{figure*}

Figure~\ref{fig:ICME} shows typical in-situ signatures of a well-defined ICME at 1~AU, revealing a simultaneous jump of all measured components (shock) with a subsequent sheath structure (compressed plasma) of increased density, speed and turbulence, that is followed by signatures of a smooth and enhanced magnetic field together with a rotation as observed in the vector components (changing from plus to minus or vice versa). The ICME magnetic structure is usually identified by that smooth field rotation (flux rope), a plasma-beta lower than 1 (referring to a dominant magnetic component), a low temperature, and a linearly decreasing proton speed \citep[see also the Living Reviews by][]{kilpua17}. Sometimes that flux rope can be associated with twisted structures observed already in white-light image data. 
Having a long-lasting southward directed magnetic field (measured in the $B_{\rm z}$ component), flux ropes are the main contribution of strong geoeffectiveness. The passage of rather isolated magnetic ejecta at 1~AU typically takes about 1 day \citep[cf.][]{RC10}\footnote{A near-Earth ICME catalogue is given under: \url{http://www.srl.caltech.edu/ACE/ASC/DATA/level3/icmetable2.htm}}. Hence, geomagnetic disturbances may last for many hours. Flank hits, interacting CMEs and complex ejecta, can have passage durations of about 3 days at 1~AU \citep[see][]{Burlaga02,Xie06,Marubashi2007Long-durationModels,Mostl2010STEREO2010}, affecting the Earth's atmospheric layers over a much longer time range, and, hence, causing stronger geomagnetic effects (see also Section~\ref{sec:preconditioning}).

By combining remote sensing and in-situ data using multi-spacecraft reconstruction methods, it is revealed that from in-situ measurements we observe localized variations of the magnetic field behavior that may not be representative of the global structure \citep[see][]{mostle08,moestl09_APJL,rouillard10,farrugia11,DeForest2013TrackingEjection}. Studies using multi-spacecraft encounters separated in radial distance and longitude give insight on the magnetic coherence of ICMEs on various scales and with that raise questions on the inner structure of CMEs as well as their interaction processes with the interplanetary magnetic field \citep[see e.g.,][]{good18,lugaz18}. Using flux rope reconstructions methods applied on in-situ measurements \citep[see e.g.,][]{al-haddad13} a comparison between the physical parameters derived close at the Sun with those measured in-situ can be performed enabling to interpret changes in the mass, flux, etc. due to the interaction with the interplanetary solar wind \citep[see e.g.,][]{bisi10,Temmer17_flar-cme,temmer21}. Especially the reconnection of the magnetic flux rope with the interplanetary magnetic field is found to lead to either a loss of magnetic flux (so-called erosion) or adding of magnetic flux \citep[see e.g., ][]{Dasso07,manchester14_CMEerosion,Ruffenach15}. Removing or adding magnetic flux may lead to a change in the ICME propagation behavior. Filament material is found less often from in-situ measurements (identified by low charge state species) despite the fact that most CMEs are accompanied by filament eruptions. Heating mechanisms or simply missing the cold filament material due to the localized in-situ measurements might be a reason for that \citep{filippov02}. This is supported by findings that magnetic ejecta are only partly filled with hot plasma related to heating by the flare \citep[e.g.,][]{gopalswamy13_filament}.

More details on the relation between white light remote sensing image data and in-situ measurements, including proper nomenclature, is given by \cite{rouillard11}. A review on multi-point ICME encounters before and during the early years of STEREO is given by \cite{kilpua11}. The recent review by \cite{luhmann20} comprises a thorough overview on the ICME propagation in the inner heliosphere.

\section{Solar Energetic Particles (SEPs)}\label{sec:4}
\subsection{General characteristics}
SEP events are observed in-situ as enhanced electron, proton and heavy ion flux (and as increased level of cosmic rays on ground) largely exceeding the thermal energy levels, ranging from keV to GeV. Strong fluxes of energetic protons (so-called proton events) cause strongest geoeffective phenomena. High energy SEPs in the range of GeV reach the Earth within less than 10 minutes and may produce ground level enhancements (GLE; measurable in neutron monitors at Earth surface), that are of special interest as they have major effects on crewed spaceflight and aircraft due to the increased radiation exposure \citep[e.g.,][]{Malandraki18_book}. 

\begin{figure*}
  \includegraphics[width=\textwidth]{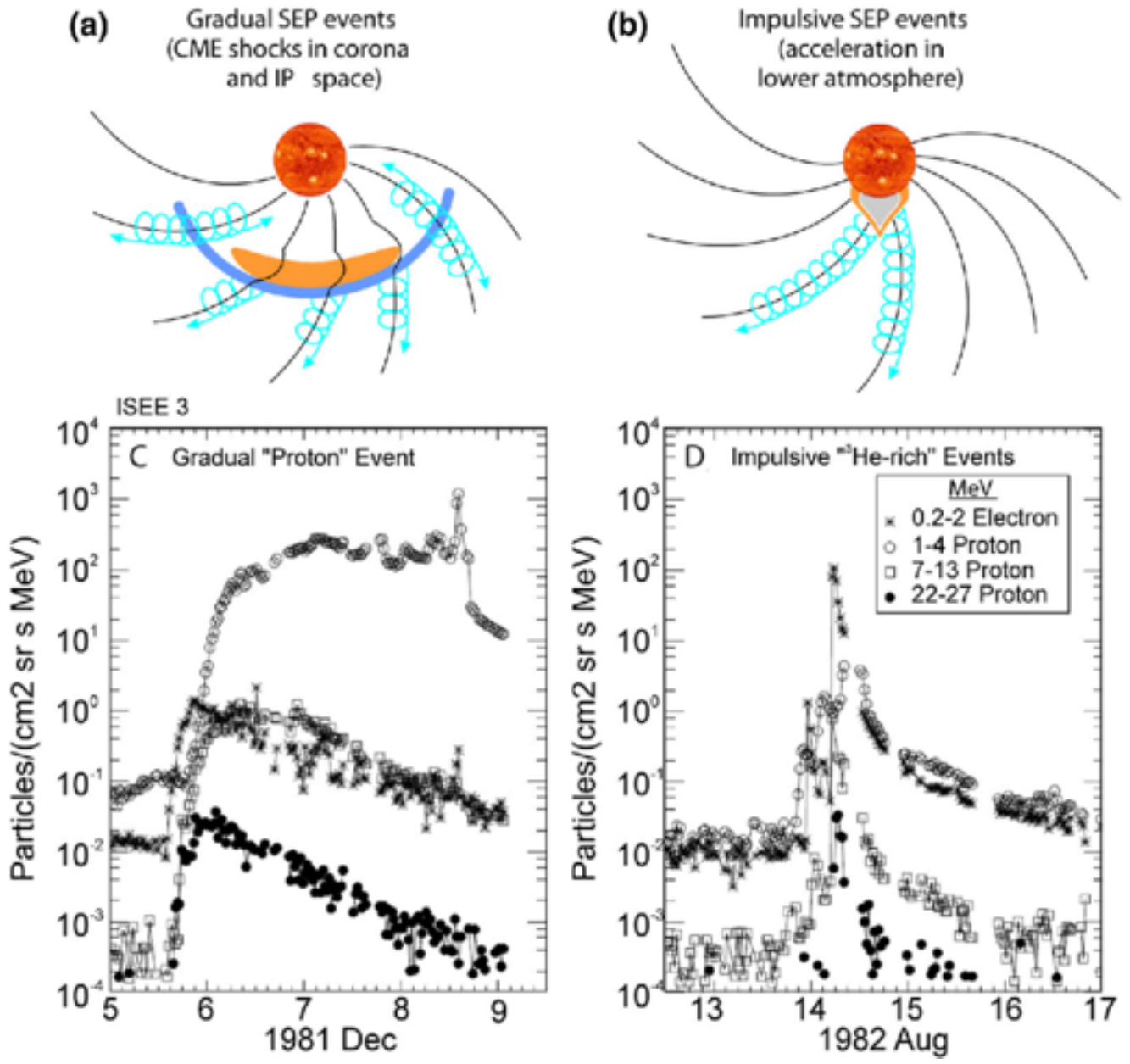}
\caption{Separation into gradual and impulsive SEP events and their suggested driving mechanisms. a) Gradual SEP events result from diffusive acceleration of particles by large-scale shocks produced by CMEs and populate interplanetary space over a wide range of longitudes. b) Impulsive SEP events result from acceleration during magnetic reconnection processes in solar flares and are observed well when magnetically connected to the flare site. Intensity-time profiles of electrons and protons in c) gradual and d) impulsive SEP events. Taken from \cite{Desai16} after \cite{Reames99}.
}
\label{fig:reames99}       
\end{figure*}

In general, there are two populations of SEP events, gradual and impulsive ones \citep[e.g.,][cf. Figure~\ref{fig:reames99}]{Reames99}. It is the different temporal scaling which is disentangling those two populations. Driving agents acting on longer time scales are related to CME shock acceleration mechanisms. However, gradual events seem to be accompanied as well by an impulsive part which is thought to be related to short-time magnetic reconnection processes, as observed in flares. Obviously gradual events are caused by both driving agents prolonging the acceleration process but on a less energetic level \citep[see also][]{Anastasiadis19}. Impulsive events are also obtained to be related to those SEP events where the location of the particle accelerator is magnetically well connected to the observer. SEP/GLE events are found to have the hardest spectra and the largest initial acceleration \citep{Gopalswamy16_backsideSEP}. There are still many open issues about the processes leading to energetic particles as well as about their (suprathermal) seed populations in the corona and interplanetary space \citep[e.g.,][]{Mason99,Desai06,Mewaldt12}. Clearly, the primary condition for the production of SEPs is the opening of magnetic field lines into interplanetary space (as for eruptive events) and that accelerated particles have access to that open field lines. It is confirmed that for confined flares no SEPs are observed \citep[e.g.,][]{Trottet15}. As shocks play an important role in the acceleration of particles, coronal shock waves on the solar surface and interplanetary space related to CMEs as well as interacting CMEs are investigated in relation to SEPs \citep[see e.g.,][and related MHD modeling results, e.g., \citeauthor{Pomoell08}, \citeyear{Pomoell08}]{Park13,Lario14,Miteva14}.

SEP events that become Space Weather effective are controlled by many factors, such as the source region location of the eruption (longitude and latitude) and width of the CME, background solar wind, seed populations, multiple CMEs and their interaction, or magnetic field configuration near the shock. For example, narrow CMEs (slow or fast) do not efficiently accelerate particles \citep[see e.g.,][]{kahler19}, and CMEs originating from the eastern hemisphere are less likely to create a SEP event near the Earth because of the weak magnetic connection between Sun and the Earth. The highest-energy particles are most likely accelerated close to the shock nose where the shock is strongest, while the lower energy particles are accelerated at all regions \citep[see e.g.,][]{bemporad11,gopalswamy18}. Hence, also the ecliptic distance to the shock nose, i.e., the event source region latitude, is an important parameter for SEP prediction \citep[see e.g.,][]{gopalswamy13}. 

\begin{figure*}
  \includegraphics[width=1.\textwidth]{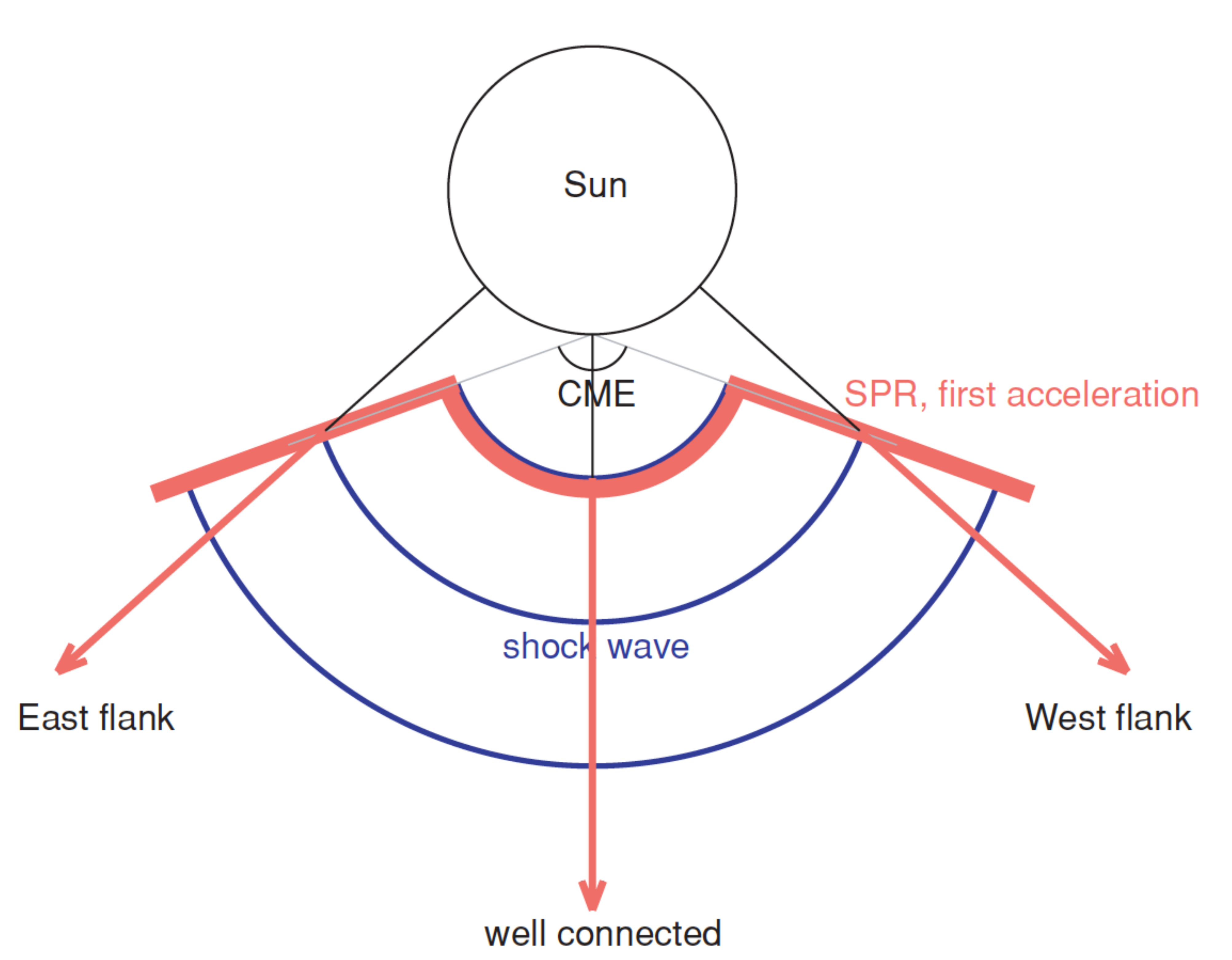}
\caption{Cartoon showing a possible acceleration scenario for SEPs. The radial field lines (black lines) are hit by the CME shock front (blue) at different heights for the nose and the flank. The solar particle release (SPR) likely begins at a 2--4 solar radii (marked by the red region) for the apex or higher up for the flanks. Taken from \cite{reames09}.}
\label{fig:reames09}       
\end{figure*}

Knowledge about the magnetic connectivity is a crucial parameter in order to detect SEPs in-situ and to relate them to the proper driving agent \citep{reames09}. Figure~\ref{fig:reames09} depicts the propagating idealized circular-shaped CME shock front in relation to the radially oriented magnetic field lines near the Sun. The cartoon describes a scenario in that a narrow range of heights (2--4 solar radii) exists where compression is sufficient for effective particle acceleration \citep[e.g.,][]{Cliver04} and for having a good connection to the observer, with heights increasing at the eastern and western flanks. Concluding, the connectivity changes with distance from the Sun. On the other hand, flare locations lying close to open structures like coronal holes, have different magnetic configuration and facilitate the acceleration of particles into the heliosphere \citep{cane88,Reames96,shen06}. In that respect, the interplay between open and closed magnetic field is important to know and due to the lack of observations needs to be supported by reliable coronal modeling.

\begin{figure*}
\includegraphics[width=\textwidth]{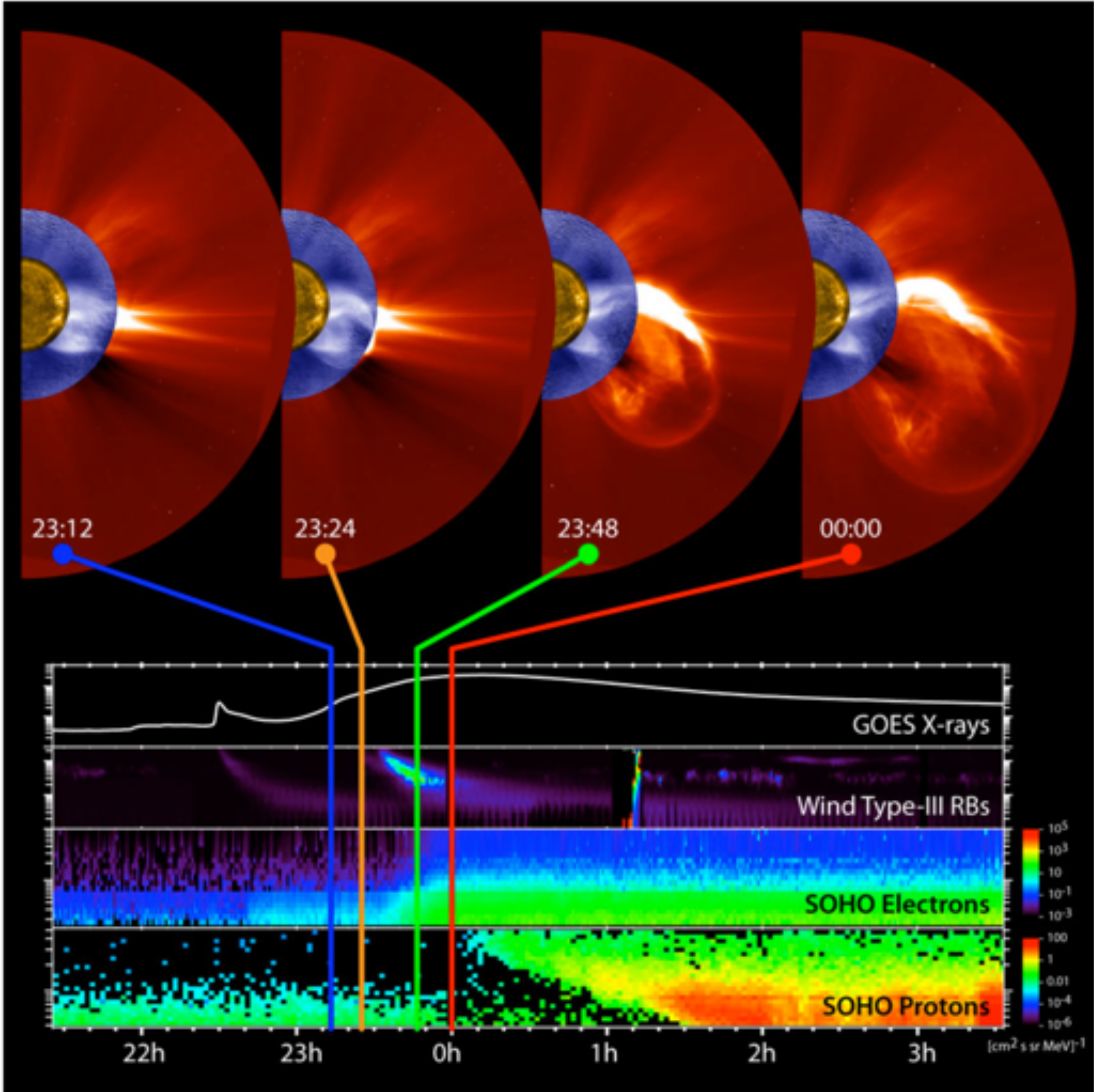}
\caption{In the top row, the January 1, 2016, eruptive event appears in the sequence of images from SDO AIA (gold), MLSO K-Cor (blue), and SOHO/LASCO (red). A fast CME associated with an SEP event detected near Earth, is seen appearing off the southwest limb of the Sun. The time profiles reveal that data from the ground-based K-Cor coronagraph could be used for a timely warning of particle events as described in that case study. Taken from \cite{stcyr17}. 
}
\label{fig:stcyr17}       
\end{figure*}

For Space Weather forecasting purposes, it is desired to derive clear signatures showing that particles associated to eruptive events were able to escape to the high corona and interplanetary space. Therefore, the monitoring of possible radio emission (from decimetre and longer waves) is found to be of utmost importance \citep[e.g.,][]{Klein10}. Observations of flares in the high energy range provide additional information on the location, energy spectra, and composition of the flare accelerated energetic particles at the Sun that can be compared to 1AU SEP events \citep{lin06}. RHESSI imaging capabilities could show that flare $\gamma$-ray sources are not co-spatial with flare HXR sources \citep{fletcher11}. \cite{Laurenza09} developed a technique for short-term forecasting of SEPs based on flare coordinates and flare flux together with the time‐integrated intensity of SXRs and type III radio emission ($\sim$1 MHz). Similar, the forecast of the occurrence of SEP events could be determined using the peak ratios in flare fluxes measured over (0.05–0.4 nm)/(0.1–0.8 nm) as described by \cite{kahler18}. For improved SEP forecasting, it is suggested to take into account parameters from both driving agents, flares and CMEs \citep[see][]{Klein17}. Figure~\ref{fig:stcyr17} shows for an eruptive event the timing between flare SXR emission (GOES flux), radio type III burst (WIND), electron and proton spectra measured at 1AU (SOHO), and combined white-light image data from the ground-based K-Cor coronagraph and LASCO/C2 instrument. Especially, observations of the early evolution of CMEs and derivation of shock formation heights (see also Sect.~\ref{shock-form}) as well as distribution of Mach numbers along the shock surface \citep{rouillard16} might give some lead time for SEP forecasting. Monitoring the generation of flare associated coronal surface waves \citep[EIT/EUV waves; see][]{Thompson98}, gives additional hints on shocks ignited by the CME lateral expansion (for confined events no coronal waves are observed).

\subsection{SEPs observed from multiple viewpoints}

\begin{figure*}
  \includegraphics[width=1.\textwidth]{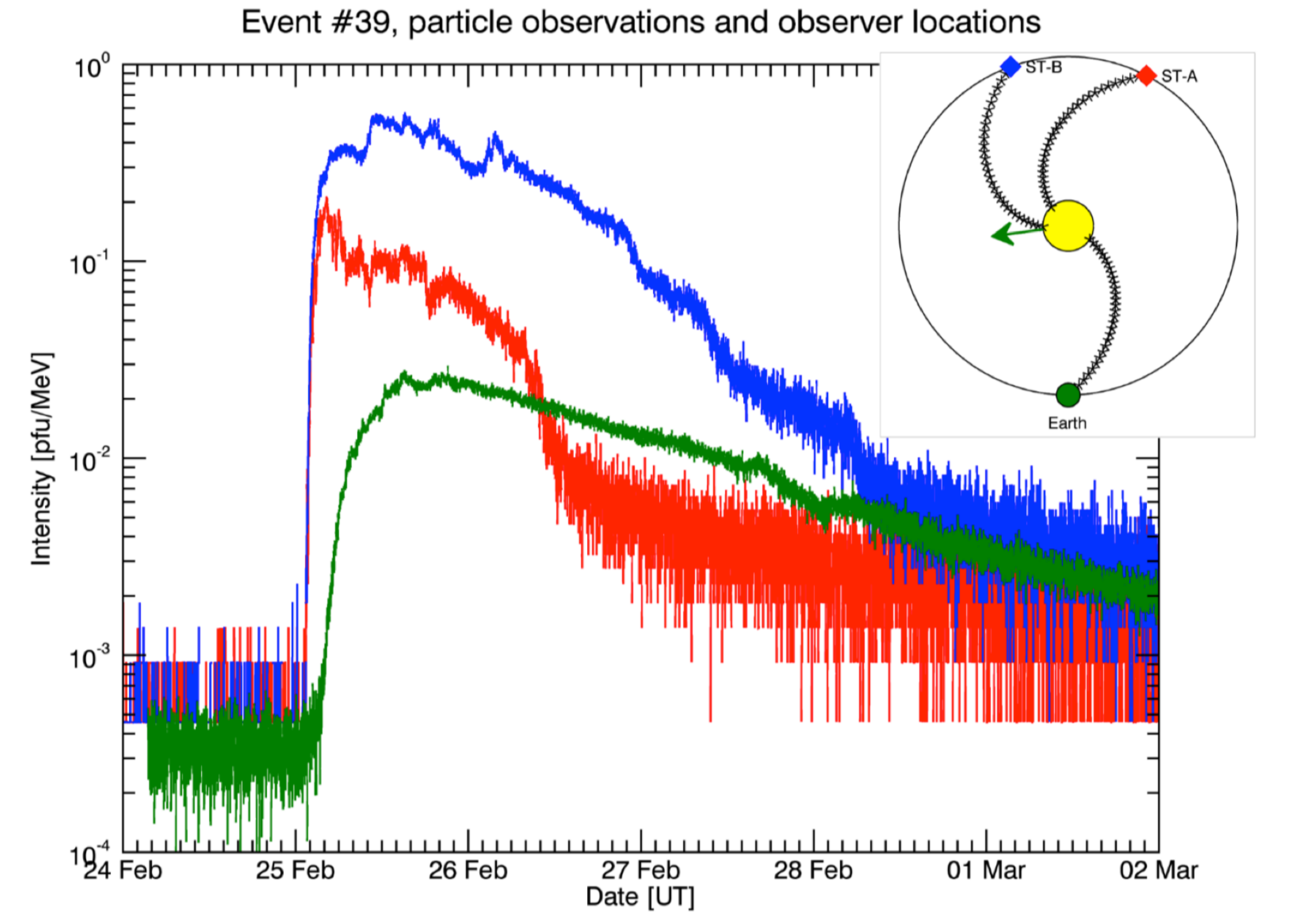}
\caption{February 25, 2014 SEP event and detected proton intensities (red = STEREO-A/HET, blue = STEREO-B/HET, green = SOHO/ERNE). Inset: relative locations of the STEREO spacecraft and the Earth during the event. The arrow pointing out from the Sun shows the location of the SEP producing active region (at longitude E82), and the asterisks mark the nominal Parker spiral magnetic field lines connecting each observer to the Sun. Taken from \cite{Paassilta18}.}
\label{fig:sep-stereo}       
\end{figure*}

Using STEREO data, the observation of wide-spread SEP events shed new light on the possible generation mechanisms together with the lateral expansion of CMEs and their interaction with other coronal structures \citep[see][]{rouillard12}. Figure~\ref{fig:sep-stereo} gives for the February 25, 2014 SEP event intensity profiles as measured by different spacecraft that are separated from Earth by 152 degrees (STEREO-A) and 160 degrees (STEREO-B). The SEP producing eruptive event is located at E82 (marked by the green arrow). As STEREO-B is closest to the SEP source, that flux profile reveals the highest intensity. The related CME was observed as halo event from Earth, having a projected speed of more than 2000~km/s and before that, other CMEs were launched from that region. SEPs may be directed to wide-spread angles by field line draping around the closed magnetic field of the CME and/or complex magnetic fields due to CME-CME conglomerates, or CME-CIR interaction \citep[e.g.,][]{Dresing16,Dresing18,Gomez-Herrero11,Gomez-Herrero17,Xie17,Guo2018ModelingMeasurement}. For more details on wide-spread SEP events, including a comprehensive catalogue see \cite{Paassilta18}.

\begin{figure*}
 \includegraphics[width=1.\textwidth]{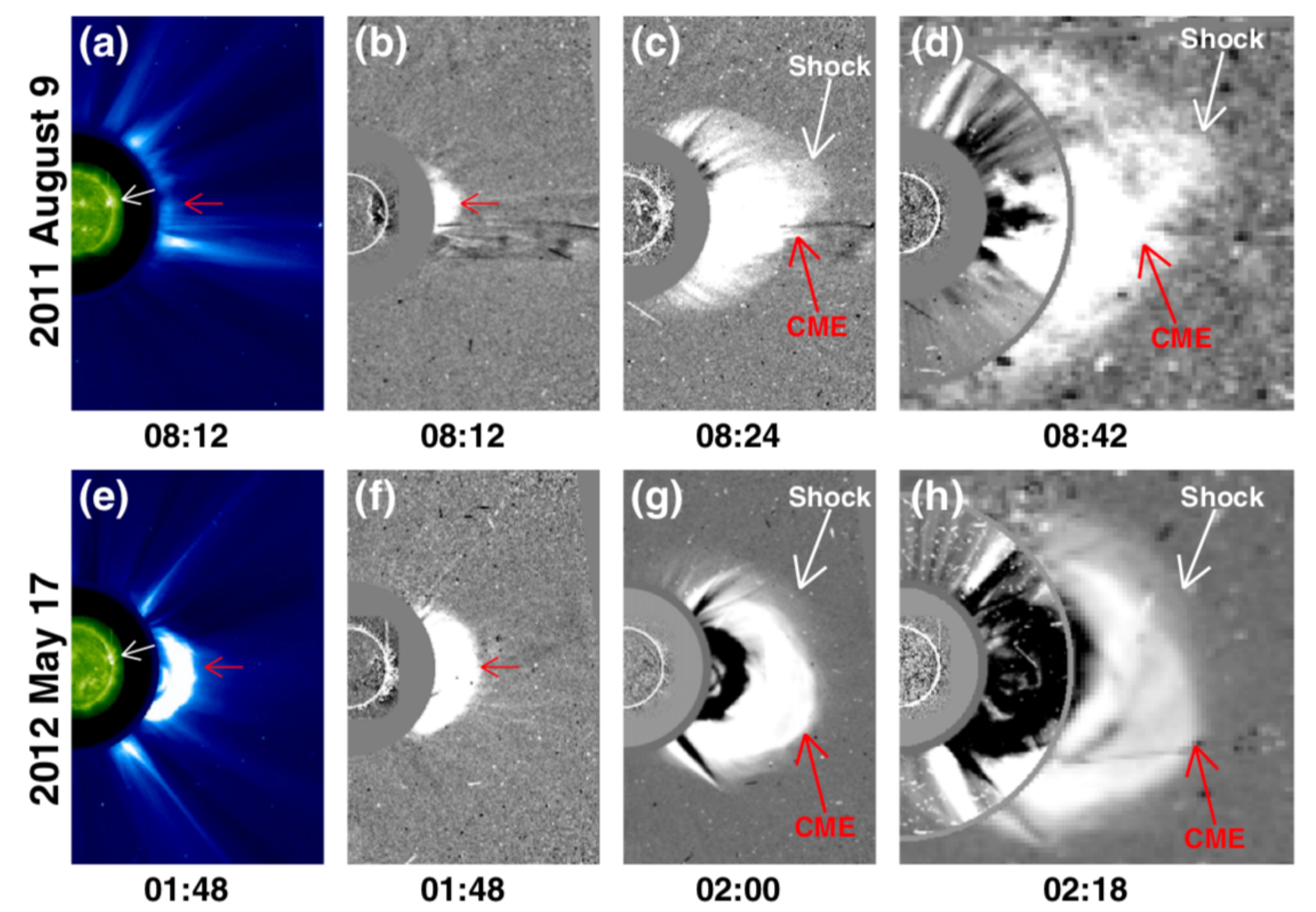}
\caption{Coronagraph images from SOHO/LASCO (left panel: direct image; other panels to the right: difference images) showing the evolution of two CMEs from August 9, 2011 and from May 17, 2012 (GLE event). Inlay images are from SDO/AIA 193\AA~showing the solar sources. Red arrows point to the CME nose. The May 17, 2012 event remains bright over the LASCO field of view and its shock structure is located close to the CME ((g), (h)). The August 9, 2011 event reveals a smaller CME main body and a wider shock structure ((c), (d)), which is a sign of a weak shock. Taken from \cite{gopalswamy13}.}
\label{fig:gopal_GLE}       
\end{figure*}

During solar cycle 24 strong SEP events produced only two GLEs, that could be related to CMEs launched from the Sun on May 17, 2012 and September 10, 2017. \cite{gopalswamy13} did a comparative study between the GLE May 17, 2012 and a non-GLE CME event from August 9, 2011 using STEREO image data as given in Figure~\ref{fig:gopal_GLE}. The study revealed for the GLE event a shock formation height very low in the corona (1.38 solar radii from solar center) and that the shock remained closer to the driver structure over the coronagraphic field of view. This is indicative of a stronger shock that is driven over a longer time and, hence, can produce very energetic particles. \cite{rouillard16} modelled for the May 17, 2012 event the background topology of the magnetic field using multiple viewpoints to derive the geometry of the shock front and to find where particles get accelerated most efficiently \citep[see also][]{Plotnikov17,Kouloumvakos19}. Particles that get magnetically trapped in between CME structures pose a particle reservoir that may play a key role in the late acceleration of gradual SEPs related to CME-CME interaction events \citep[e.g.,][]{lugaz17}. Despite these tremendous enhancements in our knowledge gained from stereoscopic observational data, still a major drawback in unraveling the SEP nature is the unknown configuration of the interplanetary magnetic field along which SEPs propagate and internal distribution \citep{Kahler13}. Therefore, the acceleration process of SEPs seems not to be spatially limited but happens over a wide range of longitudes including transport before being injected at distant longitudes \citep[e.g.,][]{Vlahos19_SEP,Kozarev17,Malandraki18_book}.

For recent reviews on SEP events covering in detail the production and acceleration processes I refer to the Living Reviews by \cite{Desai16} or the book by \cite{Reames17_book}. The review by \cite{lugaz17} is focusing on SEPs with respect to CME-CME interaction events and the Living Reviews by \cite{kilpua17} on particle acceleration due to ICME shocks.

\section{Energy budget between flares, CMEs, and SEPs}\label{sec:budget}

\begin{figure*}
  \includegraphics[width=\textwidth]{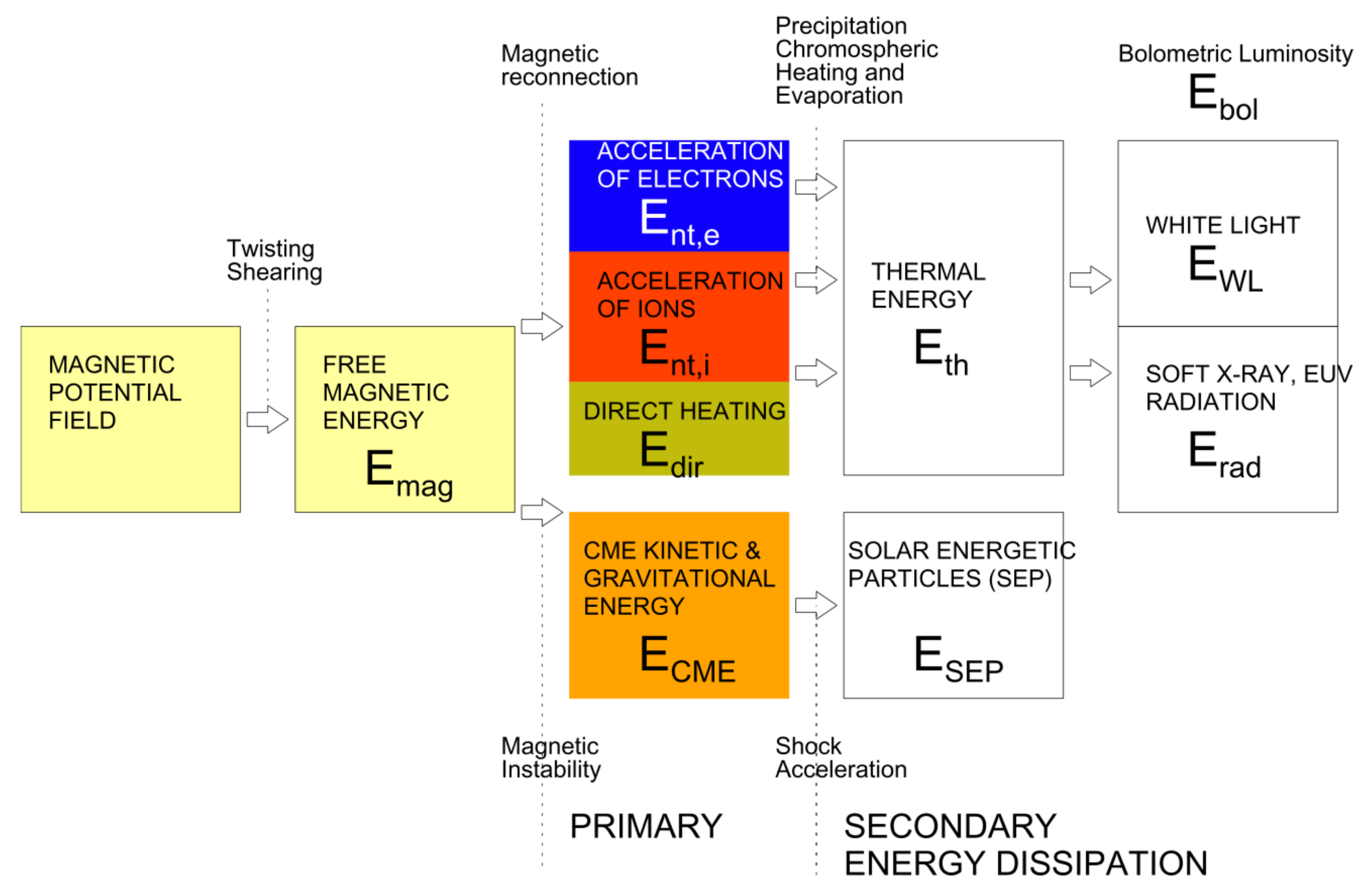}
\caption{Schematic diagram showing the different energy dissipation processes of the observed activity phenomena. The chart covers the energy input (light yellow shaded boxes) and energy dissipation via primary (colored boxes) and secondary processes (white boxes). Those are the major processes identified for studying their energy closure relationship. Taken from \cite{Aschwanden17}.}
\label{fig:energy_budget}       
\end{figure*}

Figure~\ref{fig:energy_budget} schematically shows the relevant components of energy build-up and dissipation processes. How much of the free energy is actually released and to which parts that released energy partitions into primary and secondary processes can only be answered by statistics. In cutting-edge studies performed by \cite{emslie04,Emslie12}, a sample of eruptive events was investigated with respect to the energy release and its distribution into different components, concluding that about a third of the total available energy might be released in an eruptive event. Similar conclusions are drawn by \cite{gopalswamy17_extreme} from calculating for extreme eruptive events the ratio of the total reconnected flux to the available active region flux. However, discrepancies were found from simulation studies \citep[e.g.,][]{reeves10}. Recent extensive statistics by \cite{Aschwanden17}, yield that $\sim$87\% of the magnetic energy is released. About 10\% of the released energy drives the CME and $\sim$80\% goes into particle acceleration. In total about 10\% percent of the free magnetic energy goes particularly into SEP acceleration. With respect to the CME, SEPs dissipate about 3\% of the CME kinetic energy \citep[similar as derived by][]{Emslie12}. The CME velocity shows strongest correlations with SEP characteristics and all that is consistent with CME-driven shock acceleration  \citep[see][]{mewaldt06,Papaioannou16}. 

The discrepancies in the results show that there might be processes that cannot be disentangled from each other, cover energy conversion (e.g., non-thermal into thermal due to cooling), or are simply not well observed. Nevertheless, the conclusion is that the free magnetic energy of an active region is sufficient to generate flare-CME-SEP events and with that confirms their common magnetic origin. For the interested reader I refer to the book by \cite{aschwanden19}.

\section{Structuring of interplanetary space: the solar wind}\label{sec:SWstructure}

The solar wind is the major hub in interplanetary space dictating how fast disturbances may evolve and guiding the motion of accelerated particles. Knowledge about the prevailing structure of the solar wind in terms of plasma and magnetic field distribution is therefore of utmost importance in order to obtain reliable Space Weather forecasts. In turn, this also leads to a better understanding and interpretation of the propagation behavior of CMEs as well as the occurrence and energetics of SEPs. Interplanetary space is strongly shaped by the interplay between slow and fast solar wind flows, causing stream interaction and compression regions, transient disturbances such as shocks and closed magnetic structures (flux ropes) of evolving CMEs. These structures pose magnetic barriers that are able to change the propagation characteristics of a specific CME and affect SEP fluxes. In the following I will focus on solar wind structures relevant for CMEs and SEPs. For more details on the heliospheric magnetic field I refer to the Living Review by \cite{owens13} and for solar wind stream interaction regions throughout the heliosphere to the Living Review by \cite{Richardson18}.

\subsection{General characteristics}
\begin{figure*}
  \includegraphics[width=1.\textwidth]{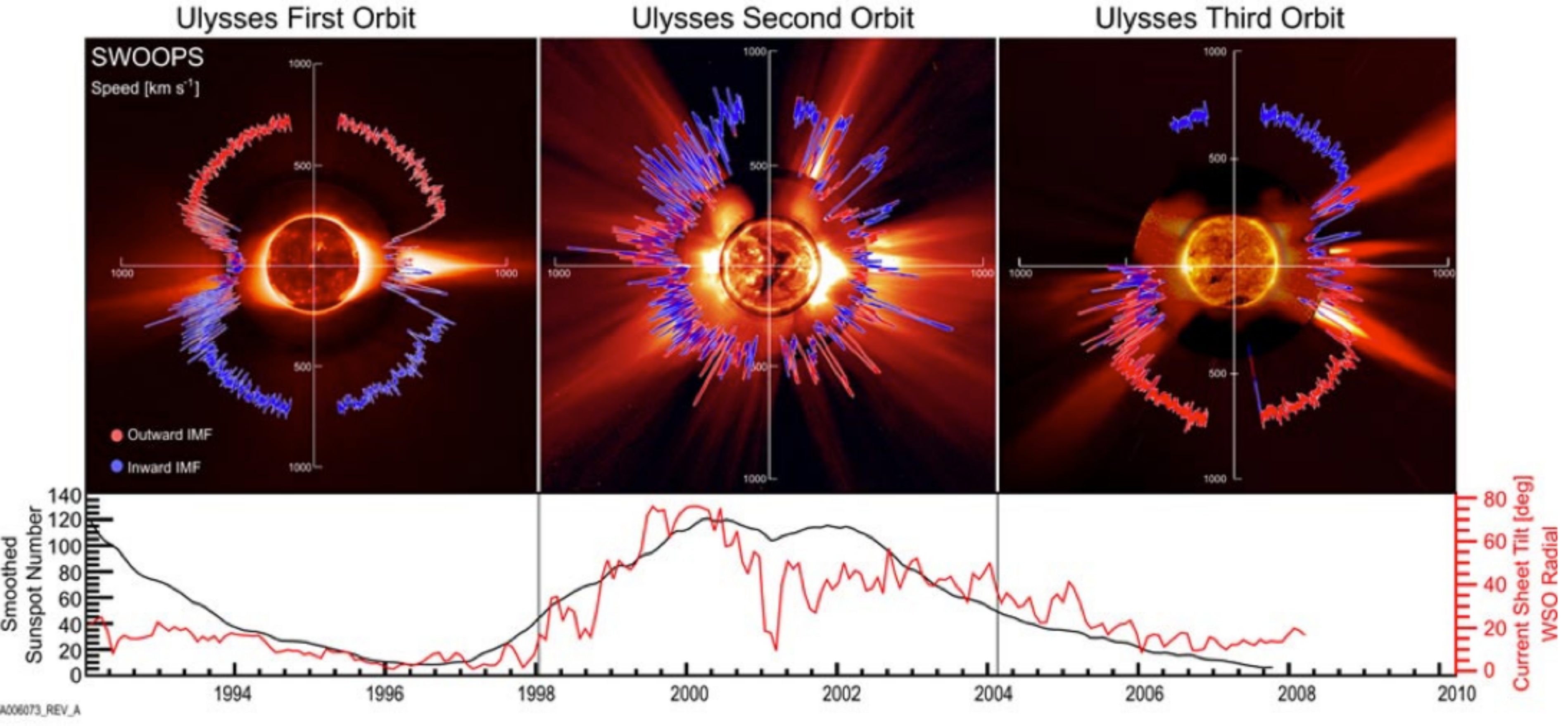}
\caption{Solar wind variation over the solar cycle. Outward interplanetary magnetic field in blue, inward interplanetary magnetic field in red. The bottom panel shows the timeline and line graphs of the relative smoothed sunspot number. Note also the inverted interplanetary magnetic field lines between the first and third orbit due to the reversal of the magnetic field. Courtesy: ESA}
\label{fig:ulysses}       
\end{figure*}

The solar wind is a continuous flow of charged particles propagating radially outward from the hot solar corona into interplanetary space. The characteristics of the solar wind are measured in-situ at specific locations such as the Lagrangian point L1 close to Earth (ACE, WIND, DSCOVR), from satellites orbiting planets (e.g., BepiColombo, VEX, MESSENGER, MAVEN), STEREO having varying longitudinal separation close to Earth's orbit, or PSP and Solar Orbiter with special mission trajectories in the inner heliosphere (cf. Figure~\ref{fig:insitu}). Figure~\ref{fig:ulysses} shows the solar wind speed measurements from spacecraft Ulysses which had the goal to examine the poles of the Sun \citep{Wenzel92_ULYSSES}. In total, Ulysses performed three polar orbits, with each one taking six years to complete, over different phases of the solar cycle 22 and 23. The first one covers the solar minimum phase revealing slow solar wind streams over the equator and a fast wind over the poles where CHs are situated. The second orbit happened over solar maximum activity and shows the intermix of fast and slow winds at all latitudes. Three quarters of the third orbit were completed during the minimum of the next solar cycle.

\begin{figure*}
  \includegraphics[width=1.\textwidth]{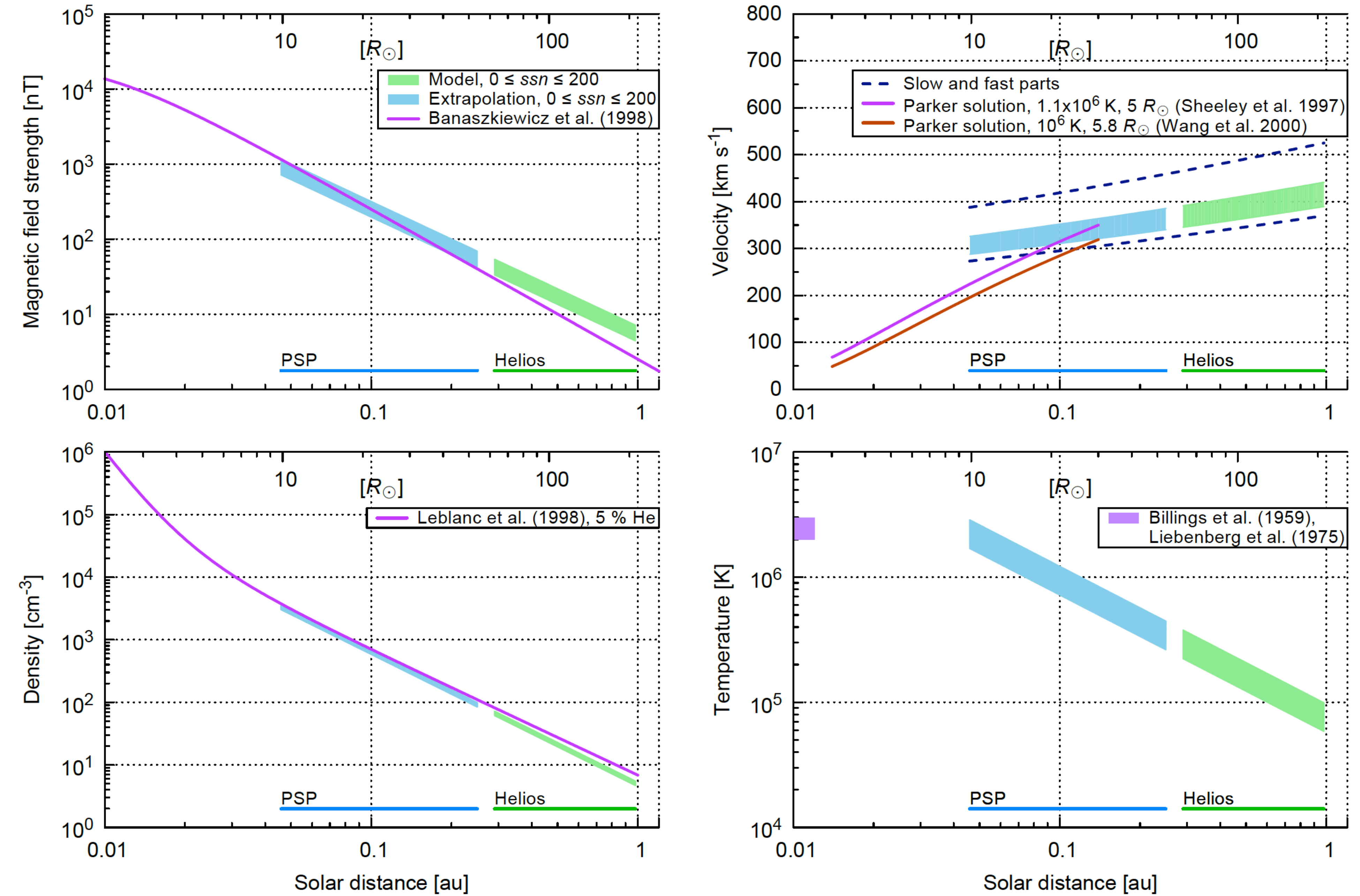}
\caption{Radial profiles of solar-wind parameters for the Sun-Earth distance. Median values are obtained from Helios and OMNI measurements and are extrapolated to the PSP orbit region as close as 10 solar radii. The lower edges of the shaded areas correspond to solar minimum and the upper edges to solar maximum. As comparison, overplotted are model results from \cite{Banaszkiewicz98}, from \cite{Sheeley97} and \cite{WangY00} for the slow solar wind speed, from \cite{Leblanc98} for the density and the range of temperature measurements given by \cite{Billings59} and \cite{Liebenberg75}. Taken from \cite{Venzmer18}.}
\label{fig:sw_venzmer}       
\end{figure*}

Besides the observed changes over latitude, the solar wind characteristics differ over distance. Early missions like Helios 1 and 2 \citep{Schwenn75,Rosenbauer77}, achieved a perihelion of 0.29~AU and gathered valuable information about the solar wind characteristics close to the Sun. Figure~\ref{fig:sw_venzmer} shows the radial dependence of the solar wind parameters over the Sun-Earth distance derived from HELIOS (1974--1981) and OMNI (1963--2016) observations at 1~AU. The study by \cite{Venzmer18} extrapolates that information to regions as close as 10 solar radii, based on an empirical solar-wind model for the inner heliosphere in dependence on the solar cycle. Solar wind measurements from PSP at a distance of about 35 solar radii basically confirm the results from these earlier missions and their derived radial scalings, but also obtain that the magnetic field is very strongly fluctuating \citep{Bale19,kasper19}\footnote{For recent PSP result see also special issues of the Astrophysical Journal Supplement series and Astronomy \& Astrophysics under  \url{https://iopscience.iop.org/journal/0067-0049/page/Early_Results_from_Parker_Solar_Probe} and \url{https://www.aanda.org/component/toc/?task=topic&id=1326}.}.

\begin{figure*}
  \includegraphics[width=1.\textwidth]{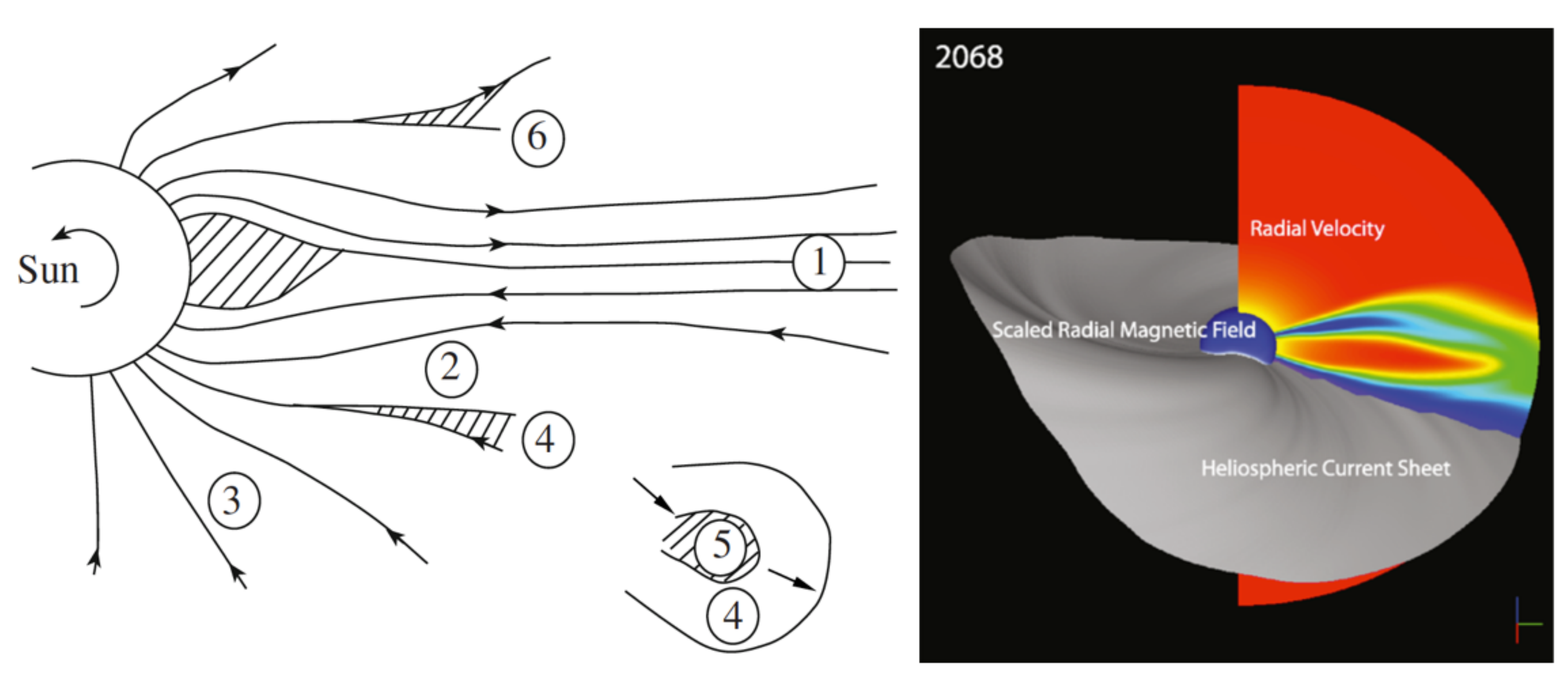}
\caption{Left: schematic drawing of the different large scale structures in interplanetary space \citep[taken from][]{Yermolaev09}. Digits designate (1) heliospheric current sheet, (2) slow streams from coronal streamers, (3) fast streams from coronal holes, (4) compressed plasma (CIR on the front of fast and slow streams, and sheath region before the leading edge of a “piston”), (5) “pistons” (such as a magnetic cloud or ejecta), (6) rarefaction region. Right: Large-scale properties of the inner heliosphere (out to 1 AU) for Carrington Rotation 2068 from a global MHD solution. The two meridional slices in each panel show the radial velocity and radial magnetic-field strength, scaled to 1 AU. The slice in the equatorial plane shows the scaled number density. The sphere at 30 solar radii shows the scaled radial magnetic-field strength. Taken from \cite{Riley10}.}
\label{fig:yermo_riley}       
\end{figure*}

The left panel of Figure~\ref{fig:yermo_riley} schematically depicts various large-scale structures in the interplanetary space. We differentiate between the heliospheric current sheet, separating opposite magnetic field, solar wind streams of different speeds, and closed magnetic fields of CMEs that, due to their rapid expansion, act as pistons creating shocks. The interaction between fast and slow streams (plasma volume with frozen-in magnetic field) leads to compression, forming so-called stream interaction regions (SIRs) to the West, and rarefaction regions to the East. On the large scale it is assumed that the interplanetary dynamics can be described by ideal MHD equations. The right panel of Figure~\ref{fig:yermo_riley} gives the results from a simulation using a 3D MHD model \citep[see more details on coronal and solar wind MHD modeling in the Living Reviews by][]{gombosi18}. The model results obtain a warping of the heliospheric current sheet and show the latitudinal dependence of the fast solar wind stream in the radial direction \citep{wilcox80}. All that reveals the complexity and interplay between open and closed magnetic field structures, occurring with different dynamics. For more details on the multi-scale nature of the solar wind I refer to the Living Reviews by \cite{verscharen19}.

\begin{figure*}
    \centering
\includegraphics[width=1.\textwidth]{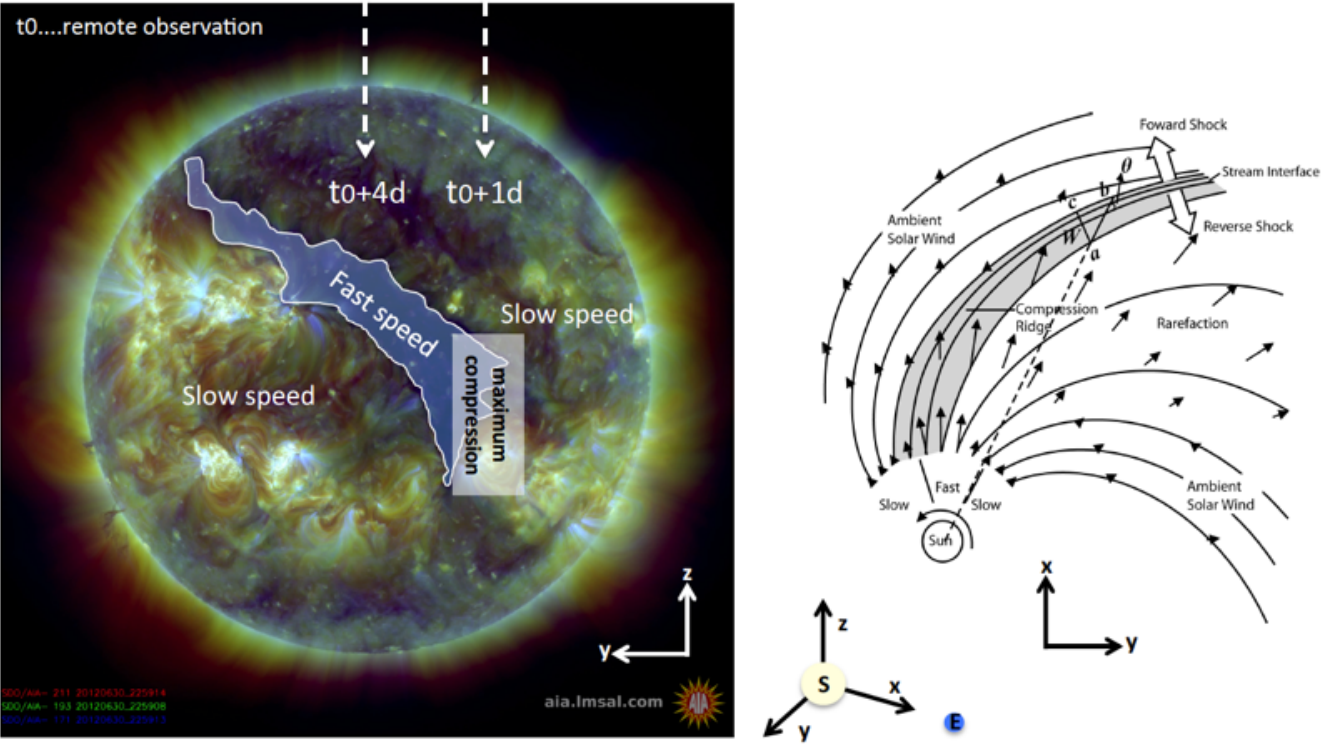}
\caption{Left: SDO/AIA composite image of the wavelength channels 211-193-171\AA~from June 30, 2012 showing the reduced density region of a coronal hole (shaded area). At the time t0, the coronal hole reaches a central position. From in-situ data at 1 AU about 1 day later the maximum in the density/magnetic field is measured and about 4 days later the maximum in the speed/temperature. Right: Fundamental processes involved in the 3D dynamics of stream evolution, adapted from \cite{pizzo78}.}
\label{fig:CH-CIR}       
\end{figure*}

The observed large-scale structures of the solar wind are intimately connected to the coronal magnetic field originating from the solar photosphere. The quasi-steady fast wind ($>$ 450 km/s) emanates from coronal holes, locations of predominantly open magnetic field, while the variable slow component is believed to originate mostly from closed magnetic field configurations around the streamer belt \citep{McComas00_firstOrbit}. Coronal holes are observed as low density and low temperature structures, and therefore appear as dark areas in the wavelength ranges of EUV and SXR, imaging coronal temperatures of a few million Kelvin. Figure~\ref{fig:CH-CIR} depicts the interplay between slow and fast solar wind streams. After a coronal hole passed the central part of the solar disk, in-situ measurements reveal about 1--2 days later an increase in the density and magnetic field, and about 3--4 days later in the plasma speed \citep[][]{vrsnak07_sw}. Since coronal holes, and with that the fast component of the solar wind, are long-lived structures, SIRs can often be observed for several solar rotations, and are correspondingly called co-rotating interaction regions (CIRs) when observed more than once. The leading edge of a CIR represents a forward pressure wave and the trailing edge of a CIR a reverse pressure wave \citep[cf.\, right panel of Figure~\ref{fig:CH-CIR}; for more details see the review by][]{cranmer17}. These waves may develop into shocks, and as such, large periodically recurrent coronal holes may cause geomagnetic storms roughly appearing with the frequency of the solar rotation, i.e., every 27 days \citep[e.g.,][]{Rotter12}. During times of low solar activity, induced storms by recurrent CIRs may put equally much energy into the Earth's magnetosphere-ionosphere system as CMEs \citep[e.g.,][]{Richardson01,tsurutani06}. On average, the strongest geomagnetic storms due to CIRs occur during the early declining phase of a solar cycle \citep[e.g.,][]{Verbanac13,Grandin19}. Compared to CMEs, CIRs may drive prolonged geomagnetic activity and cause strong high energy particle enhancements in the Earth’s radiation belts \citep[e.g.,][]{Reeves03,Miyoshi13,Kilpua15}. As the solar wind parameters vary with the level of solar activity, so does their geoeffectiveness \citep[see e.g.,][]{Jian11,Richardson12,Watari18}. There is substantial effort in the solar and heliospheric physics community to improve the understanding and modeling of the spatial and temporal distribution of solar wind plasma and magnetic field properties \citep[see][]{Cranmer19}. 

Since the source regions of slow and fast streams on the Sun, namely closed and open magnetic field, are different, their intermix affects detailed analyzes of the solar wind. Therefore, it is suggested that solar wind studies should be organized by the origin of the solar wind plasma \citep{schwenn83,Zhao09,Borovsky19}. As minimum requirement, it is accepted to distinguish between the different solar wind structures by their in-situ measured plasma (density, speed, temperature) and magnetic field characteristics. A categorization scheme developed by \cite{xu15} can be applied to separate the solar wind plasma into four types, namely, coronal-hole-origin (fast solar wind), streamer-belt-origin (slow solar wind), sector-reversal-region (plasma from top of helmet streamers), and ejecta (solar transients, such as CMEs).

\begin{figure*}
    \centering
\includegraphics[width=0.7\textwidth]{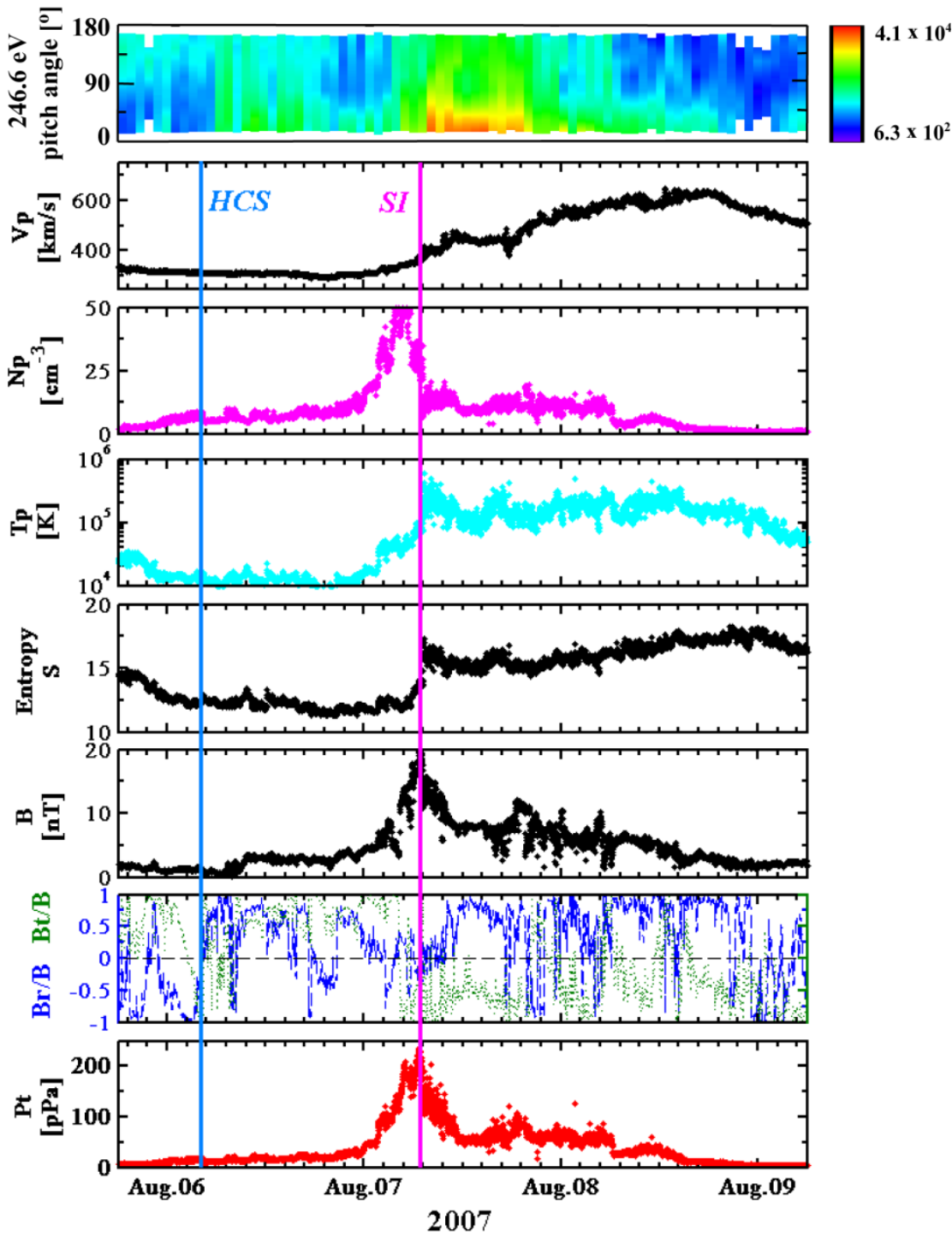}
\caption{STEREO-A in-situ solar wind measurements of a SIR and identification of the stream interface (SI; given by a magenta vertical line marking the peak of the total perpendicular pressure). As blue vertical line, the heliospheric current sheet (HCS) is marking the sector structure of different interplanetary magnetic field polarity. Top to bottom panels: pitch-angle distribution data of suprathermal electrons, solar wind proton bulk speed, proton number density, proton temperature, entropy, total magnetic field intensity, the ratios of the radial and transversal component of the magnetic field, total perpendicular pressure. Taken from \cite[][]{jian09}. }
\label{fig:jian09}       
\end{figure*}
 

Figure~\ref{fig:jian09} shows typical plasma and magnetic field characteristics for a well defined SIR at 1~AU. The SIR measurements reveal an abrupt drop in density, a simultaneous rise in the proton temperature, and an East-West flow deflection \citep[see e.g.,][]{Jian06,Jian19}. The stream interface (SI), given by a rather symmetric profile in the total perpendicular pressure peaking shortly after the density maximum, separates originally slow, dense plasma from originally fast, thin plasma back at the Sun \citep[e.g.,][]{Wimmer-Schweingruber97}. The behaviour of the suprathermal electrons gives additional information about the topology, hence, connectivity of the interplanetary magnetic field lines between the Sun and the observer. That also allows to investigate, e.g., the Parker spiral versus non‐Parker spiral orientation of the magnetic field and the related open flux versus closed or disconnected flux natures of the magnetic field \citep[see][]{owens13}. SIR signatures differ clearly from those of ICMEs (cf.\,Figure~\ref{fig:ICME}) and as such, the specific characteristics can be applied to identify SIRs and CMEs from visual inspection of the in-situ solar wind plasma and magnetic field measurements. Together with the information of transit times (can be roughly derived from the average in-situ speed of the specific structure) we can link the in-situ signatures back onto the solar surface to study their relation. Table~\ref{tab:1} gives average values and standard deviations for the different structures observed, such as slow and fast speed solar wind streams, CME sheath and magnetic ejecta. The time range used for that statistics covers the years 1976--2000 \citep[see][]{Yermolaev09}. As can be seen, the values reveal large standard deviations reflecting the variations of the interplanetary conditions over the solar cycle. Similar as the activity level changes over time with respect to the occurrence of sunspots, active regions, flares and CMEs (see also Section\ref{subsec:cme}), also the occurrence of coronal holes varies having implications on the global structuring of interplanetary space \cite[e.g.,][]{harvey02, Heinemann2019StatisticalCATCH}.

\begin{table}
\caption{Properties of the different solar wind types derived from OMNI data analyzed over the period 1976–-2000. Average values and standard deviations are given for the flow speed ($v_{\rm p}$), proton density ($n_{\rm p}$), total magnetic field ($B$), and proton temperature ($T_{\rm p}$)}. In addition, the geoeffectiveness of the different solar wind types is given as measured by the disturbance storm time (Dst). Taken from Table 3 in \cite{Yermolaev09}.
\label{tab:1}       
\begin{tabular}{lllll}
\hline\noalign{\smallskip}
& fast wind & slow wind & CMEs (shock-sheath) & CME (magnetic ejecta)  \\
\noalign{\smallskip}\hline\noalign{\smallskip}
$v_{\rm p}$, km~s$^{-1}$ & $>$ 450--500 & $<$ 400--450 & $\sim$450$\pm$110 & $\sim$410$\pm$110 \\
$n_{\rm p}$, cm$^{-3}$ & 6.6$\pm$5.1 & 10.8$\pm$7.1 & 14.3$\pm$10.6 & 10.1$\pm$8.0 \\
$B$, nT & 6.4$\pm$3.5 & 5.9$\pm$2.9 & 8.5$\pm$4.5 & 12.0$\pm$5.2 \\
$T_{\rm p}$ $\times10^4$ K & 13.1$\pm$11.8 & 4.4$\pm$4.4 & 12.9$\pm$17.6 & 4.5$\pm$6.6 \\
Dst, nT & $-$28.7$\pm$25.9 & $-$10.7$\pm$18.2 & $-$21.5$\pm$33.0 & $-$52.1$\pm$45.8 \\

\noalign{\smallskip}\hline
\end{tabular}
\end{table}

\subsection{Background solar wind}
Open magnetic flux from the solar surface structures interplanetary space and with that influences CME and SEP propagation behavior. We assume that the majority of open flux originates within coronal holes. This is supported by an empirical relation linking the size of coronal holes observed on the solar surface to the in-situ solar wind plasma and magnetic field measured at 1AU a few days later \citep[][]{vrsnak07_sw}. In detail, the width of a coronal hole, i.e., its longitudinal extension, is found to be strongly related to the in-situ measured peak speed \citep{garton18}. Deriving the solar wind characteristics stemming from a specific coronal hole, is found to be tricky as each coronal hole evolves rather individually and the surrounding solar surface structures may play a major role in shaping the interplanetary solar wind \citep{Heinemann20}. Still, an open discussion is the discrepancy between estimates of open solar magnetic flux from remote photospheric and in-situ spacecraft observations that may differ by as much as a factor of two \citep[e.g.,][]{Arden14,Linker17,wallace19}. This suggests a fundamental issue in our understanding about the topology of the coronal magnetic field and the energization of plasma, hence, the acceleration of fast solar wind flows. Recent studies find that the open magnetic field within coronal holes is predominantly concentrated in unipolar magnetic flux tubes with high outflow velocities but covering only a fraction of about 10\% of the coronal hole area \citep[see][]{akiyama13,wiegelmann14,Hofmeister2017Characteristics24}. For more details on modeling the coronal magnetic field and open solar flux see e.g., the Living Reviews by \cite{mackay12} and \cite{lockwood13}.

\begin{figure*}
  \includegraphics[width=1.\textwidth]{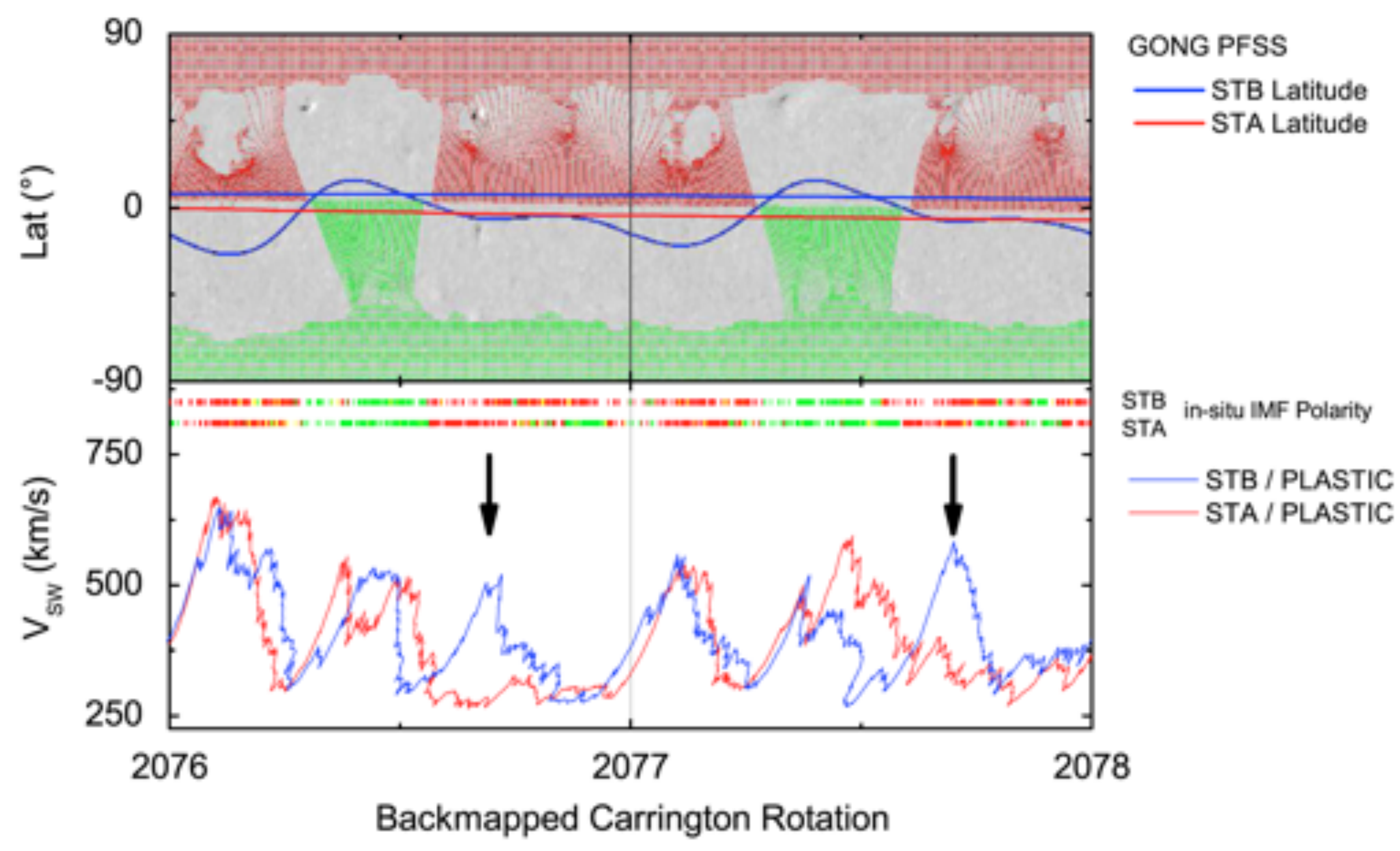}
\caption{Top panel: synoptic map from GONG overlaid with magnetic field extrapolation results from the PFSS \citep[potential field solar surface; see e.g.,][]{Schrijver03} model showing the location of open field lines at the ecliptic plane. Bottom panel: solar wind bulk speed measured by PLASTIC instruments aboard STEREO-A (red) and STEREO-B (blue). Colored bands on the top part of this panel represent the interplanetary magnetic field polarity. Black arrows mark the solar wind stream observed over two rotations (CR 2076 and 2077) by STEREO-B, but missed by STEREO-A due to the latitudinal separation of the spacecraft. Taken from \cite{Gomez-Herrero11}.}
\label{fig:gomez11}       
\end{figure*}

Due to the rather slow evolution of coronal holes, a legitimate assumption is that the solar wind parameters do not vary strongly over the duration of an entire solar rotation. Based on that, to forecast the occurrence of high speed solar wind streams at Earth, L1 spacecraft data may simply be forward shifted over one Carrington rotation period (27.28 days). These so-called persistence models are found to work remarkably well \citep[see e.g.,][]{Owens13_persistence,Reiss2016VerificationModels}. However, spacecraft used for forecasting may be located over different latitudes and in-situ measurements gather the characteristics of rather localized structures within that large scale three dimensional objects. The spatial restrictions of such localized structures are demonstrated in Figure~\ref{fig:gomez11}, showing how solar wind streams appear differently as measured by STEREO-A and STEREO-B spacecraft that are separated in latitude by $\sim$10 degrees. The data cover two Carrington rotations (2076 and 2077) revealing a solar wind stream well observed by STEREO-B, but clearly missed by STEREO-A. For comparison, the top panel of Figure~\ref{fig:gomez11} gives the photospheric magnetic field configuration on the Sun using magnetic field extrapolations from GONG data marking open and closed field lines. With such inevitable differences in the latitude between the measuring spacecraft, rather large uncertainties in the 1~AU measured speed of a particular coronal hole are obtained \citep[see][]{Hofmeister18,Owens19}. Recent studies suggest that solar wind forecasting based on persistence models may work best when using a combination of spacecraft located behind Earth (L5 position or STEREO data from time-varying spacecraft position) and empirical or numerical solar wind modeling \citep[see e.g.,][]{Opitz10_Multispacecraft,Temmer18,Owens19,Bailey20}.

\subsection{Solar wind structures affecting CME and SEP evolution}\label{sec:affect}
\begin{figure*}
  \includegraphics[width=1.\textwidth]{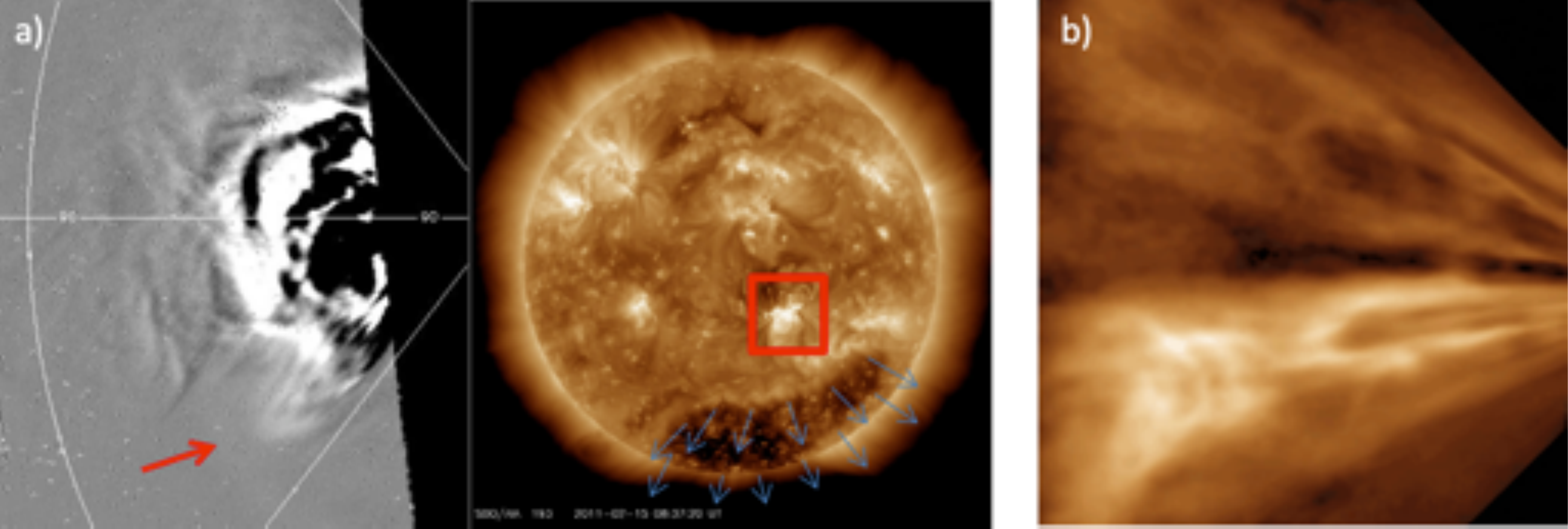}
\caption{a) Left panel shows a STEREO/HI difference image with the distorted CME front (red arrow). Right panel gives SDO/AIA 193\AA~EUV image with the CME source region (red box) and the southern coronal hole. The fast solar wind stream out of the coronal hole (depicted with blue arrows) deforms the CME front. b) Highly structured solar wind flow close to the Sun from a STEREO/HI image after computer processing (credit: NASA's Goddard Space Flight Center/Craig DeForest, SwRI).}
\label{fig:sw-structure}       
\end{figure*}

CMEs are known to change shape, accelerate or decelerate, depending on the background solar wind flow properties, as well as may change their propagation direction if encountering other magnetic structures (cf.\,Section \ref{subsec:cme}). From recent results it is found that many coronal fine structures and small scale magnetic flux ropes are embedded in the solar wind \citep[first PSP/WISPR results see][]{howard19}. Likewise, PSP in-situ measurements show that low-latitude coronal holes might be an additional source of slow solar wind, causing close to the Sun strong fluctuations in the solar wind flow \citep[][]{Bale19}. Such local solar wind dynamics, comprising of numerous voids, compact small- to large-scale sized structures (``woodgrain'' appearance), clearly have impact on the evolution of a CME and with that complicates forecasting. As example,  Figure~\ref{fig:sw-structure} shows a disrupted CME front close to the Sun as consequence of an interaction with a high speed solar wind stream emanating from the southern polar coronal hole. The increased speed of the ambient solar wind flow and stretched magnetic field, causes a weaker compression and, hence, deviations from the shape of a circular/elliptic CME front. The structured solar wind in the outer corona is also revealed from specially noise reduced post-processed STEREO/HI image data applying an algorithm to dim the appearance of bright stars and dust \citep[right panel of Figure~\ref{fig:sw-structure}; cf.,][]{DeForest18}. See also  recent results from PSP image data resolving small-scale flux ropes, density structures and fluctuations in solar wind streamers \citep{howard19}. Besides adjusting to the ambient solar wind flow speed, CMEs also tend to rotate for adjusting to the ambient magnetic field \citep[e.g.,][]{Yurchyshyn01,Yurchyshyn09,Vourlidas11,Isavnin14}. 

SEPs propagate along the magnetic field lines, hence are directly guided by the interplanetary magnetic field structure. During CME events, SEP path lengths are found to be much longer compared to quiet solar wind conditions \citep[e.g.,][]{masson12}. For SEPs, which are injected into the legs of a CME, the magnetic path lengths from the Sun to Earth may vary largely, up to a factor of two, depending on the specific width of the CME \citep{reames09}. CME geometry as well as kinematics (propagation of shock front) and time-dependent changes play an important role with respect to particle acceleration processes. In that respect, SEPs can be taken as probes as they map the interplanetary magnetic field structure \citep{Reames17_book}.

\subsection{Preconditioning of interplanetary space}\label{sec:preconditioning}
The correct simulation of the prevailing background solar wind structures in interplanetary space is important for a reliable CME-SEP forecasting. Using MHD modeling to simulate the solar wind distribution works well for low solar activity. However, increased solar activity changes the magnetic field in the photosphere, which serves as main observational model input, more quickly and propagating CMEs strongly disturb the interplanetary background solar wind making models tend to fail \citep[see e.g.,][]{Lee09,Gressl14}. This effect is commonly known as \textit{preconditioning} of interplanetary space, which alters the initial conditions for CME and SEP evolution. It is found that single CME events may disturb the slowly evolving solar wind flow for 2--5 days \citep{Temmer17_preconditioning,janvier19} and most strong preconditioning effects are obtained due to CME-CME or CME-CIR interacting events as they form complex magnetic structures \citep[e.g.,][]{gopalswamy00_accelCME,Burlaga02,Harrison12,dumbovic19}. Multiple CME activity is supposed to cause a decrease of density in interplanetary space and a radial stretching of the interplanetary magnetic field. That leads to a low drag force acting on subsequently propagating CMEs \citep[see e.g.,][]{Farrugia04,Maricic14}. The STEREO-A directed July 23, 2012 event was one of the fastest CMEs ever recorded and propagated over a 1AU distance in less than 21 hours \citep{liu14_nature}. It could be shown that the drag parameter, due to a preceding CME from July 19, 2012, was lowered by one order of magnitude \citep{temmer15}. If the super-fast CME from July 2012 would have been Earth directed, it would have caused an extreme Space Weather event with an estimated Dst of $-$600 to $-$1100~nT \citep[e.g.][]{ngwira14,baker13}. Due to increased fluctuations and extended periods of negative $B_{\rm z}$ most intense geomagnetic storms occur for complex interacting CMEs \citep[e.g.,][]{wang03,Farrugia06,Xie06,dumbovic15,scolini20}.

For SEPs, the preconditioning that comes from a preceding CME has consequences in the seed population. The presence of a previous CME is found to increase the probability for the subsequent fast CME to be SEP-rich \citep{Gopalswamy02,Gopalswamy04,Kahler05}. The amount of particles that can get accelerated is increased for multiple CMEs and the increased turbulence in the interaction region is likely to accelerate the particles more efficiently to higher energies \citep[so-called twin-CME scenario as proposed by][]{Li05}. Furthermore, the magnetic structure configuration in the sheath region for ICMEs is changed as the CME propagates in interplanetary space, by which the magnetic connectivity is altered \citep[see also review by][on CME sheath regions]{kilpua17}. Interacting CMEs are also found to be more often related to widespread SEPs that can be observed all around the Sun using multiple viewpoints \citep[e.g.,][]{Dresing16,Gomez-Herrero17}. For more details on CME-CME interaction and SEPs, I refer to the review by \cite{lugaz17}.

Over the solar cycle, the CME occurrence rate lies on average in the range of 0.3 per day during solar minimum phase and about 4--5 per day during solar maximum phase \citep[e.g.,][]{St.Cyr00}. With average CME transit times from Sun to 1AU of the order of 1--4 days we can assume that during times of increased solar activity CME-CME interaction happens rather frequently. Reliably modeling these dynamic conditions in interplanetary space is therefore key for improving space weather forecasting capabilities.

\section{The chain of action on the example of the September 2017 events}\label{sec:active}

The September 2017 activity phenomena are so far the most well studied strong Space Weather events in our modern space research era. Therefore, the chain of action can be described in great detail, especially with respect to the solar surface signatures and deduced parameters. The multiple disturbances can be well connected from Sun to Earth (and also up to Mars) and show how complex interactions lead to preconditioning effects and strong geoffectiveness.

\begin{figure}
  \includegraphics[width=\textwidth]{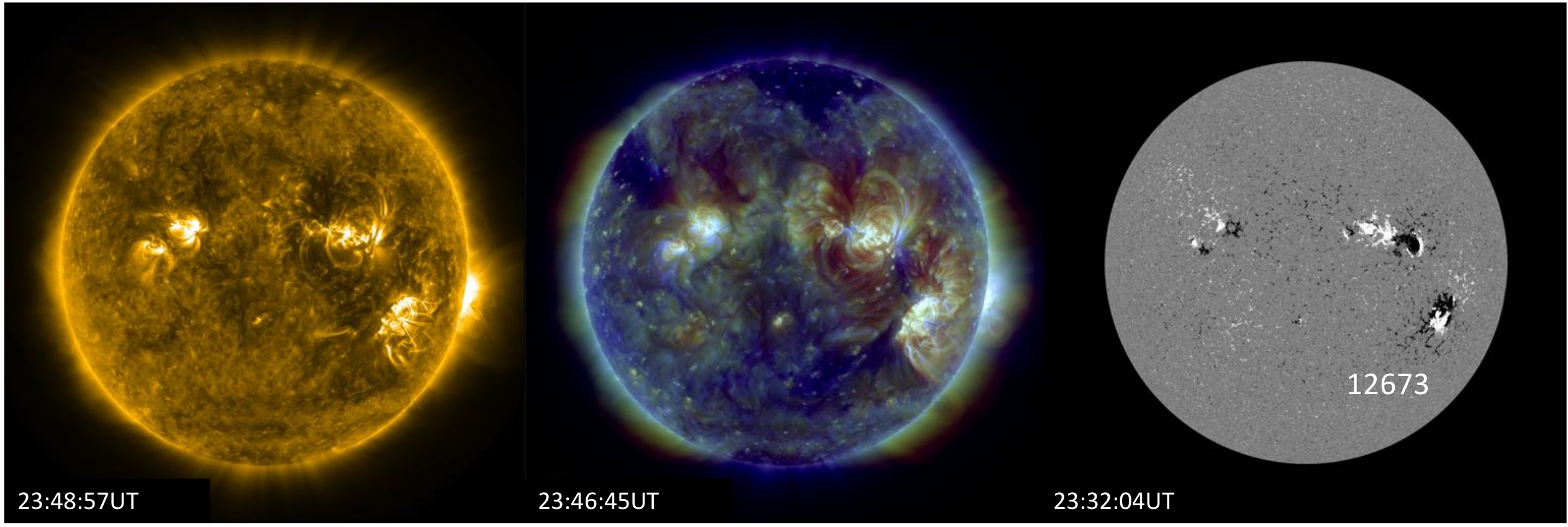}
\caption{High-resolution SDO/AIA multi-wavelength imagery from September 6, 2017 (left: 171\AA, middle: composite from 211-193-171\AA) and SDO/HMI line-of-sight magnetic field (right panel). The prominent active region NOAA 12673 caused the strongest eruptions during solar cycle 24 including SEPs. Courtesy of AIA team, adapted from \url{http://suntoday.lmsal.com}.} 
\label{fig:sdo}        
\end{figure}

A strong emergence of magnetic flux in the southern hemisphere of the Sun rapidly led to the development of active region NOAA 12673 into a complex $\alpha\beta\gamma$ magnetic field configuration. As consequence, between September 4--10, 2017 that active region released a series of major flare events, which were actually the largest in more than a decade. These multiple events caused very strong geomagnetic disturbances with a minimum Dst of $-$142 nT on September 7, 2017. Additional minor storms were produced by high speed solar wind streams that arrived together with the transient events. Figure~\ref{fig:sdo} gives EUV image data taken with SDO/AIA on September 6, 2017 in different wavelength ranges showing the hot corona from about 0.6 to 2~MK. The wavelength ranges cover 171\AA\ (left panel), and with a triple-filter 211\AA\ (red), 193\AA\ (green), and 171\AA\ (blue) to highlight different temperatures (middle panel). The line-of-sight magnetogram for the same day taken with SDO/HMI reveals the magnetic field in the photosphere (right panel).  

\begin{figure}
  \includegraphics[width=1.9\textwidth,angle=90]{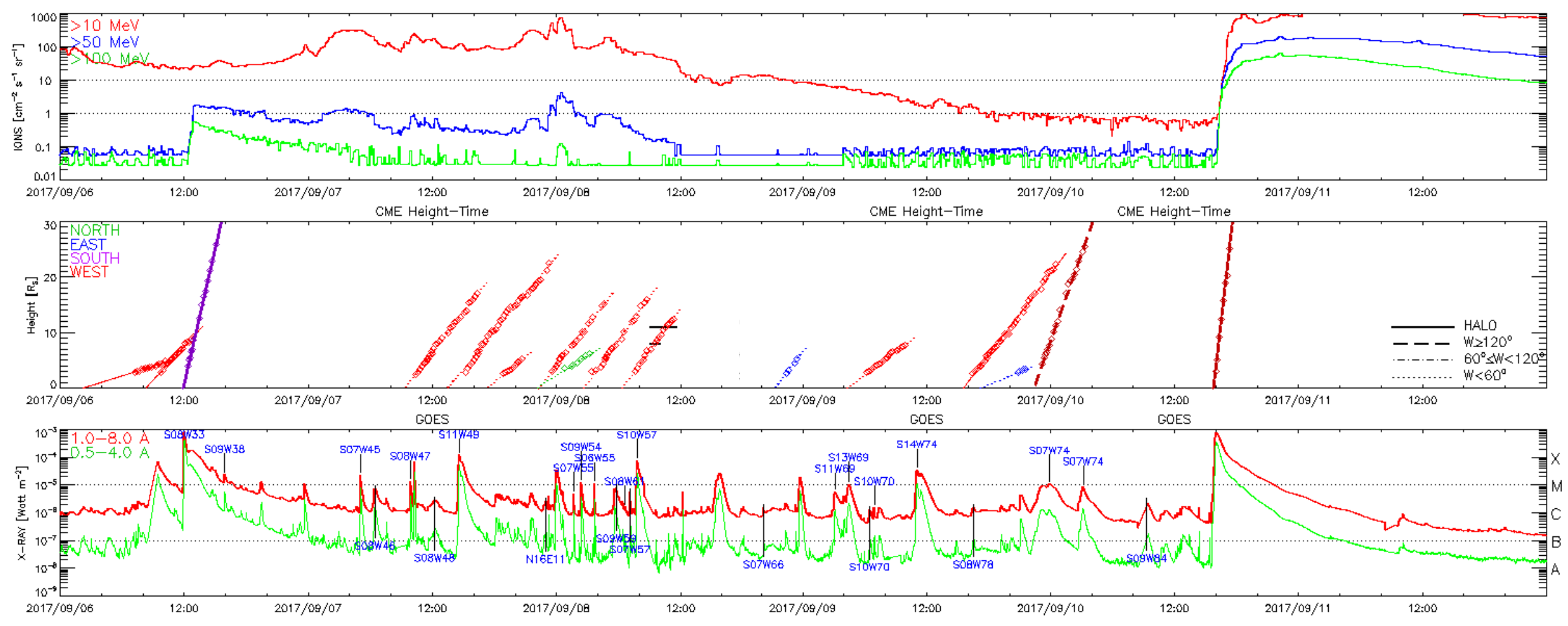}
\caption{September 2017 flare-CME-SEP series taken and adapted from the CDAW catalogue (proton-height/time-X-ray plots - PHTX - can be found under \url{https://cdaw.gsfc.nasa.gov/CME\_list}). Top panel shows solar energetic particle events for protons in the GOES energy channels $>$10, $>$50 and $>$100~MeV. Middle panel gives the CME height-time profile as measured from LASCO (colors give the main propagation direction - see legend to the left in the middle panel). Bottom panel gives the GOES flare SXR emission disk integrated over the wavelength ranges 0.5--4.0 and 1.0--8.00~\AA. }
\label{fig:cdaw}       
\end{figure}

Figure~\ref{fig:cdaw} gives a combined overview on the series of activity pulses covering the two major flare-CME-SEP events during September 6--10, 2017. In total, active region NOAA 12673 produced an intense solar storm period revealing five X-class flares and 39 M-class flares (including the two largest flares from solar cycle 24, the X9.3 flare on September 6, 2017 and the X8.2 flare on September 10, 2017). The first SEP event was measured in the GOES channels at 1AU over September 6, 2017 12:15UT--September 7, 2017 23:25UT and is related to the halo CME that occurred on the Sun on September 6, 2017 12:24UT (first observation in LASCO/C2) with a projected speed of 1570~km/s over the coronagraph field of view. The CME has the source region location coordinates S08W33 and is launched together with a flare that started on September 6, 2017 11:53UT and reached X9.3 class in the measured GOES SXR flux. The second flare-CME-SEP event occurred September 10, 2017 (SEP: September 10, 2017 16:25UT--September 11, 2017 11:40UT; halo CME: September 10, 2017 16:00 with 1490~km/s; flare: X8.2 class on September 10, 2017 15:35UT) with the source region located behind the west limb. For both events long-duration high-energy gamma-ray emission was detected by the Fermi-Large Area Telescope, having durations exceeding 15 hours and with that being the third and fifth largest among all detected \citep[][]{longo17,gopalswamy18,omodei18}.

\begin{figure}
  \includegraphics[width=\textwidth]{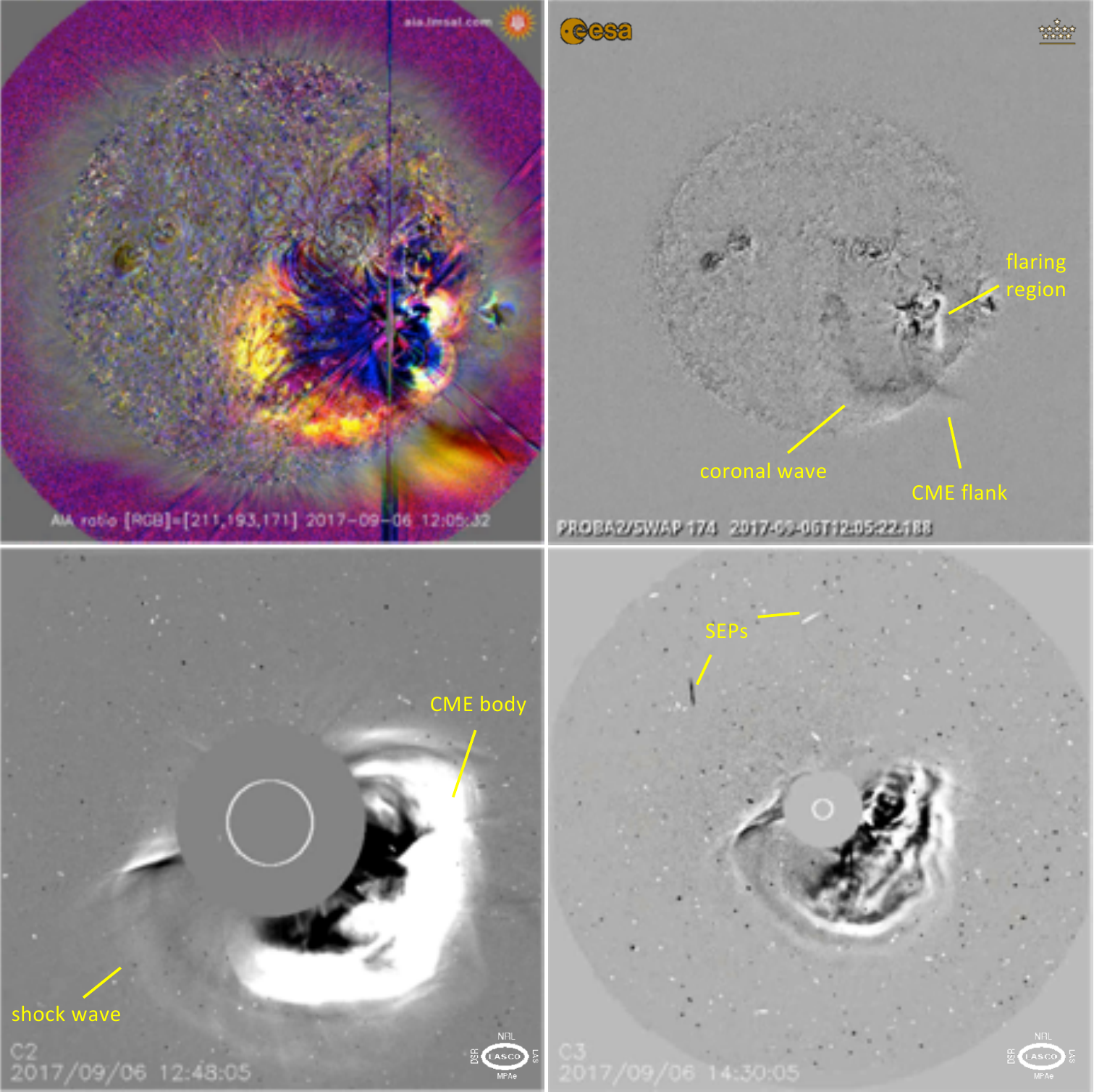}
\caption{Top panels: EUV images from September 6, 2017 observed with SDO AIA (running ratio of composite image data) and Proba-2/SWAP (difference image of 171\AA\ data). Bottom panels: LASCO/C2 and C3 coronagraphs covering a field of view up to 30 solar radii. The generated SEPs are accelerated to relativistic speeds producing spikes in the image data (``snowstorm'' effect). This event was the first flare event in a sequence of X-class flares on 6, 7, and 10 September 2017 causing strong disturbances at Earth and Mars. Note that the field of view of Proba-2 is larger compared to AIA and that due to different CCD techniques no saturation effects are visible in Proba-2. Movies for each panel are available online.}
\label{fig:sep17}       
\end{figure}

Figure~\ref{fig:sep17} impressively shows the manifestation of the eruptive event from September 6, 2017 around 12~UT in solar observations using EUV image and white-light coronagraph data. To make changes visible, different techniques are used, such as a running ratio images as applied on the SDO/AIA composite image and difference images as applied on the Proba-2/SWAP 171\AA\ image or on the LASCO white-light images. SDO and Proba-2 processed images nicely reveal the flare component (X9.3 class), a coronal surface wave that was ignited, and the CME flank on-disk as well above the limb visible in EUV due to compression and heating of the plasma. The off-disk counterpart of the coronal wave is observed in white-light as the shock wave surrounding the CME body. In white-light coronagraph data from LASCO C2 and C3 the CME is detected as partial halo event. The projected CME speed over the field of view up to 20 solar radii was measured from LASCO data with $\sim$1500~km/s. SEPs that are produced by relativistic protons hit the CCD camera of LASCO within minutes and generate the well-known ``snowstorm'' effect. Heavy snowstorms strongly disturb the image quality which may complicate the analysis of halo events (especially during times of increased solar activity).

\begin{figure}
  \includegraphics[width=\textwidth]{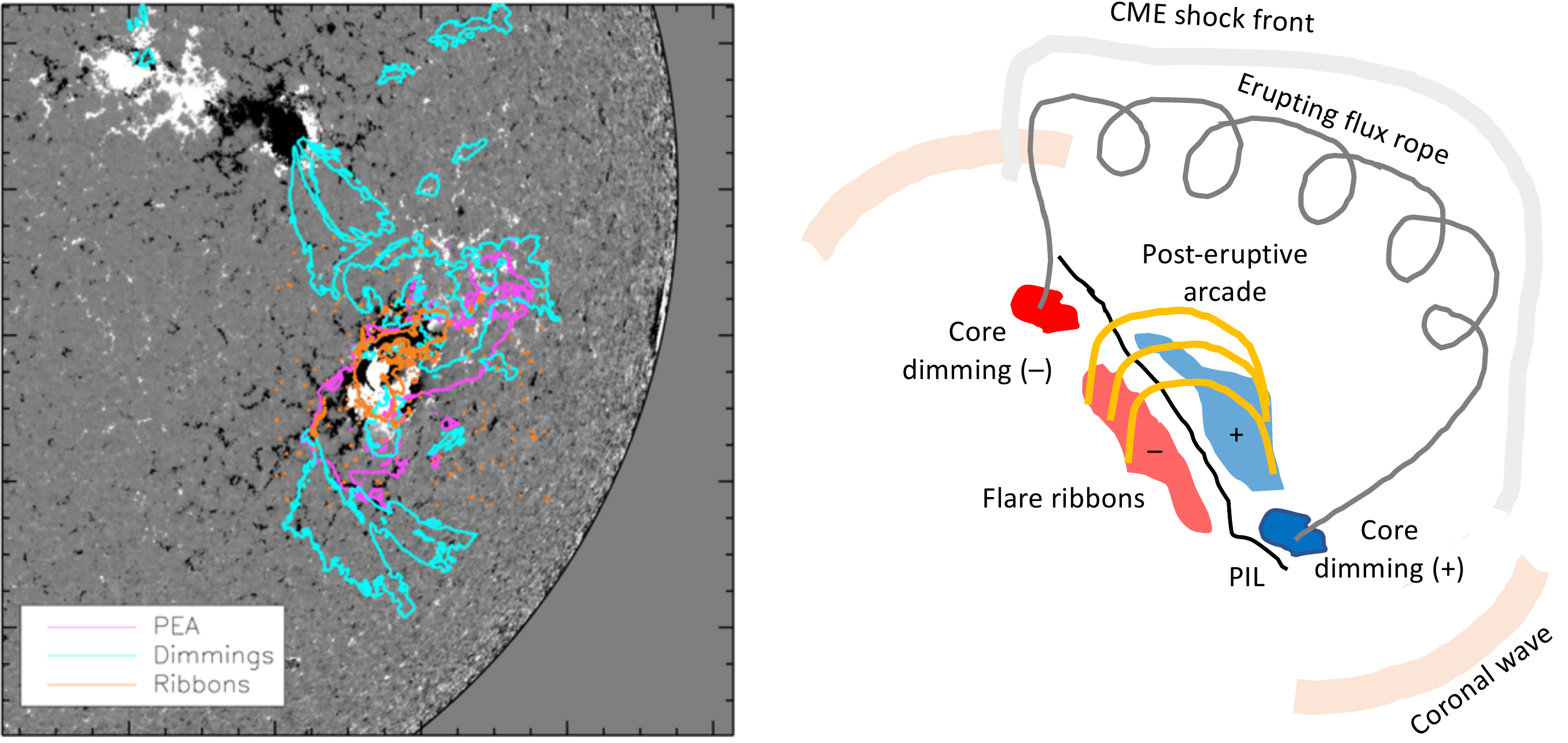}
\caption{Left panel: location of flare ribbons (orange contour), dimming areas (cyan contour) and post-eruptive arcade (PEA, magenta contour) derived for the eruptive event from September 6, 2017. The greyscale background HMI line-of-sight magnetogram is scaled $\pm$100~G with black and white representing negative and positive polarities, respectively \citep[adapted from][]{scolini20}. Right panel: cartoon giving the relation between flare, CME erupting flux rope, coronal wave and dimming areas. }
\label{fig:scolini}       
\end{figure}

Figure~\ref{fig:scolini} shows for the September 6, 2017 event the associated dimming region and post-eruptive arcade together with the magnetic field. The core dimmings are accompanied by secondary dimmings covering larger areas. The cartoon to the right in Figure~\ref{fig:scolini} depicts the relation between different features of an eruptive event and associated solar surface signatures. E.g., measurements of the post-eruptive arcade (PEA) areas and their underlying magnetic field were taken to derive the reconnected flux to feed CME propagation models using magnetized CMEs \citep[][]{scolini20}. The orientation of the PEA with respect to the underlying magnetic field, revealed the handedness of the flux rope. Coronal waves initiated close to the eruption side, hint towards the main compression direction of the eruption, hence, a southward propagation direction of the associated CME (confirmed by the coronagraph images; cf.\,Figure~\ref{fig:sep17}). Therefore, monitoring surface structures gives additional (and early) information about a potential Space Weather event that might affect Earth, and moreover, provides valuable input for modeling efforts.

\begin{figure}
  \includegraphics[width=\textwidth]{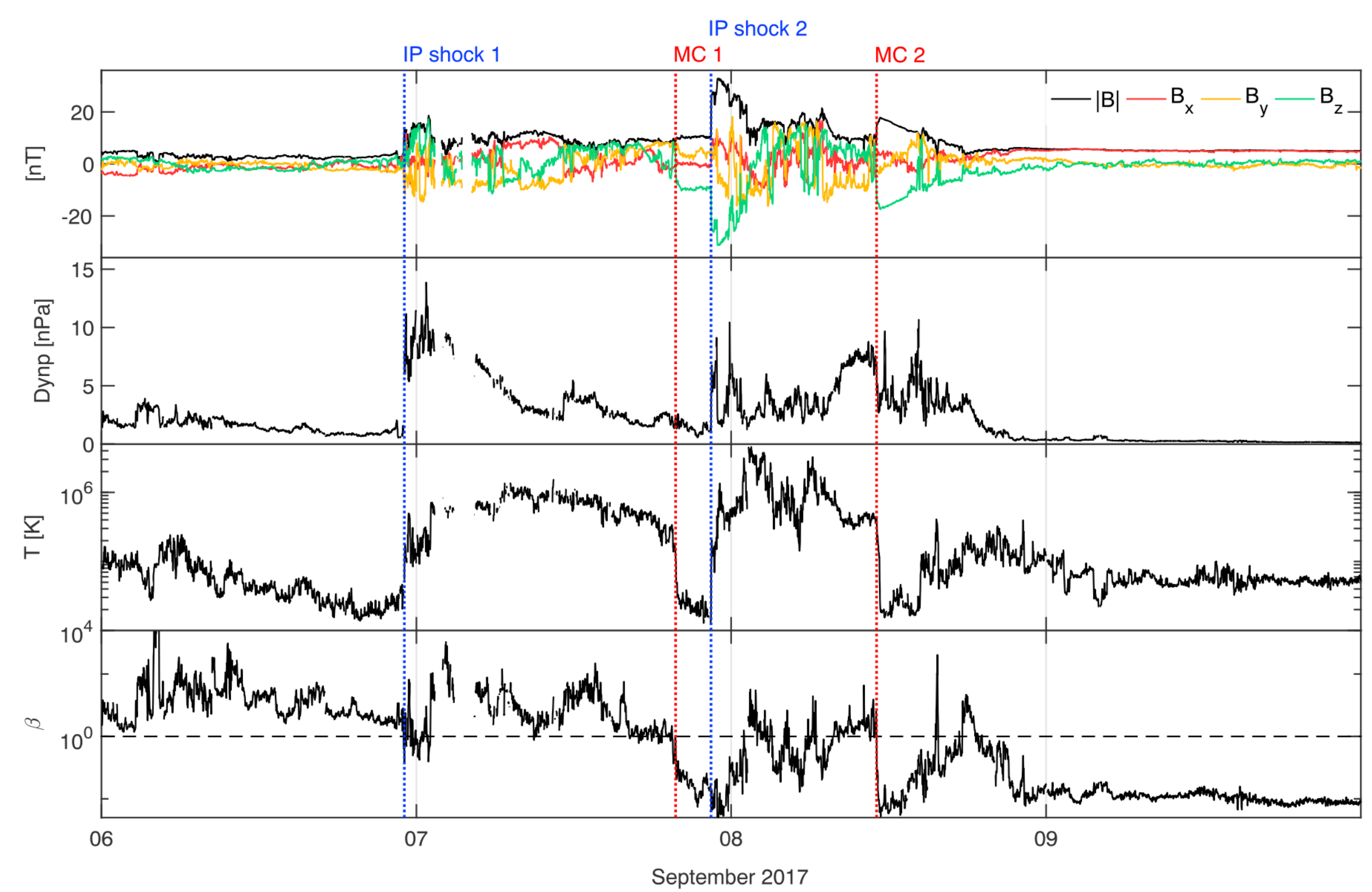}
\caption{Interacting in-situ signatures for the multiple CME events from September 6--9, 2017. Magnetic field and plasma data are given from WIND at L1. Top to bottom: Total magnetic field and vector components, dynamical pressure, proton temperature, and plasma-beta. Taken from \cite{Werner2019ModelingWSA-ENLIL+Cone}. }
\label{fig:werner}       
\end{figure}

The multiple CME events from September 6--9, 2017 preconditioned interplanetary space and were interacting, which intensified their geomagnetic effects. Figure~\ref{fig:werner} shows the complex in-situ signatures revealing multiple shocks and magnetic ejecta regions. Note that the shock of ICME2 propagated into the magnetic structure of ICME1. These so-called ``shock-in-a-cloud'' events are found to cause stronger geomagnetic responses than isolated geoeffective CMEs \citep{lugaz15}. For the September 2017 events, the shock compression might have enhanced the geoeffectiveness by a factor of 2 \citep{shen18}.

\begin{figure*}
  \includegraphics[width=\textwidth]{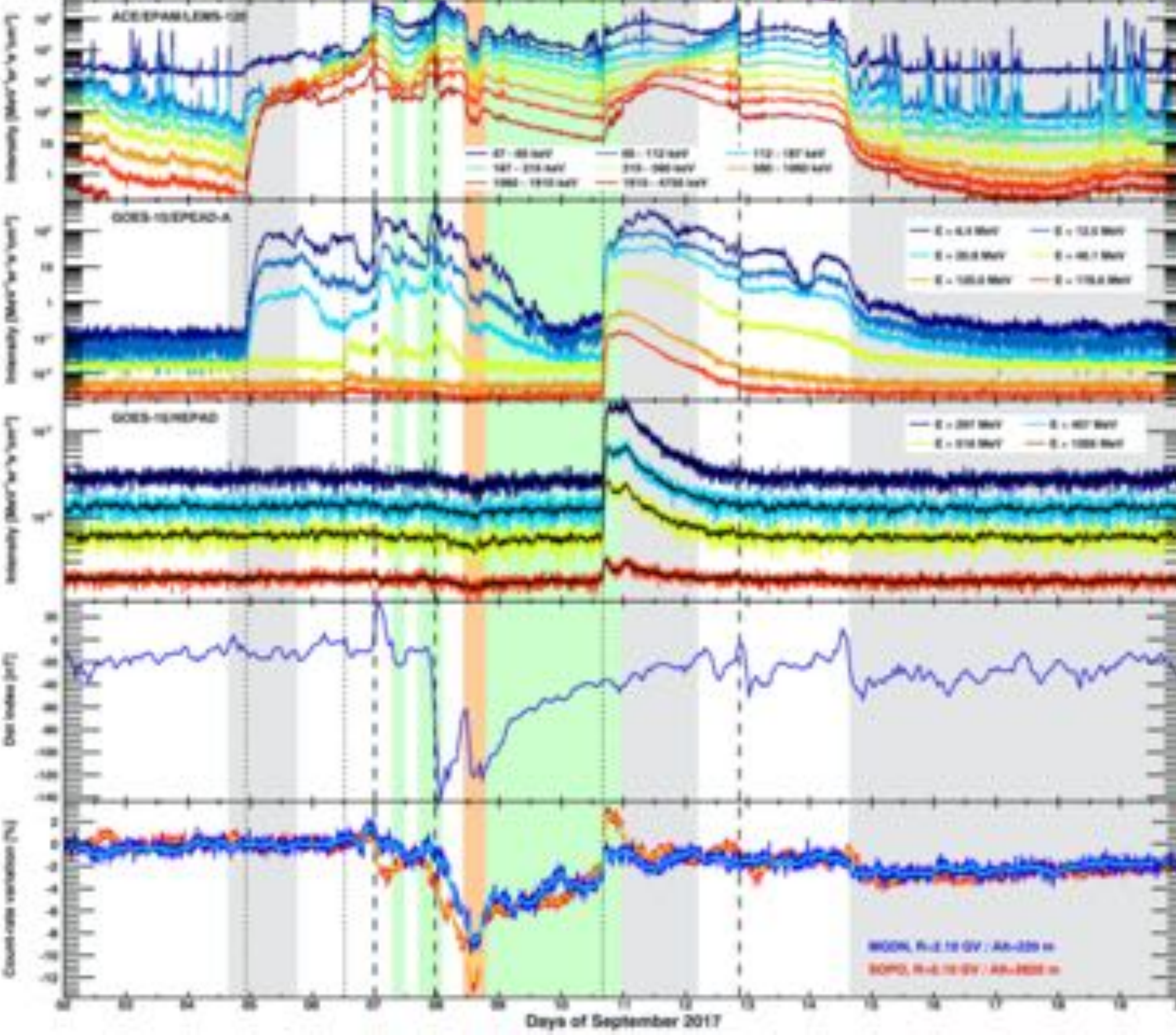}
\caption{Proton intensities measured by ACE/EPAM, GOES/EPEAD, and GOES/HEPAD. Dst index and count rate variations registered by SOPO and MGDN neutron monitor stations. The vertical dotted and dashed lines mark the onset of the SEP events and the time of the shocks, respectively. The green, orange, and gray areas indicate the periods of the ICME, magnetic cloud, and high speed streams, respectively. Taken from \cite{bruno19}.}
\label{fig:bruno19}       
\end{figure*}

As the region rotates close to the West limb of the Sun, the event from September 10 occurred as partially occulted and, because of the favorable magnetic connectivity, generated a GLE. Figure~\ref{fig:bruno19} gives the SEP proton fluxes measured near Earth (ACE, GOES) together with the geomagnetic Dst index and neutron monitor profiles showing the Forbush decrease related to the arrival of the closed magnetic structure \citep[for more details see][]{bruno19}. Due to the eruption site located close to the limb, measurements of the CME kinematics were less strongly affected by projection effects. From EUV observations of the wide field of view Solar Ultraviolet Imager on board the GOES-16 spacecraft exceptionally high values for the CME acceleration and speed were derived, revealing the huge expansion in both, radial and lateral direction \citep{Gopalswamy18_sept17,seaton18,veronig18}. This caused the initiation of a coronal wave propagating over the entire solar surface \citep{liu18}.

The September events did not only affect the Earth, but were also registered at Mars. Energetic particles were observed by MAVEN orbiter, and on the surface by the Curiosity Mars rover \citep[e.g.,][]{hassler18}. Many further aspects of the September 2017 activity period are well represented in a collection of publications in the AGU Space Weather journal\footnote{\url{https://agupubs.onlinelibrary.wiley.com/doi/toc/10.1002/(ISSN)1542-7390.SW-SEPT2017}}.

\section{Space weather forecasting models}\label{sec:6}
``Since all models are wrong the scientist must be alert to what is importantly wrong. It is inappropriate to be concerned about mice when there are tigers abroad.'' by \cite{box76}. \\

For comprehensive investigations of the structure of interplanetary space, propagating transient events, and magnetic connectivity, we need to rely on modeling efforts. Having only scarcely distributed single point in-situ measurements and the lack of plasma and magnetic field information derived by applying remote sensing imaging techniques, we are clearly limited in our ability to assess the performance and reliability of these models hindering improvements. Nevertheless, the close collaboration between model developers and observational community is a need to push forward our understanding of CME-solar wind coupling and interaction processes in interplanetary space. Also on kinetic scales, particle acceleration processes and trigger mechanisms of flare-CME-SEP events need to be supported by modeling, as from observational data the information are not sufficient to make conclusive interpretations about the underlying physics. In recent years substantial progress has been made in efficiently combining models with observations. Data assimilation techniques are known to hugely improve operational forecasting models, as observational data are incorporated in a self consistent way into numerical models for increasing the model accuracy. Examples about the application to solar data can be found in e.g., \cite{Schrijver03,arge10}. In heliospheric physics this method is difficult to apply due to the relatively sparse observations. Recent efforts comprise in-situ measurements from 1~AU to update and improve inner-boundary conditions of solar wind models \citep[see e.g.,][]{lang19}.

Forecasting flares and SEPs with a lead time of at least a few hours is related to the forecasting of the evolution of active regions producing eruptive solar flare events. However, that means to estimate the emergence of magnetic flux from below the photosphere that is not accessible from direct observations. Therefore, statistical relations between photospheric magnetic field characteristics of active regions and flare occurrence are usually used for solar flare forecasting. More precisely, it is the time evolution of the magnetic parameters that plays a major role, but is very complex to derive. The prediction of major flares seems to be more easily achievable compared to flares of medium to low energy, however, the uncertainties coming from the statistical approach are rather large preventing from more accurate single event prediction \citep[see e.g.,][]{schrijver07,georgoulis07,bloomfield12}. Real-time solar flare forecasting is provided by e.g., the EU project Flarecast\footnote{\url{http://flarecast.eu}} with a fully automated system and the NASA/CCMC flare forecasting scoreboard provides a platform to test different forecasting methods fostering further development of the available algorithms\footnote{\url{https://ccmc.gsfc.nasa.gov/challenges/flare.php}}. In general, having standardized metrics is a necessity in order to reliably validate and cross-check the performance of the different flare forecasting tools \citep{barnes08}. See also a review on the origin, early evolution and predictability of solar eruptions by \cite{green18}.

For forecasting CME arrival times and impact, a vast amount of models is currently available in various levels of complexity \citep[see, e.g., Table~1 in][]{riley18}. Simple models, such as empirical relations between CME transit time and speed from statistical results provide tools to derive an average behavior of CME propagation in interplanetary space \citep{Gopalswamy01}. The forecasting accuracy can be significantly improved, when using observational data that can track the CME kinematics to beyond a distance of 50 solar radii (e.g., using image data from STEREO/HI, or radio IPS) where the CME is assumed to evolve in a linear way \citep[see][]{colaninno13,rollett16,hess17}. Several analytical models include the physics of drag force (viscous, aerodynamic, hybrid - that means a linear or quadratic relation between solar wind and CME speed difference) that a CME experiences in interplanetary space. A widely used analytical model is the drag-based-model \citep[DBM; see][]{vrsnak13}. It applies the aerodynamic drag as analogon for the MHD drag force exerted on the CME embedded flux rope. 
More sophisticated are numerical MHD models such as e.g., EUHFORIA \citep{pomoell18_euhforia}, ENLIL \citep{odstrcil99_enlil}, CORHEL \citep{Riley12} or SUSANOO \citep{shiota16}. Besides simulating transient events, MHD models also cover the variation in the background solar wind \citep[see also][]{arge00}. A compilation of data, services and tools can be found at CDPP (Plasma physics data center in France)\footnote{\url{http://cdpp.irap.omp.eu}}. For CMEs, the disentanglement between shock and magnetic ejecta is an important issue highly relevant for forecasting. Numerical models usually use a simple pressure pulse to ignite a CME shock front, but do not include the magnetic structure. There are recent efforts of magnetized CMEs incorporating the observed reconnected magnetic flux at the Sun as model input parameter for the CME flux rope \citep[e.g.][]{Scolini19,singh19}. 

For validation purposes and for increasing the awareness of the limitations of the accuracy of the model results, the uncertainties of observational model input data need to be quantified. \cite{riley18} summarized hit and miss statistics from the CCMC scoreboard\footnote{\url{https://kauai.ccmc.gsfc.nasa.gov/CMEscoreboard/}}, a CME prediction board that gives the possibility to use different models in real-time forecasting (crude facts approach that challenges models and their users). Metadata and metrics are suggested to give the community a common base for an objective inter-comparison of their models \citep{Verbeke19}. A review on the current status and open issues on CME propagation and forecasting methods is given by \cite{vourlidas19}.

Compared to the prediction of CME arrival times and impact speeds, SEP forecasting is a more tricky issue as the accelerated particles propagate with fractions of the speed of light and the lead time is of only a few minutes. Several models are available, mostly using empirical relations based on statistical relations with CME-flare locations or type II radio burst occurrence at decametric–hectometric (DH) wavelengths. Current models cover e.g., PROTONS \citep{balch08}, PPS \citep{kahler07}; ESPERTA \citep{Laurenza09}, FORSPEF \citep{anastasiadis17} as well as physics based models such as e.g., SOLPENCO \citep{Aran06} or HESPERIA \citep[][and references therein]{Malandraki18_book}. Incorporating CME characteristics is difficult as the real-time coronagraph image data, from which usually the CME parameters are derived, are delivered with some delay. Nevertheless, the SEPForecast tool resulting from the EU project COMESEP\footnote{\url{http://comesep.aeronomy.be}} also uses CME parameters as model input \citep{Dierckxsens15}. The application of solar surface proxies for some CME parameters might overcome that drawback. A detailed comparison of false alarm rates for SEP forecasting is presented in e.g., \cite{Alberti17}. The current status of forecasting and nowcasting of SEPs and open questions is given in a review by \cite{Anastasiadis19}.

\begin{figure*}
  \includegraphics[width=1.\textwidth]{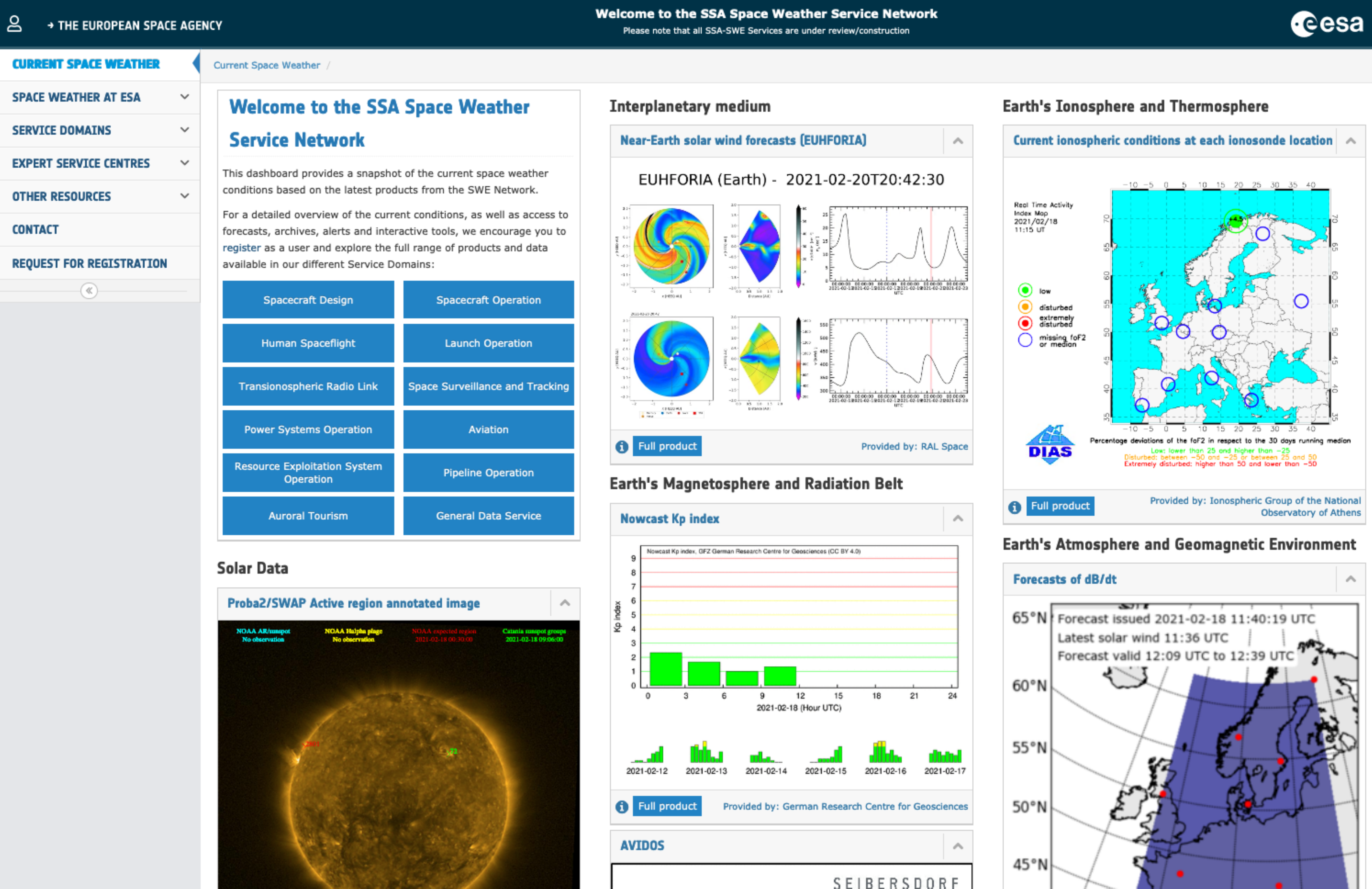}
\caption{Webpage of the SSA Space Weather Service Network where Expert Service Centers provide their tools and services covering Space Weather forecast and overview from the solar-heliosphere perspective to space radiation, ionospheric and geomagnetic conditions. ESA/SSA: \url{http://swe.ssa.esa.int}}
\label{fig:ssa}       
\end{figure*}

Further approaches for improving forecasting purposes also cover ensemble models incorporating the uncertainties in the observational data \citep[e.g.,][]{lee13,mays15,Dumbovic2018ThePropagation,Amerstorfer2018EnsembleImagers}, and machine learning techniques \citep[see e.g.,][]{camporeale19}. In addition to methods covering large statistics, case studies that model Space Weather events from Sun to Earth, give a wealth of detailed information from which we can hugely improve our understanding of flare-CME-SEP events. Community centers like ESA/VSWMC\footnote{\url{https://esa-vswmc.eu}} or NASA/CCMC\footnote{\url{https://ccmc.gsfc.nasa.gov}} cover the increased need of computational power and appropriate IT infrastructure and provide a platform for models to be tested and actually used. Such platforms are also the driveway for R2O (research to operation) activities and where scientists and users meet. A collection of Space Weather tools is presented at the ESA/SSA website, including services from the European Expert Service Centers and their individual Expert Groups (cf.\,Figure~\ref{fig:ssa}). 

\section{Concluding remarks}
Nowadays, the term Space Weather covers basic research as well as application and is a platform for strong interdisciplinary research bringing the domains of solar-, heliospheric, and geo-physics closer together. The real-time forecasting results of flare-CME-CIR-SEP events at Earth using operational tools reveal that we still face large uncertainties. The reason is on the one hand the huge complexity of the solar phenomena and their unknown coupling processes and propagation behavior in interplanetary space. On the other hand, there is a clear lack of accurate enough measurements to properly feed the available models. The source of high errors and inaccuracies in the measurements is due to extraction of remote sensing data which are affected by projection effects, idealized assumptions of magnetic field extrapolations (lack of coronal magnetic field measurements) and 3D reconstructions, as well as localized in-situ measurements making it tricky to validate all the results. For understanding and forecasting the geoeffectiveness of Earth affecting Space Weather events, a far bigger challenge for the future is the modeling of the interplanetary magnetic field orientation and variation.  


A clear advantage is given by the combination of several instruments having multiple views on the Sun covering different longitudes and latitudes. STEREO-A is still operational providing data material from varying angles. PSP will be the first spacecraft that comes as close to the Sun as 10 solar radii and aims to answer questions about the origin of the solar wind, how is it accelerated or about SEP acceleration and transport processes. The mission already gathered unprecedented data enabling new interpretation and finding on the source regions of the solar wind. Solar Orbiter revealed first data in July 2020 and will study the solar surface and its magnetic field in great detail. The spacecraft will have a highly elliptical orbit to progressively move to a more inclined orbit out of ecliptic. That perspective will give new insight onto the polar regions of the Sun and is expected to improve magnetic field modeling and to better understand the solar wind from coronal holes in the polar regions. 
Conjunctions with other active missions (such as STEREO, L1 missions, MAVEN), complemented by ground-based instruments, will give unique possibilities to investigate the evolution of solar flares, CMEs, or SEPs, as well as coronal holes and solar wind high speed streams. Future missions, such as the ESA \textit{Lagrange} mission to L5 (launch is planned for 2027) or polar missions, as the proposed Solaris Solar Polar Mission \citep{hassler20}, will provide more valuable data and fill the gaps in our understanding of magnetic field connectivity and coupling processes between open and closed magnetic field structures in interplanetary space.

We are facing challenging and exciting times having a wealth of data at hand and promising future missions yet to come. Many of the currently available data are well prepared in ready-to-use catalogues, waiting to be explored. However, we still need to obtain data from L4/L5 Lagrange points revealing the necessary side views on a regular basis to fully track eruptions with low projection effects and for gaining a better understanding of their characteristics that affect Earth. Moreover, we need to improve solar wind models to be used in MHD simulations for properly determining CME propagation. With that we will successfully enhance the reliability of Space Weather forecasting tools and models.

\begin{acknowledgements}
I would like to thank Stephan G.\,Heinemann for his support in generating the bibtex entries. 

Some images were created using the ESA and NASA funded Helioviewer Project. LASCO Images courtesy of SOHO/LASCO consortium. SOHO is a project of international cooperation between ESA and NASA. SDO Images courtesy of SDO/AIA consortium. STEREO Images courtesy of STEREO/SECCHI consortium. SWAP Images courtesy of PROBA2/SWAP consortium. PROBA2 is a project of the Centre Spatial de Li{\`e}ge and the Royal Observatory of Belgium funded by the Belgian Federal Science Policy Office (BELSPO) and ESA. Images reproduced with permission ??

\end{acknowledgements}

%
\section*{Conflict of interest}
The authors declare that they have no conflict of interest.

\bibliographystyle{spbasic_FS}      


\end{document}